\documentclass[iop,apj,numberedappendix]{emulateapj}

\usepackage{wasysym}
\usepackage{rotating}
\usepackage{afterpage}
\usepackage{xcolor}
\usepackage{txfonts}
\usepackage{graphicx}
\usepackage{environ}
\NewEnviron{myresizeenv}{\resizebox{\linewidth}{!}{\BODY}}

\usepackage[]{natbib}
\bibpunct{(}{)}{;}{a}{}{,}

\newcommand{\um}{$\mu$m}

\newcommand{\kms}{km\thinspace s$^{-1}$}

\def\utw{\smash{\rlap{\lower5pt\hbox{$\sim$}}}}
\def\udtw{\smash{\rlap{\lower6pt\hbox{$\approx$}}}}

\def\farcs{\hbox{$.\!\!^{\prime\prime}$}}

\def\Lsun{\hbox{\it L$_\odot$}}

\def\Teff{\hbox{\it T$_{\rm eff}$}}

\def\Msun{\hbox{\it M$_\odot$}}

\def\Mbol{\hbox{\it M$_{bol}$}}
\def\mbolint{\hbox{\rm m$_{\rm bol}^{\rm int}$}}
\def\mbolbc{\hbox{\rm m$_{\rm bol}^{\rm BC_K}$}}

\newcommand{\Ks}{{\it K$_{\rm s}$}}

\newcommand{\Aks}{{\it A$_{\rm K_{\rm s}}$}}

\newcommand{\Av}{{\it A$_{\rm V}$}}

\def\BCK{\hbox{\it BC$_{\rm K}$}}

\def\ew{\hbox{\rm EW}}
\def\ewmg{\hbox{\rm $[{\rm Mg~I,1.71}]$}}
\def\ewca{\hbox{\rm $[{\rm Ca~I,2.26}]$}}
\def\ewna{\hbox{\rm ${[\rm Na~I,2.21}]$}}
\def\ewal{\hbox{\rm $[{\rm Al~I,1.67}]$}}
\def\ewco{\hbox{\rm $[{\rm CO}]$}}
\def\ewcodue{\hbox{\rm $[{\rm CO,2.29}]$}}
\def\ewcoucs{\hbox{\rm $[{\rm CO,1.57}]$}}
\def\ewcousd{\hbox{\rm $[{\rm CO,1.62}]$}}
\def\ewsiucn{\hbox{\rm $[{\rm Si~I,1.59}]$}}
\def\ewsiusq{\hbox{\rm ${[\rm Si~I,1.64}]$}}
\def\ewCaT{\hbox{\rm $[{\rm CaT}]$}}
\def\ewcau{\hbox{\rm $[{\rm Ca1}]$}}
\def\ewcad{\hbox{\rm $[{\rm Ca2}]$}}
\def\ewcat{\hbox{\rm $[{\rm Ca3}]$}}
\def\ewJ12J21J22{\hbox{$\rm [{\rm J12}$+$\rm {\rm J21}$+$\rm {\rm J22}]$}}
\def\ewJudJduJdd{\hbox{$\rm [{\rm J12}$+$\rm {\rm J21}$+$\rm {\rm J22}]$}}

\def\ewJ8J9J10{\hbox{$\rm [{\rm J8}$+$\rm {\rm J9}$+$\rm {\rm J10}]$}}
\def\ewJoJnJd{\hbox{$\rm [{\rm J8}$+$\rm {\rm J9}$+$\rm {\rm J10}]$}}

\def\ewJ2J4J6{\hbox{$\rm [{\rm J2}$+$\rm {\rm J4}$+$\rm {\rm J6}]$}}

\def\ewJ2J4J8{\hbox{$\rm [{\rm J2}$+$\rm {\rm J4}$+$\rm {\rm J8}]$}}
\def\ewJdJqJo{\hbox{$\rm [{\rm J2}$+$\rm {\rm J4}$+$\rm {\rm J8}]$}}

\def\ewJ12J17J18{\hbox{$\rm [{\rm J12}$+$\rm {\rm J17}$+$\rm {\rm J18}]$}}
\def\ewJudJusJuo{\hbox{$\rm [{\rm J12}$+$\rm {\rm J17}$+$\rm {\rm J18}]$}}

\def\ewJ21J22{\hbox{$\rm [{\rm J21}$+$\rm {\rm J22}]$}}
\def\ewJduJdd{\hbox{$\rm [{\rm J21}$+$\rm {\rm J22}]$}}

\def\ewJuno{\hbox{\rm $[{\rm J1}]$}}
\def\ewJdue{\hbox{\rm $[{\rm J2}]$}}
\def\ewJ3{\hbox{\rm $[{\rm J3}]$}}
\def\ewJtre{\hbox{\rm $[{\rm J3}]$}}
\def\ewJquattro{\hbox{\rm $[{\rm J4}]$}}
\def\ewJcinque{\hbox{\rm $[{\rm J5}]$}}
\def\ewJ6{\hbox{\rm $[{\rm J6}]$}}
\def\ewJsei{\hbox{\rm $[{\rm J6}]$}}
\def\ewJsette{\hbox{\rm $[{\rm J7}]$}}
\def\ewJotto{\hbox{\rm $[{\rm J8}]$}}
\def\ewJnove{\hbox{\rm $[{\rm J9}]$}}
\def\ewJ10{\hbox{\rm $[{\rm J10}]$}}
\def\ewJdieci{\hbox{\rm $[{\rm J10}]$}}
\def\ewJundici{\hbox{\rm $[{\rm J11}]$}}
\def\ewJdodici{\hbox{\rm $[{\rm J12}]$}}
\def\ewJutre{\hbox{\rm $[{\rm J13}]$}}
\def\ewJuquattro{\hbox{\rm $[{\rm J14}]$}}
\def\ewJucinque{\hbox{\rm $[{\rm J15}]$}}
\def\ewJusei{\hbox{\rm $[{\rm J16}]$}}
\def\ewJusette{\hbox{\rm $[{\rm J17}]$}}
\def\ewJuotto{\hbox{\rm $[{\rm J18}]$}}
\def\ewJunove{\hbox{\rm $[{\rm J19}]$}}
\def\ewJdzero{\hbox{\rm $[{\rm J20}]$}}
\def\ewJ21{\hbox{\rm $[{\rm J21}]$}}
\def\ewJdu{\hbox{\rm $[{\rm J21}]$}}
\def\ewJ22{\hbox{\rm $[{\rm J22}]$}}
\def\ewJdd{\hbox{\rm $[{\rm J22}]$}}
\def\ewJdtre{\hbox{\rm $[{\rm J23}]$}}
\def\ewJdquattro{\hbox{\rm $[{\rm J24}]$}}
\def\ewJdcinque{\hbox{\rm $[{\rm J25}]$}}

\def\Vlsr{\hbox{V$_{\rm LSR}$}}

\def\tcombi{\hbox{\it T$_{\rm combi}$}}
\def\ttio{\hbox{\it T$_{\rm TiO}$}}
\def\spcombi{\hbox{Sp$_{\rm combi}$}}
\def\indwater{\hbox{\rm ind$_{\rm H_2O}$}}

\shorttitle{Infrared spectral indices for cold stars}
\shortauthors{Messineo et al.}

\usepackage[utf8]{inputenc}

\usepackage{lineno}
\RequirePackage{rotating}
\begin{document}


\title{New infrared spectral indices of luminous cold stars: from early K to M-types.}


\author{Maria~Messineo\altaffilmark{1}, 
        Donald F. Figer\altaffilmark{2},
        Rolf-Peter ~Kudritzki\altaffilmark{3},\\
        Qingfeng Zhu \altaffilmark{1},
        Karl M. Menten \altaffilmark{4},
        Valentin D. Ivanov \altaffilmark{5},
        C.-H. Rosie Chen   \altaffilmark{4}
}

\altaffiltext{1}{Key Laboratory for Researches in Galaxies and Cosmology, University of Science and Technology of China, 
Chinese Academy of Sciences, Hefei, Anhui, 230026, China
\email{messineo@ustc.edu.cn}
}

\altaffiltext{2}{Center for Detectors, Rochester Institute of Technology, 54 Memorial Drive, Rochester, NY 14623, USA	}

\altaffiltext{3}{Institute for Astronomy, University of Hawaii, 2680 Woodlawn Drive, Honolulu, HI 96822	}

\altaffiltext{4}{Max-Planck-Institut f\"ur Radioastronomie, Auf dem H\"ugel 69, D-53121 Bonn, Germany	}


\altaffiltext{5}{European Southern Observatory, Karl Schwarzschild-Strasse 2, D-85748 Garching bei Munchen, Germany	}

\begin{abstract} 
We present infrared spectral indices (1.0-2.3 \um)  of Galactic 
late-type giants and red supergiants (RSGs). 
We used existing and new spectra obtained at resolution 
power R=2000 with SpeX on the IRTF telescope. 
While a large CO equivalent width (\ew), at 2.29 \um\ 
(\ewcodue$\ga$45 \AA) is a typical signature of RSGs later 
than spectral type M0, \ewco\ of K-type RSGs and giants 
are similar. In the  \ewcodue\ versus 
\ewmg\ diagram,  RSGs of all spectral types 
can be distinguished from red giants, because 
the Mg I line  weakens with increasing temperature and decreasing 
gravity. We find several lines that vary with luminosity, 
but not temperature: Si I (1.59 \um), Sr (1.033 \um), 
Fe+Cr+Si+CN (1.16 \um), Fe+Ti (1.185 \um), Fe+Ti (1.196 \um), 
Ti+Ca (1.28 \um), and Mn (1.29 \um). Good markers of CN enhancement 
are the Fe+Si+CN line at 1.087 \um\ and CN line at 1.093 \um. 
Using these lines, at the resolution of SpeX, it is possible 
to separate RSGs and giants. Contaminant O-rich Mira and S-type 
AGBs are recognized by strong molecular features due to water 
vapor features, TiO band heads, and/or ZrO absorption. 
Among the 42 candidate RSGs that we observed, all but one were 
found to be late-types. 21 have EWs consistent with those of RSGs, 
16 with those of O-rich Mira AGBs, and one with an S-type AGB. 
These infrared results  open new, unexplored, potential for 
searches at low-resolution of RSGs in the highly obscured 
innermost regions of the Milky Way. 
\end{abstract}





\section{Introduction}

Stars with initial masses from  8 to  35 \Msun\
evolve through the red supergiant (RSG)  phase
\citep[e.g.][]{chieffi13}.

Stellar evolution models suggest that the upper mass 
limit for RSGs is 25 \Msun\ for rotating stars and 40 \Msun\ 
for non-rotating stars \citep[e.g.,][]{limongi17}. 
Massive stars with luminosities from 10,000 to 400,000 
\Lsun\ enter this phase when their envelopes expand and 
their effective temperatures drop below 4500 K.
Because of their intrinsically high luminosity and red colors, 
they can be seen at large distances and through large columns of dust, 
making them good tracers of galactic disk morphology and kinematics. 
The spectra of RSGs are rich in molecular absorptions and atomic lines; 
for example, Mg, Fe, Al, Si, Ti, Cr, Mn, and Sr lines. 
These lines may serve to map metallicity 
in  galactic disks  at large 
extragalactic distances \citep{davies17,lardo15}, 
and to constrain the uncertain upper part of the stellar mass function 
(poorly populated because these stars are short-lived). 
Detections of Galactic RSGs are hampered by our position in the disk, 
the patchiness of interstellar extinction, unknown distances, and the 
difficulty of separating giants from supergiants \citep[e.g.][]{messineo17}.

 About 500 spectroscopic RSGs are listed in optical catalogs, 
 and 300 in infrared catalogs, see for example, \citet{messineo19}, 
\citet{skiff14}, \citet{humphreys78}, \citet{elias85}, 
\citet{jura90}, \citet{figer06}, \citet{davies07}, \citet{clark09}, 
\citet{negueruela11}.
 
 It appears that we know only about 10\%
of this important Galactic population. 
\citet[][]{gehrz89} estimated the existence of 5000 RSGs
in the Milky Way and this number is well consistent 
with the numbers of RSGs
recently discovered in the M31 ($>6412$) and  M33 ($>2858$) 
spiral galaxies
\citep[][]{massey20}. 

Improved searches  are of primary importance for Galactic studies,
The narrow range of ages spanned by RSGs (4-30 Myr) and their youth 
may help to trace Galactic structure \citep[][]{massey20}, 
as RSGs are useful for tracing the spiral
structure, to estimate the star formation rate at the 
epoch of their formation, 
and to model the history of Galactic chemical enrichment 
with their yields. 

We are living in a golden age of Galactic astronomy. 
Gaia is scanning the entire Galaxy,  measuring parallaxes, 
radial velocities, and spectral types for 1.3 billion stars 
\citep{gaiadr3}. 
Galactic multi-wavelength photometric surveys with sub-arcsec 
spatial resolution are already available and several 
high-resolution spectroscopic atlases are expected soon. 
\citet{park18} presented an atlas of $\approx 80$ spectra covering 
$H$- and $K$-bands at R=45,000, containing dwarfs, giants, and 
supergiants with spectral types from O to M. More recently, 
an ESO/Xshooter Atlas of 754 stars at R=10,000 covering 300 
nm to 2480 nm has been released by \citet{gonneau20}. 
The atlas includes hot and cool stars spanning a wide range 
of gravity. The ongoing GALAH survey with the HERMES 
instrument on the Anglo-Australian Telescope will 
release more than 1 million spectra of Galactic southern 
stars at R=28,000 in the 4700 to 7600 nm range for 
chemically tagging the Milky Way \citep[e.g.][]{sharma20}. 
Starting in 2023, 4MOST on the ESO/Vista telescope will 
release 20 million spectra at R$\approx$5000 (from 390 to 1000 nm) 
and R$\approx$20,000 (from 395 to 456.5 nm \& from 587 to 673 nm) 
of southern stars \citep[e.g.,][]{dejong12}. 

The high-resolution spectra already available have allowed 
us to better characterize RSG stars with chemical features, 
abundance and temperature determinations. New spectral modeling 
identifies bands and lines suitable for parameter determinations. 
Many recent studies use infrared lines to accurately estimate 
temperature, gravity, and metallicity 
\citep{taniguchi20,park18,patrick17,davies17,davies15,patrick15,lardo15,gazak15,gazak14,gazak14b}. 
Despite ongoing revision of the 
temperature scale, based on the TiO molecular bands 
\citep[][]{levesque05}, temperatures can already 
be inferred with an accuracy of about 150 K. Spectroscopic 
monitoring for detecting binary RSGs has also been started 
\citep[e.g.][]{jesus19}, while Gaia data allow us 
to identify binary RSGs with proper motion studies \citep[][]{kervella19}.  

Low-resolution infrared spectroscopy is  also 
important for 
i) efficiently identifying good targets for high-resolution follow-up, 
achieved by separating rare RSGs from the bulk of giant stars; 
ii) estimating parameters; 
and iii)  delineating the morphology of 
the inner regions of the Galaxy.  The first quantitative analysis 
of infrared spectra was reported by \citet[][$K$-band]{kleinmann86}, 
\citet[][$H$-band]{origlia93}, \citet[][$K$-band]{blum96} 
and \citet[][$J$-band]{joyce98}, with line identifications, 
index definitions,  and relationships between line strengths 
and stellar physical parameters. These early studies used 
infrared spectra to estimate luminosity classes and spectral 
types within a few subtypes. Low resolution spectroscopic 
libraries have been available since the beginning of 
this new millenium, e.g.\ \citet[$H$-band]{meyer98}, 
\citet[0.5-2.5 \um, at R=1100]{lancon00}, 
of \citet[$H$- and $K$-bands at R $\approx$ 2000--3000,][]{ivanov04}. 
Unfortunately, indices were only sufficient to 
separate dwarfs from giants/supergiants. 
The library of cool stars of \citet{rayner09} 
made with SpeX \citep{rayner03} on the 
NASA InfraRed Telescope Facility (IRTF) covers 0.8 to 2.4 
\um\ at R$\approx 2000$ with an unprecedented high throughput. 
The 210 IRTF spectra have extensively been used as an 
empirical library for building indices useful for simple 
stellar population models \citep[e.g.][]{morelli20,cesetti13}; 
 \citet{davies10} analyzed the $J$-band portion of the spectra of 
RSGs as a metallicity diagnostic. We use the library to confirm 
the detection of new RSGs, and how to more precisely distinguish 
them from other Galactic cool stars. 

Usually, obscured RSGs are classified by their broad CO band heads at 
2.293 \um\ and by the shape of their stellar continuum. 
The equivalent width (EW) of the CO bands is an estimator of 
stellar temperature; however, the \ew s of CO of giants and 
supergiants follow two different relations, with the supergiants 
spanning a larger range   \citep{blum03}. Unfortunately, 
this infrared technique allows us only to identify RSGs with 
spectral types later than M0. 

Infrared spectra from the IRTF library suggest that RSGs 
and giants occupy different regions in  \ewcodue\ 
(thereafter  we refer to the \ew\ of a specific line
by enclosing  the element or molecule name and wavelength within brackets)
versus \ewmg\ diagram down to early K-type stars 
\citep[][]{messineo17}. However, only five RSGs of early 
K-type were available in the IRTF library. 
In this paper, we report observations with SpeX of 
known K- and M-type RSGs and candidate RSGs, 
 and use them to define infrared diagnostics for 
assigning spectral types and luminosity classes. 
Only empirical quantities are presented. 
Quantities to be inferred with a comparison to synthetic spectra,
 e.g. metallicities,
will be presented elsewhere.
In Sect. \ref{obs}, we describe the  spectroscopic data, in Sect. \ref{index} 
the adopted infrared  spectral indices. 
In Sect.\  \ref{secgravity} 
we focus on the stellar luminosity of RSGs and infrared indices that correlate with it.
In Sect. \ref{magnesiumsec}, we describe  the \ewcodue\  versus \ewmg\  diagram
to separate giants and RSGs, while in Sect.\
\ref{sectemperature} we derive temperature estimates from $HK$ indices.
In Sec. \ref{metsec} we briefly describe what is known on  metallicity of  galactic RSGs. 
A few comments on specific sources are given in Sect.
\ref{notes}.
The new targets are spectroscopically classified 
in Sect. \ref{classtarget}.
Finally, in Sect. \ref{result}, we provide a 
summary and remarks  on the usefulness of these 
indices as a luminosity diagnostic to identify  RSGs from K0 
to late M-types and to spectroscopically distinguish them 
from giants.

\section{Spectroscopic infrared data with SpeX}
\label{obs}

In this work, we present new spectra taken with SpeX of 72 bright late-type stars.
To increase the statistical significance, 
we analyse the new SpeX spectra along with those spectra of bright 
late-type stars
included in the SpeX libraries of \citet{rayner09} and \citet{villaume17}.
These two libraries include 24 K-M stars of class I.

\begin{figure*}
\begin{center}
\resizebox{0.9\hsize}{!}{\includegraphics[angle=0]{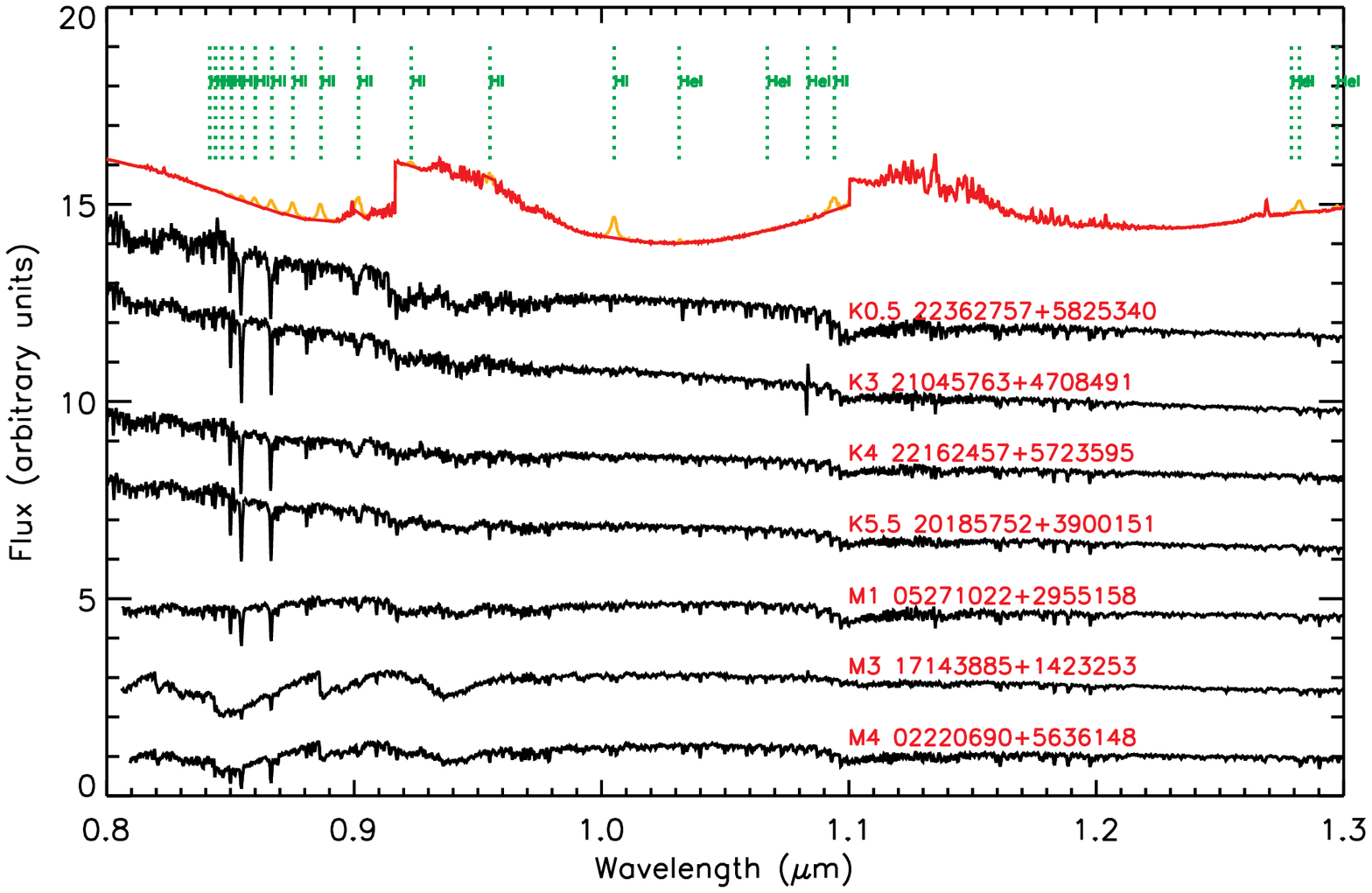}}
\caption{ \label{examplespectra} Some examples of spectra of 
RSGs taken with SpeX  (in black). 
Over-plotted in red (clean) and orange (with H and He I lines still not subtracted) 
is an example of the instrumental response and atmospheric transmission curve.
The locations of the $H$ I and $He$ I lines are marked with the green 
dotted-vertical lines.  The spectra are corrected for reddening 
and normalized to the mean flux density within the range from 1.1 \um\ to 1.25 \um.
The SpeX spectra of stars in Tables \ref{aliaskrsg},
\ref{aliasnewrsg}, and \ref{aliasgiants}
are displayed in  Appendix  Fig.  \ref{appendixspectra}. }
\end{center}
\end{figure*}

\addtocounter{figure}{-1}
\begin{figure*}
\begin{center}
\resizebox{0.9\hsize}{!}{\includegraphics[angle=0]{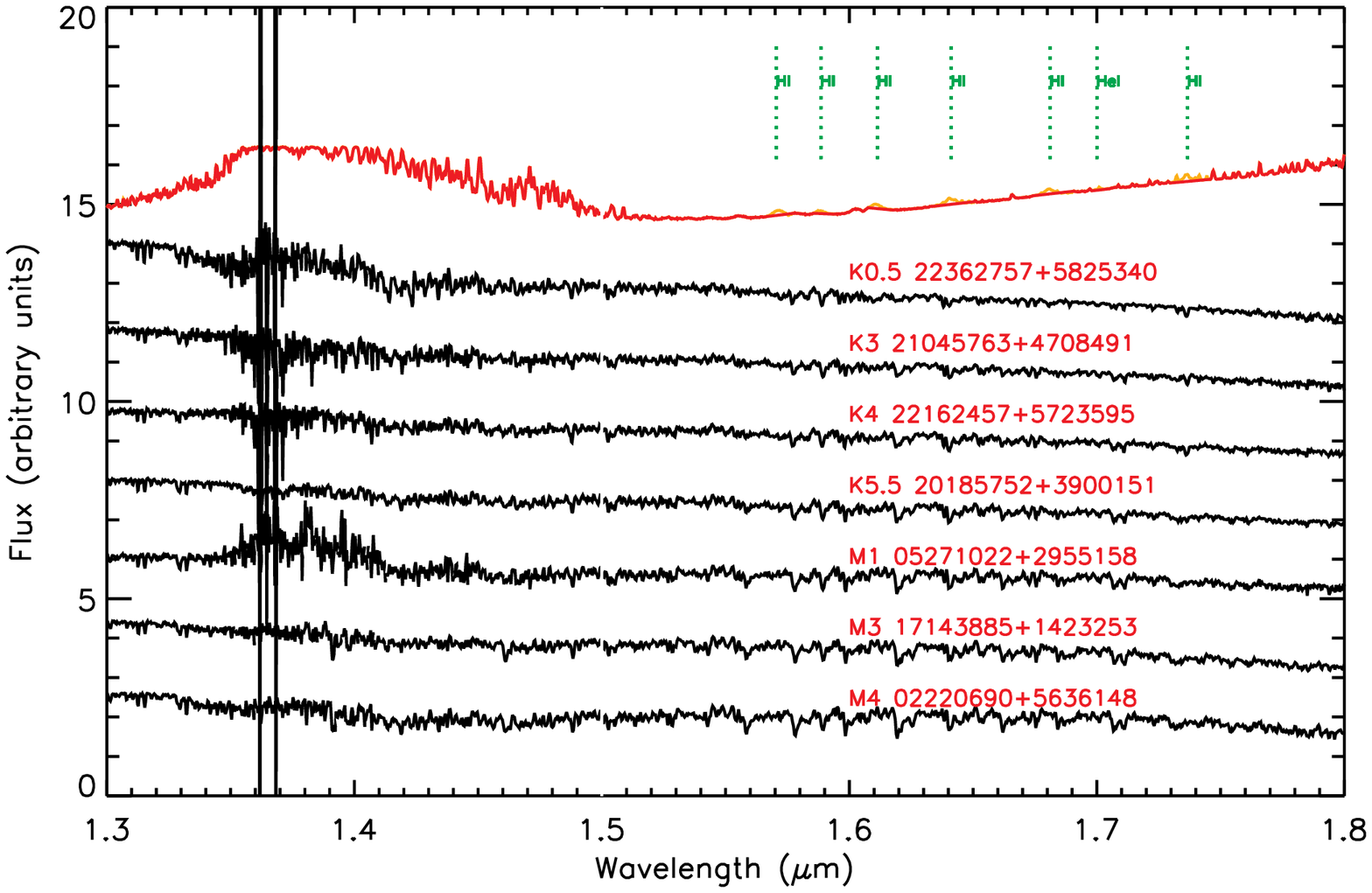}}
\caption{ Continuation of Fig. \ref{examplespectra}.  Here, the spectra are 
normalized to the mean flux density from 1.5 \um\ to 1.6 \um. } 
\end{center}
\end{figure*}

\addtocounter{figure}{-1}
\begin{figure*}
\begin{center}
\resizebox{0.9\hsize}{!}{\includegraphics[angle=0]{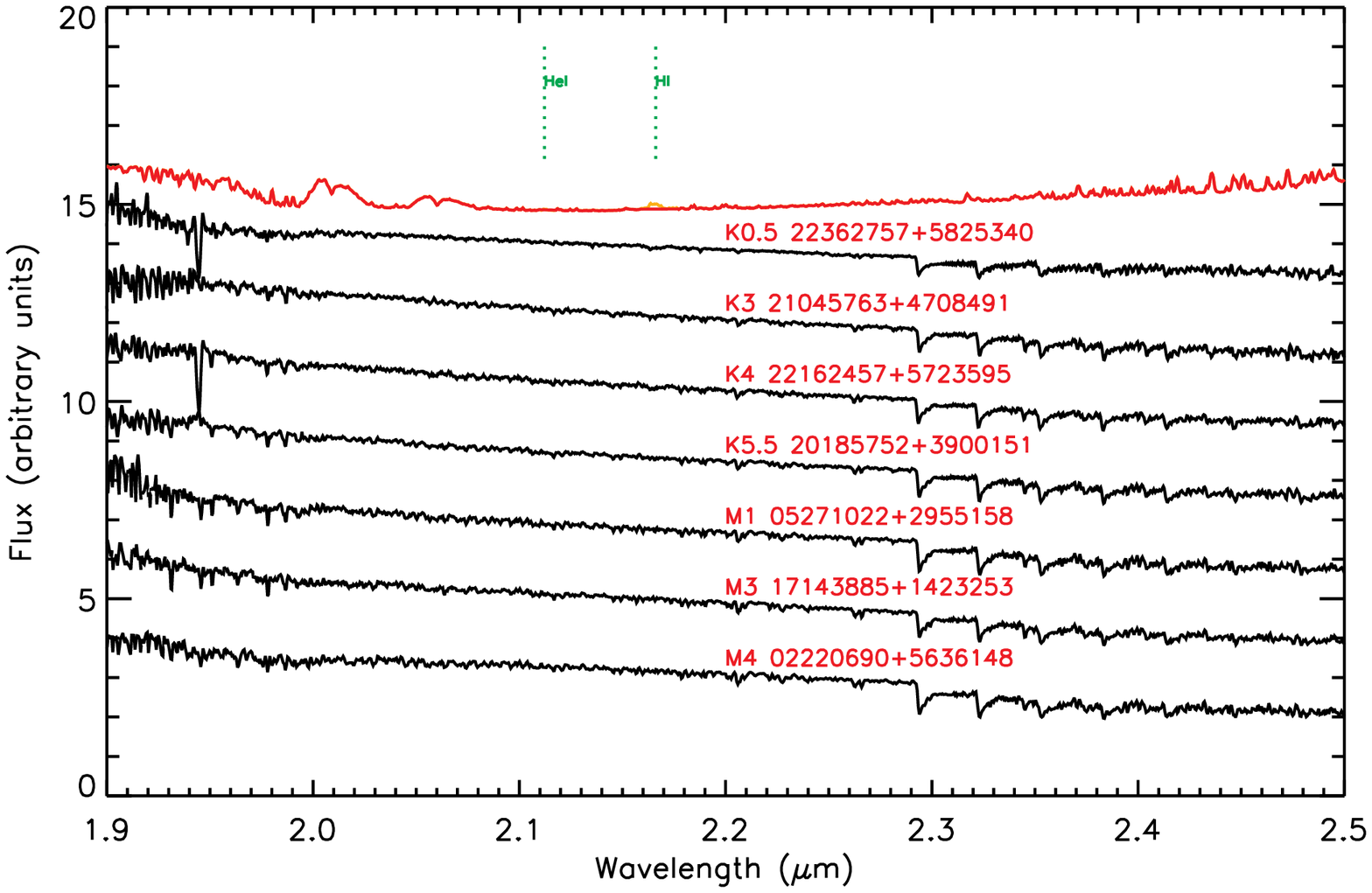}}
\caption{ Continuation of Fig. \ref{examplespectra}.  Here, the spectra are 
normalized to the mean  flux density at 2.2-2.25 \um.}  
\end{center}
\end{figure*}

\subsection{New observations and data reduction}

Two half-nights of observations with the 3-m IRTF telescope 
were awarded to this  program (number 047) in Period 19A,
 allocated on 2019 June 23 and 24.
 We observed  for about 2.5h (out of the 9h awarded) 
 due to poor weather.
The observational program was continued in Period 20A 
(number 045, 24 h allocated) from  2020 May 22 to June 25.

We used the upgraded SpeX spectrograph in 
cross-dispersed mode  \citep[SXD;][]{rayner03}.
The 0\farcs3  slit (the plate scale is 0\farcs1 per pixel) 
was used, providing a resolving power,
R, of $\sim 2000$. The observations covered from 0.7 to 2.55 \um.
For comparison, the IRTF library of \citet{rayner09} 
was performed with the pre-upgraded SpeX in SXD mode, having the
same  resolution (R=2000), but slightly smaller wavelength range of
0.80-2.55 \um.

The classical nodding technique was used 
to obtain a good sky-subtracted frame.
Frames were taken in the A and B positions, 
with the telescope  nodding between the two 
positions  from 3 to 10 times (AB cycles). 
The rotator angle was synchronized to the parallactic angle and the AB distance optimized  
in order to secure the target in the slit while nodding. 
For each position, the number of non-destructive reads (NDR)  
was set to a maximum of 32, and
from 1 to 8 coadds were taken with a basic integration time  from 0.4 to 90 s.
Our targets were bright enough to use as guide stars in 
 the SpeX slit viewer. 
Flats and arcs were taken  during the night, every two to four targets.

\begin{sidewaystable*} []
\vspace*{+8.3cm} 
\caption{\label{aliaskrsg} List of already known K-M supergiants  with IRTF spectra.}
{ \centering
\begin{myresizeenv}
\begin{tabular}{@{\extracolsep{-.05in}}ll|llllrrrrrr|lr|llllllllllll}
\hline
\hline
2MASS-ID               & Lib-   &  Sp        & Teff$^a$ & \Aks   & dist(cl)$^{aa}$ & Ref      & dist(EDR3)$^b$& \Vlsr$^c$  & dist(Kin)$^d$ & \mbolint$^e$ &\mbolbc$^f$ &   Sp$^g$ &Ref     &Sp       & Temp      & Temp    &Temp    & Temp   &Temp    & Temp   &   Comments$^h$\\
                       & rary          & (adopt)    &  [K]     &   [mag] & [kpc]   & $\rm \tiny (cl)$&[kpc]    & [\kms]     & [kpc]         &[mag]         &   [mag]    &          &  (Sp)  & (RSG)   & (CO)      & (Ca)    &(Na)    & (Mg I) &(Si I ) & (CO)   &           \\
                       &           &            &          &        &         &    &          &        &                 &      &                & &         &(CO-2.29)& (2.29)    & (2.26)  & (2.21) &(1.71)  & (1.64 )& (1.62) &           \\ 
                       &           &            &          &        &         &    &          &        &                 &      &                & &         &         &  [K]      &   [K]   & [K]    &[K]     &[K]     & [K]    &           \\          
\hline
    00511639+6148196  &            1   &                 K5.5    &3847.00$\pm$  46.00    &   0.12  $\pm$   0.19  &    & $..$    &     0.99  $^{   0.04  }_{  -0.04    }   $& -13.94$\pm$   0.18  &   0.83  $^{   0.39  }_{  -0.39  }$&   4.01$\pm$   0.79    &   4.09$\pm$   0.30    &                       K2Ib-II  &                                       5  &             K4  & 3926  & 3902  & 3815  & 3948  & 3732  & 3891  &~CN    \\%
    01075303+6325116  &            1   &                 K4.5    &3889.00$\pm$  31.00    &   0.25  $\pm$   0.02  &    & $..$    &     3.18  $^{   0.20  }_{  -0.17    }   $& -38.70$\pm$   0.14  &   2.77  $^{   0.45  }_{  -0.43  }$&   6.48$\pm$   0.76    &   6.57$\pm$   0.07    &                         K3Iab  &                                       3  &           K4.5  & 3888  & 3900  & 3911  & 3939  & 3797  & 3910  &~CN    \\
\hline
\end{tabular} 
\end{myresizeenv}
}
\begin{list}{}
\item {\bf Libraries:} 1=This work; 2=\citet{rayner09}; 3=\citet{villaume17}.
{\bf References:} 1= \citet{levesque05}; 2=\citet{keenan89}; 3=\citet{dorda18};
4=\citet{messineo17}; 5=\citet{negueruela16}; 6=\citet{messineo14b}; 7=\citet{skiff14};
8=\citet{negueruela12}; 9=\citet{negueruela10}; 10=\citet{alexander09}; 11=\citet{davies08};
12=\citet{davies07}; 13=\citet{medhi07}; 14=\citet{figer06}; 15=\citet{ginestet99}; 16=\citet{jura90};
17=\citet{fawley74}; 18=\citet{boulon63}; 19=\citet{perraud61}; 20=\citet{wilson50}; 21=\citet{bidelman57}; 
22=Heard,1956; 23=\citet{ian55}; 24=\citet{nassau55}; 
25=\citet{adams35}; 26=\citet{deburgos20};
27=\citet{cantat18}; 28=\citet{clark09}; 29=\citet{humphreys20}.
 ($^a$) Estimated \Teff\ with the adopted 
 spectral types and  temperature scale of \citet{levesque05}.~
 ($^{aa}$)  Distance via a membership to associations or 
 clusters \citep[as listed in][]{messineo19} (see also Ref. column).~
 ($^b$)  Gaia  distances  from EDR3 with 
  parallactic errors larger than 25\% are given between brackets.~ 
  This distance is the inverse of the parallax 
 (corrected with the zero point of 0.017 mas).
 ($^c$) \Vlsr\ is the radial velocities. 
 Values from Gaia DR2 \citep{gaiadr3} are adopted when available, 
 or from \citet{kharchenko07},
 otherwise the values measured from the SpeX spectra ($\sigma= 22$ \kms).
 For the four RSGC members 2MASS J18391505-0605191, 2MASS J18392036-0601426, 
 2MASS J18375890-0652321, and 2MASS J18451939-0324483, 
 cluster velocities are listed \citep{davies07,davies08,origlia15}.~
 ($^d$)  Distance inferred with the adopted radial 
 velocity and the Galactic rotation curve. 
 We used the code of \citet{reid09}.~ 
($^e$) Apparent bolometric magnitudes,\mbolint, 
are estimated by integration with  data from 1 \um\ to 24 \um.
For the four RSGC members 2MASS J18391505-0605191, 
2MASS J18392036-0601426, 2MASS J18375890-0652321, and 
2MASS J18451939-0324483,\mbolint\ and \Aks\ are as in \citet{humphreys20}.~ 
($^f$) Apparent bolometric magnitudes,\mbolbc, are estimated with the \BCK\ of \citet{levesque05} and \citet{neugent20}.~ 
($^g$) Spectral types as listed in the References.~
($^h$) CN= strong CN band heads at 1.09 \um.~ 
TIO0.88= detected TiO bands at 0.88 \um.~
TiO0.93= detected TiO bands at 0.93 \um.~ 
TiO1.1= detected TiO bands at 1.1 \um.~
ZrO0.93= visible ZrO band heads at 0.93 \um.~
VO=  VO absorption.~
H2O=  water vapor absorption.
($^{++}$)  2MASS J17143885+1423253/HD 156014/$\alpha$ Her 
(M5 Ib-II) was found to be a luminous AGB star by \citet{moravveji13},
in Sect. \ref{notes}.
($^{**}$) 2MASS J18345133$-$0713162  and 
 2MASS J18345840$-$0714247 are likely AGBs as explained in Sect. 
\ref{notes}.
\end{list}
\end{sidewaystable*}

\begin{sidewaystable*} []
\vspace*{-8.cm} 
\caption{\label{aliasnewrsg} List of candidate RSGs observed   with IRTF/SpeX.}
\begin{myresizeenv}
\begin{centering}
\begin{tabular}{@{\extracolsep{-.05in}}ll|llllrrrr|lr|lllllllllll}
\hline
\hline
2MASS-ID               & Lib-   &  Sp        & Teff$^a$ & \Aks   & dist(cl)$^{aa}$ & Ref      & dist(EDR3)$^b$& \Vlsr$^c$  & dist(Kin)$^d$ & \mbolint$^e$ &\mbolbc$^f$ &   Sp$^g$ &Ref     &Sp       & Temp      & Temp    &Temp    & Temp   &Temp    & Temp   &   Comments$^h$\\
                       & rary          & (adopt)    &  [K]     &  [mag] & [kpc]      & $\rm \tiny (cl)$&[kpc]    & [\kms]     & [kpc]         &[mag]         &   [mag]    &          &  (Sp)  & (RSG)   & (CO)      & (Ca)    &(Na)    & (Mg I) &(Si I ) & (CO)   &           \\
                       &           &            &          &        &         &    &          &        &                 &      &                & &         &(CO-2.29)& (2.29)    & (2.26)  & (2.21) &(1.71)  & (1.64 )& (1.62) &           \\ 
                       &           &            &          &        &         &    &          &        &                 &      &                & &         &         &  [K]      &   [K]   & [K]    &[K]     &[K]     & [K]    &           \\          
\hline
    18104421-1929072  &            1   &                 M5.5    &3380.00$\pm$  86.00    &   0.82  $\pm$   0.02  &    & $..$    & (   3.01  $^{   5.18  }_{  -1.17   })   $&  56.13$\pm$  22.00  &   4.72  $^{   0.78  }_{  -1.16  }$&   7.79$\pm$   0.71    &   7.09$\pm$   0.07    &                          $..$  &                                    none  &             M0  & 3793  & 3613  & 3139  & 3429  & 3551  & 3403  &Mira~AGB?~TiO0.88~TiO-ZrO0.93~TiO1.1um~VO    \\%
    18105174-1947374  &            1   &                 M5.5    &3391.00$\pm$  61.00    &   1.38  $\pm$   0.02  &    & $..$    &        $..$ &  51.00$\pm$  22.00  &   4.53  $^{   0.84  }_{  -1.30  }$&   7.18$\pm$   0.68    &   6.84$\pm$   0.07    &                          $..$  &                                    none  &           M1.5  & 3696  & 3858  & 3207  & 3435  & 3477  & 3445  &Mira~AGB~CN~TiO1.1um~VO~H2O    \\%
    18105920-1634067  &            1   &                   M5    &3424.00$\pm$  54.00    &   0.73  $\pm$   0.02  &    & $..$    &        $..$ &  50.49$\pm$  22.00  &   4.09  $^{   0.89  }_{  -1.29  }$&   7.41$\pm$   0.71    &   7.03$\pm$   0.06    &                          $..$  &                                    none  &           K5.5  & 3838  & 3494  & 3279  & 3490  & 3402  & 3523  &~CN~TiO0.88~ZrO0.93~VO    \\%
    18153105-1744228  &            1   &                   M6    &3278.00$\pm$  88.00    &   2.57  $\pm$   0.02  &    & $..$    &        $..$ &  23.13$\pm$  22.00  &   2.43  $^{   1.45  }_{  -1.74  }$&   7.04$\pm$   0.71    &   7.17$\pm$   0.07    &                          $..$  &                                    none  &           M0.5  & 3771  & 3511  & 3040  & 3294  & 3308  & 3469  &    \\%
    18192723-1707592  &            1   &                   M5    &3469.00$\pm$  71.00    &   0.74  $\pm$   0.02  &    & $..$    &        $..$ &  57.02$\pm$  22.00  &   4.32  $^{   0.82  }_{  -1.14  }$&   6.62$\pm$   0.05    &   6.85$\pm$   0.06    &                       K3I-III  &                                       4  &           K4.5  & 3907  & 3812  & 3279  & 3438  & 3559  & 3599  &Mira~AGB~TiO0.88~TiO0.93~TiO1.1um~VO~H2O    \\%
    18201613-1627415  &            1   &                 M4.5    &3505.00$\pm$  22.00    &   0.87  $\pm$   0.02  &    & $..$    & (   1.80  $^{   0.63  }_{  -0.37   })   $& -58.82$\pm$  22.00  &      $..$ &  $..$      &   6.89$\pm$   0.07    &                       K5I-III  &                                       4  &           K5.5  & 3839  & 3869  & 3509  & 3476  & 3468  & 3568  &Mira~AGB~TiO0.88~TiO0.93~TiO1.1um~VO~H2O    \\%
    18251767-1218360  &            1   &                   M6    &3276.00$\pm$ 125.00    &   1.13  $\pm$   0.02  &    & $..$    &        $..$ &  89.07$\pm$  22.00  &   5.06  $^{   0.69  }_{  -0.79  }$&   7.23$\pm$   0.69    &   7.04$\pm$   0.07    &                          $..$  &                                    none  &             M4  & 3557  & 3633  & 2940  & 3540  & 3354  & 3269  &Mira~AGB~TiO1.1um~VO~H2O    \\%
    18252725-1311090  &            1   &                   M5    &3448.00$\pm$  86.00    &   0.74  $\pm$   0.02  &    & $..$    & (   4.09  $^{   3.88  }_{  -1.34   })   $&  18.16$\pm$  22.00  &   1.60  $^{  14.25  }_{   1.47  }$&   6.53$\pm$   0.05    &   6.73$\pm$   0.07    &                          $..$  &                                    none  &             M0  & 3818  & 3329  & 3210  & 3482  & 3474  & 3627  &~CN~TiO0.88~ZrO0.93~VO    \\%
    18253802-1229509  &            1   &                 M5.5    &3384.00$\pm$  85.00    &   0.94  $\pm$   0.02  &    & $..$    & (   5.24  $^{  10.73  }_{  -2.11   })   $& -23.23$\pm$  22.00  &      $..$ &  $..$      &   7.02$\pm$   0.07    &                          $..$  &                                    none  &             M1  & 3745  & 3407  & 3132  & 3421  & 3482  & 3499  &~CN~TiO0.88~VO    \\%
    18253811-1306250  &            1   &                   M6    &3287.00$\pm$ 109.00    &   2.51  $\pm$   0.02  &    & $..$    &        $..$ &  73.57$\pm$  22.00  &   4.57  $^{   0.74  }_{  -0.91  }$&   6.95$\pm$   0.72    &   7.03$\pm$   0.07    &                          $..$  &                                    none  &             M4  & 3540  & 3651  & 2992  & 3514  & 3360  & 3281  &Mira~AGB~H2O    \\%
    18292354-1134599  &            1   &                 M5.5    &3388.00$\pm$  89.00    &   0.95  $\pm$   0.02  &    & $..$    &     1.15  $^{   0.33  }_{  -0.21    }   $&  65.50$\pm$  22.00  &   4.13  $^{   0.82  }_{  -1.01  }$&   7.32$\pm$   0.69    &   7.05$\pm$   0.06    &                          $..$  &                                    none  &           M1.5  & 3708  & 3802  & 3135  & 3484  & 3537  & 3393  &Mira~AGB~TiO0.88~TiO0.93~TiO1.1um~VO~H2O    \\%
    18314108-1059145  &            1   &                   M5    &3463.00$\pm$  93.00    &   0.89  $\pm$   0.02  &    & $..$    & (   3.40  $^{   5.84  }_{  -1.32   })   $&  48.42$\pm$  22.00  &   3.32  $^{   0.96  }_{  -1.25  }$&   6.61$\pm$   0.07    &   6.86$\pm$   0.07    &                          $..$  &                                    none  &             M1  & 3742  & 3441  & 3194  & 3609  & 3491  & 3560  &~CN~TiO0.88~ZrO0.93~VO    \\%
    18334070-0750531  &            1   &                 M5.5    &3367.00$\pm$  80.00    &   0.97  $\pm$   0.02  &    & $..$    & (   2.33  $^{   2.77  }_{  -0.82   })   $&  58.40$\pm$  22.00  &   3.62  $^{   0.91  }_{  -1.10  }$&   6.66$\pm$   0.06    &   6.92$\pm$   0.06    &                          $..$  &                                    none  &             M1  & 3741  & 3570  & 3163  & 3317  & 3463  & 3526  &Mira~AGB?~CN~TiO0.88~TiO-ZrO0.93~TiO1.1um~VO    \\%
    18341160-0940545  &            1   &                 M3.5    &3549.00$\pm$ 109.00    &   2.90  $\pm$   0.02  &    & $..$    &        $..$ &  88.94$\pm$  22.00  &   4.90  $^{   0.78  }_{  -0.84  }$&   6.18$\pm$   0.74    &   6.22$\pm$   0.07    &                          $..$  &                                    none  &             M2  & 3676  & 4032  & 3289  & 3797  & 3639  & 3472  &Mira~AGB~H2O    \\%
    18345908-0853504  &            1   &                 M5.5    &3374.00$\pm$  29.00    &   1.54  $\pm$   0.02  &    & $..$    & (   1.07  $^{   1.09  }_{  -0.36   })   $&  82.26$\pm$  22.00  &   4.63  $^{   0.81  }_{  -0.89  }$&   7.12$\pm$   0.73    &   7.18$\pm$   0.07    &                          $..$  &                                    none  &           M0.5  & 3777  & 3358  & 3299  & 3429  & 3357  & 3411  &Mira~AGB~TiO1.1um~VO~H2O    \\%
    18353576-0811451  &            1   &                   M6    &3313.00$\pm$ 101.00    &   2.04  $\pm$   0.01  &    & $..$    &        $..$ &  53.00$\pm$  22.00  &   3.38  $^{   0.95  }_{  -1.16  }$&   6.72$\pm$   0.71    &   6.82$\pm$   0.06    &                          $..$  &                                    none  &             M3  & 3598  & 3217  & 3011  & 3393  & 3451  & 3396  &    \\%

    18355534-0738197  &            1   &                 F3-6    \\%

    18360390-0740042  &            1   &                   M5    &3418.00$\pm$ 104.00    &   3.23  $\pm$   0.02  &    & $..$    &        $..$ & 121.21$\pm$  22.00  &   6.12  $^{   0.30  }_{  -0.27  }$&   6.27$\pm$   0.72    &   6.36$\pm$   0.07    &                          $..$  &                                    none  &             M4  & 3554  & 3365  & 3244  & 3673  & 3252  & 3505  &    \\%
    18360451-0747437  &            1   &                   M6    &3259.00$\pm$ 128.00    &   1.93  $\pm$   0.03  &    & $..$    &        $..$ &  48.33$\pm$  22.00  &   3.13  $^{   0.07  }_{  -0.07  }$&   7.19$\pm$   0.72    &   7.28$\pm$   0.08    &                          $..$  &                                    none  &             M3  & 3621  & 3465  & 2887  & 3317  & 3480  & 3352  &Mira~AGB~TiO1.1um~VO~H2O    \\%
    18363595-0718472  &            1   &                   M5    &3409.00$\pm$  65.00    &   1.21  $\pm$   0.02  &    & $..$    & (   2.65  $^{   4.74  }_{  -1.04   })   $&  58.98$\pm$  22.00  &   3.61  $^{   0.19  }_{  -0.19  }$&   7.24$\pm$   0.72    &   7.05$\pm$   0.06    &                          $..$  &                                    none  &           K5.5  & 3833  & 3485  & 3247  & 3361  & 3499  & 3528  &~CN~VO    \\%
    18365257-0735576  &            1   &                   M5    &3478.00$\pm$  85.00    &   2.58  $\pm$   0.02  &    & $..$    &        $..$ & 116.87$\pm$  22.00  &   5.94  $^{   0.21  }_{  -0.19  }$&   6.49$\pm$   0.72    &   6.55$\pm$   0.06    &                          $..$  &                                    none  &             M3  & 3621  & 3441  & 3244  & 3656  & 3494  & 3520  &    \\%
    18370859-0711207  &            1   &                   M5    &3442.00$\pm$  69.00    &   1.50  $\pm$   0.01  &    & $..$    & (   1.95  $^{   2.96  }_{  -0.73   })   $& 106.55$\pm$  22.00  &   5.53  $^{   0.11  }_{  -0.11  }$&   6.78$\pm$   0.71    &   6.63$\pm$   0.06    &                          $..$  &                                    none  &           M3.5  & 3592  & 3351  & 3270  & 3581  & 3396  & 3522  &    \\%
    18391007-0655526  &            1   &                 M5.5    &3382.00$\pm$ 110.00    &   1.87  $\pm$   0.02  &    & $..$    & (   0.64  $^{   9.47  }_{  -0.31   })   $&  19.43$\pm$  22.00  &   1.42  $^{   0.06  }_{  -0.07  }$&   6.94$\pm$   0.74    &   6.99$\pm$   0.06    &                          $..$  &                                    none  &           M0.5  & 3771  & 3642  & 3054  & 3457  & 3527  & 3492  &Mira~AGB~VO~H2O    \\%
    18423799-0410583  &            1   &                   M5    &3453.00$\pm$  85.00    &   2.30  $\pm$   0.02  &    & $..$    &        $..$ &  84.00$\pm$  22.00  &   4.62  $^{   0.02  }_{  -0.02  }$&   6.54$\pm$   0.73    &   6.57$\pm$   0.06    &                          $..$  &                                    none  &             M2  & 3661  & 3465  & 3244  & 3661  & 3451  & 3455  &    \\%
    18443153-0339407  &            1   &                   M6    &3248.00$\pm$ 124.00    &   1.75  $\pm$   0.02  &    & $..$    &        $..$ & 104.17$\pm$  22.00  &   5.58  $^{   0.12  }_{  -0.12  }$&   6.91$\pm$   0.72    &   7.01$\pm$   0.07    &                          $..$  &                                    none  &           M3.5  & 3592  & 3040  & 2876  & 3350  & 3372  & 3395  &    \\%
    18444023-0315329  &            1   &                 M5.5    &3362.00$\pm$  87.00    &   0.62  $\pm$   0.02  &    & $..$    &        $..$ &  99.85$\pm$  22.00  &   5.38  $^{   0.44  }_{  -0.40  }$&   6.86$\pm$   0.06    &   7.13$\pm$   0.07    &                          $..$  &                                    none  &           K5.5  & 3832  & 3530  & 3125  & 3355  & 3537  & 3431  &Mira~AGB~TiO0.88~TiO0.93~TiO1.1um~VO    \\%
    18453701-0238366  &            1   &                 M5.5    &3323.00$\pm$  71.00    &   2.08  $\pm$   0.02  &    & $..$    &        $..$ & 109.90$\pm$  22.00  &   6.02  $^{   0.23  }_{  -0.20  }$&   7.14$\pm$   0.73    &   7.23$\pm$   0.07    &                          $..$  &                                    none  &             M3  & 3597  & 3333  & 3111  & 3393  & 3372  & 3417  &    \\%
    18480312-0238003  &            1   &                   M4    &3537.00$\pm$ 100.00    &   1.13  $\pm$   0.02  &    & $..$    & (   1.96  $^{   1.83  }_{  -0.64   })   $&  68.82$\pm$  22.00  &   3.94  $^{   0.24  }_{  -0.24  }$&   6.99$\pm$   0.73    &   6.93$\pm$   0.07    &                          $..$  &                                    none  &             M2  & 3690  & 3318  & 3257  & 3738  & 3572  & 3582  &~CN    \\%
    18483751-0041320  &            1   &                   M5    &3475.00$\pm$  54.00    &   0.94  $\pm$   0.02  &    & $..$    &        $..$ & 120.02$\pm$  22.00  &   7.07  $^{   0.80  }_{  -0.80  }$&   7.14$\pm$   0.72    &   7.00$\pm$   0.06    &                          $..$  &                                    none  &           M2.5  & 3649  & 3435  & 3383  & 3612  & 3393  & 3514  &~CN~TiO0.88~ZrO0.93~VO    \\%
    18491641-0211205  &            1   &                   M5    &3471.00$\pm$  97.00    &   1.63  $\pm$   0.02  &    & $..$    & (   1.06  $^{   0.61  }_{  -0.29   })   $&  75.39$\pm$  22.00  &   4.25  $^{   0.14  }_{  -0.14  }$&   6.86$\pm$   0.73    &   6.89$\pm$   0.06    &                          $..$  &                                    none  &           M1.5  & 3708  & 3326  & 3183  & 3604  & 3518  & 3578  &~VO    \\%
    18530896+0132180  &            1   &                 M2.5    &3628.00$\pm$  59.00    &   1.67  $\pm$   0.02  &    & $..$    &        $..$ &  89.48$\pm$  22.00  &   5.20  $^{   0.01  }_{  -0.01  }$&   6.80$\pm$   0.74    &   6.67$\pm$   0.07    &                          $..$  &                                    none  &             M1  & 3730  & 3536  & 3561  & 3802  & 3543  & 3606  &    \\%
    18532140+0122107  &            1   &                 M5.5    &3397.00$\pm$  74.00    &   3.20  $\pm$   0.02  &    & $..$    &        $..$ &  84.15$\pm$  22.00  &   4.85  $^{   0.25  }_{  -0.23  }$&   6.63$\pm$   0.74    &   6.65$\pm$   0.07    &                          $..$  &                                    none  &             M1  & 3745  & 3322  & 3183  & 3500  & 3405  & 3501  &    \\%
    18550277+0214342  &            1   &                   K5    &3865.00$\pm$  26.00    &   1.47  $\pm$   0.02  &    & $..$    & (   1.84  $^{   1.38  }_{  -0.55   })   $&  60.17$\pm$  22.00  &   3.56  $^{   0.38  }_{  -0.37  }$&   6.49$\pm$   0.76    &   6.53$\pm$   0.07    &                          $..$  &                                    none  &           K3.5  & 3963  & 3812  & 3907  & 3839  & 3801  & 3912  &    \\%
    18570375+0152049  &            1   &                   M5    &3464.00$\pm$  42.00    &   3.00  $\pm$   0.02  &    & $..$    &        $..$ &  67.22$\pm$  22.00  &   3.93  $^{   0.00  }_{  -0.00  }$&   6.55$\pm$   0.74    &   6.57$\pm$   0.07    &                          $..$  &                                    none  &             M1  & 3756  & 3918  & 3567  & 3415  & 3375  & 3499  &Mira~AGB~H2O    \\%
    18585383+0500368  &            1   &                 M1.5    &3703.00$\pm$  93.00    &   0.81  $\pm$   0.02  &    & $..$    &        $..$ &  57.33$\pm$  22.00  &   3.47  $^{   0.12  }_{  -0.12  }$&   6.51$\pm$   0.75    &   6.62$\pm$   0.06    &                     K3.5I-III  &                                       4  &           K4.5  & 3917  & 3431  & 3437  & 3773  & 3730  & 3871  &~CN    \\%
    19021078+0440152  &            1   &                   M3    &3588.00$\pm$  52.00    &   1.71  $\pm$   0.01  &    & $..$    &        $..$ & 110.72$\pm$  22.00  &   6.54  $^{   1.01  }_{  -1.01  }$&   6.78$\pm$   0.74    &   6.84$\pm$   0.06    &                          $..$  &                                    none  &           M1.5  & 3699  & 3347  & 3434  & 3639  & 3616  & 3663  &    \\%
    19112742+0857518  &            1   &                 M5.5    &3361.00$\pm$  53.00    &   1.30  $\pm$   0.02  &    & $..$    &        $..$ &  57.73$\pm$  22.00  &   3.79  $^{   0.53  }_{  -0.46  }$&   6.77$\pm$   0.72    &   6.84$\pm$   0.07    &                          $..$  &                                    none  &             M2  & 3681  & 3592  & 3227  & 3460  & 3431  & 3327  &Mira~AGB~TiO1.1um~VO~H2O    \\%
    19142785+1037474  &            1   &                   M5    &3451.00$\pm$  66.00    &   1.12  $\pm$   0.02  &    & $..$    & (   2.95  $^{   2.35  }_{  -0.91   })   $&  60.62$\pm$  22.00  &   4.28  $^{   0.34  }_{  -0.29  }$&   6.87$\pm$   0.72    &   6.93$\pm$   0.06    &                          $..$  &                                    none  &             M4  & 3547  & 3393  & 3299  & 3604  & 3393  & 3510  &~CN~VO    \\%
    19200744+1358108  &            1   &                   M2    &3642.00$\pm$  39.00    &   1.82  $\pm$   0.02  &    & $..$    & (   2.53  $^{  24.72  }_{  -1.20   })   $&  75.15$\pm$  22.00  &   5.52  $^{   1.48  }_{  -1.48  }$&   6.45$\pm$   0.75    &   6.53$\pm$   0.07    &                          $..$  &                                    none  &             M0  & 3797  & 3601  & 3588  & 3744  & 3569  & 3665  &    \\%
    19293750+1758006  &            1   &                   M5    &3419.00$\pm$  49.00    &   1.11  $\pm$   0.02  &    & $..$    & (   2.02  $^{   1.15  }_{  -0.54   })   $&  24.47$\pm$  22.00  &   1.87  $^{   0.05  }_{  -0.05  }$&   6.25$\pm$   0.74    &   6.30$\pm$   0.07    &                          $..$  &                                    none  &           M1.5  & 3712  & 3775  & 3559  & 3332  & 3417  & 3370  &Mira~AGB~VO~H2O    \\%
    19362879+2241257  &            1   &                 M1.5    &3690.00$\pm$  18.00    &   0.74  $\pm$   0.01  &    & $..$    & (   5.12  $^{   2.69  }_{  -1.31   })   $&  52.47$\pm$   0.59  &   4.41  $^{   0.49  }_{  -0.49  }$&   6.37$\pm$   0.76    &   6.46$\pm$   0.06    &                          $..$  &                                    none  &             M0  & 3810  & 3595  & 3716  & 3706  & 3634  & 3705  &~CN~ZrO0.93    \\%
    19415961+2257103  &            1   &                 M1.5    &3714.00$\pm$  16.00    &   0.74  $\pm$   0.02  &    & $..$    &     2.86  $^{   0.45  }_{  -0.34    }   $&  42.86$\pm$   0.64  &   4.30  $^{   0.53  }_{  -0.53  }$&   6.65$\pm$   0.75    &   6.75$\pm$   0.07    &                          $..$  &                                    none  &             M0  & 3793  & 3622  & 3704  & 3762  & 3683  & 3707  &~CN    \\%

\hline
\end{tabular} 
\end{centering}
\end{myresizeenv}

\begin{list}{}
\item Columns are as described in Table \ref{aliaskrsg}.
\item 2MASS J18585383+0500368, 2MASS J18201613-1627415, 2MASS J18192723-1707592 were
 already observed in $K$-band by \citet{messineo17}, 
 but their class remained unknown.

\end{list}
\end{sidewaystable*}

\begin{sidewaystable*} [t]
\vspace*{+8.cm} 
\caption{\label{aliasgiants} List of observed known giants.}
\begin{myresizeenv}
\begin{centering}
\begin{tabular}{@{\extracolsep{-.05in}}ll|llllrrrr|lr|lllllllllll}
\hline
\hline
2MASS-ID               & Lib-   &  Sp        & Teff$^a$ & \Aks   & dist(cl)$^{aa}$ & Ref      & dist(EDR3)$^b$& \Vlsr$^c$  & dist(Kin)$^d$ & \mbolint$^e$ &\mbolbc$^f$ &   Sp$^g$ &Ref     &Sp       & Temp      & Temp    &Temp    & Temp   &Temp    & Temp   &   Comments$^h$\\
                       & rary          & (adopt)    &  [K]     &  [mag] & [kpc]     & $\rm \tiny (cl)$&[kpc]    & [\kms]     & [kpc]         &[mag]         &   [mag]    &          &  (Sp)  & (giant)   & (CO)      & (Ca)    &(Na)    & (Mg I) &(Si I ) & (CO)   &           \\
                       &           &            &          &        &         &    &          &        &                 &      &                & &         &(CO-2.29)& (2.29)    & (2.26)  & (2.21) &(1.71)  & (1.64 )& (1.62) &           \\ 
                       &           &            &          &        &         &    &          &        &                 &      &                & &         &         &  [K]      &   [K]   & [K]    &[K]     &[K]     & [K]    &           \\          
\hline
    00403044+5632145  &            1   &                   K1    &4120.00$\pm$  33.00    &  -0.00  $\pm$   0.14  &    & $..$    &     0.07  $^{   0.00  }_{  -0.00    }   $&   2.51$\pm$   0.02  &      $..$ &  $..$      &   2.14$\pm$   0.23    &                        K0IIIa  &                                       2  &             K2  & 4216  & 4141  & 4101  & 4185  & 4034  & 4159  &    \\%
    13492867+1547523  &            1   &                 K4.5    &3886.00$\pm$  38.00    &  -0.05  $\pm$   0.17  &    & $..$    &        $..$ &   4.27$\pm$   0.19  &      $..$ &  $..$      &   3.06$\pm$   0.26    &                       K5.5III  &                                       2  &             M2  & 4033  & 3920  & 3959  & 3852  & 3793  & 3940  &    \\%
    15005772+3122384  &            1   &                 M1.5    &3726.00$\pm$  95.00    &   0.08  $\pm$   0.16  &    & $..$    &     0.89  $^{   0.02  }_{  -0.02    }   $&   2.30$\pm$   0.88  &   0.23  $^{   0.06  }_{  -0.06  }$&   6.05$\pm$   0.77    &   6.19$\pm$   0.26    &                       M2-4III  &                                   19,20  &             M4  & 3963  & 3897  & 3961  & 3519  & 3641  & 3782  &~TiO0.88~ZrO0.93    \\%
    15061509+1521575  &            1   &                 K4.5    &3884.00$\pm$  67.00    &   0.08  $\pm$   0.01  &    & $..$    &     1.23  $^{   0.03  }_{  -0.03    }   $& -80.75$\pm$   0.25  &      $..$ &   7.90$\pm$   0.76    &   8.03$\pm$   0.06    &                         K5III  &                                      24  &             M2  & 4025  & 3959  & 4055  & 3735  & 3835  & 3910  &    \\%
    15062101+2626136  &            1   &                   K2    &4035.00$\pm$  28.00    &   0.09  $\pm$   0.02  &    & $..$    &     0.38  $^{   0.00  }_{  -0.00    }   $&  18.59$\pm$   0.14  &   1.27  $^{   0.01  }_{  -0.01  }$&   6.89$\pm$   0.77    &   7.11$\pm$   0.07    &                         K4III  &                                    22,7  &             K4  & 4148  & 4009  & 4087  & 4023  & 3960  & 4071  &    \\%
    16112324+2458016  &            1   &                   M0    &3784.00$\pm$  68.00    &   0.01  $\pm$   0.18  &    & $..$    &     0.76  $^{   0.01  }_{  -0.01    }   $& -40.39$\pm$   0.22  &      $..$ &   6.43$\pm$   0.77    &   6.53$\pm$   0.38    &                         M1III  &                                       7  &           M2.5  & 4013  & 3863  & 3944  & 3621  & 3739  & 3831  &~TiO0.88    \\%
    16120838+1432561  &            1   &                   K3    &3992.00$\pm$  34.00    &   0.11  $\pm$   0.01  &    & $..$    &     0.48  $^{   0.01  }_{  -0.01    }   $&  52.45$\pm$   0.14  &   3.23  $^{   0.01  }_{  -0.01  }$&   7.16$\pm$   0.77    &   7.35$\pm$   0.06    &                         K4III  &                                      25  &           K5.5  & 4112  & 3988  & 4069  & 3946  & 3923  & 4031  &    \\%
    20224530+4101338  &            1   &                 K4.5    &3897.00$\pm$  20.00    &   0.15  $\pm$   0.19  &    & $..$    &     0.30  $^{   0.00  }_{  -0.00    }   $&  19.67$\pm$   0.19  &   1.61  $^{   0.41  }_{  -0.41  }$&   4.31$\pm$   0.79    &   4.43$\pm$   0.33    &                         K7III  &                                       2  &           M0.5  & 4076  & 3856  & 3920  & 3835  & 3917  & 3915  &~CN    \\%
    23300740+4907592  &            1   &                 K2.5    &4018.00$\pm$  14.00    &   0.19  $\pm$   0.23  &    & $..$    &     0.31  $^{   0.00  }_{  -0.00    }   $&   1.20$\pm$   0.20  &      $..$ &   5.07$\pm$   0.80    &   5.28$\pm$   0.41    &                         K3III  &                                       2  &           K4.5  & 4133  & 3999  & 4024  & 3997  & 3992  & 4057  &~CN    \\%

\hline
\end{tabular} 
\end{centering}
\end{myresizeenv}
\begin{list}{}
\item Columns are as described in Table \ref{aliaskrsg}.
\end{list}
\end{sidewaystable*}

Data were processed with the SpeXtool software \citep{cushing04}\footnote{see 
the IRTF webpage: 
{\bf \url{http://irtfweb.ifa.hawaii.edu/$\sim$ spex/observer/}}}.

For each pair, the B image was subtracted from the A image,
and the resulting image was flat-fielded and checked 
for non-linearity. Wavelength calibration was performed using  arc lamps.
Two aperture positions per frame were marked   
(the positive and the negative),
and the corresponding stellar traces defined. 
Each optimally extracted spectrum was background-subtracted using a constant 
value from nearby pixels (same width as the aperture).
Typically, for each target and  each order 10 spectra were extracted and
combined. 

Telluric standard stars were selected with  B or A0 types  and observed 
at an airmass within 0.2 from that of the target.
The H  lines  
were  removed with Gaussian fits and linear interpolations.
After having removed the H I lines, we inspected the spectra of the early B 
telluric stars  and removed the detected He I lines at  
7065.0 \AA, 10141.2 \AA, 10314.2 \AA, 10670.7 \AA, 10833.2 \AA, 
12788.4 \AA, 12972.0 \AA, 17007.3 \AA,
and 21123.8 \AA. 
Eventually, to obtain the final instrumental response and atmospheric 
transmission, the standard spectrum extracted from each order 
was divided by a normalized black body 
with temperature as that of  the standard star.

The observed spectrum, order by order, was divided by the 
instrumental response and atmospheric transmission curve.
The target and standard spectra were checked for small shifts in wavelength 
by cross-correlating spectral regions of high atmospheric absorption.
The final spectrum was obtained by merging the segments  
extracted from the various orders.
Eventually, the spectra were corrected for interstellar extinction 
with the \Aks\ from Sect. \ref{seclum}
and the extinction curve  of \citet{messineo05}
(basically the law of \citet{cardelli89} 
with an infrared power law of index = $-1.9$).\footnote{
We used  average colors $(J-$\Ks)=1.05 mag  and ($H-$\Ks)=0.235 mag
to initially correct the spectra for interstellar extinction,
corresponding to an initial  uncertainty $\Delta$ \Aks\ within 0.17 mag. 
After  having measured the \ew s and 
determined the individual spectral types (Sect. \ref{sectemperature}), 
we recomputed the \Aks\ values (Sect. \ref{seclum}). 
The initial $\Delta$ \Aks\ causes fluctuations 
of 2-3 \permil\ in the \ew s
and 10 K in the stellar temperatures.}

A few SpeX spectra are shown in Fig.\ \ref{examplespectra}.
We obtained spectra for 72 objects (one object was observed twice).

Table \ref{aliaskrsg} lists the  
21 known RSGs which we observed (marked with the column Library=1)
and which are not included in the IRTF library, as well 
as other known class I stars from the IRTF libraries 
(marked with the column Library=2 and 3).

Table \ref{aliasnewrsg} lists the 42 new candidate RSGs, 
and Table \ref{aliasgiants} the
9 known giants  (with three of them 
2MASS J23300740+4907592/HD 221246, 
2MASS J20224530+4101338/HD 194193, 
2MASS J13492867+1547523/HD 120477 
already listed in the IRTF library). 

In Sect. \ref{index} to \ref{metsec}, we analyze the spectral 
properties of the previously known RSGs.  
The analysis of the candidate RSGs is presented later in Sect. \ref{classtarget}, 
i.e., after having established a set of 
appropriate indices for assigning spectral types and classes.

\begin{figure*}
\begin{center}
\resizebox{0.8\hsize}{!}{\includegraphics[angle=0]{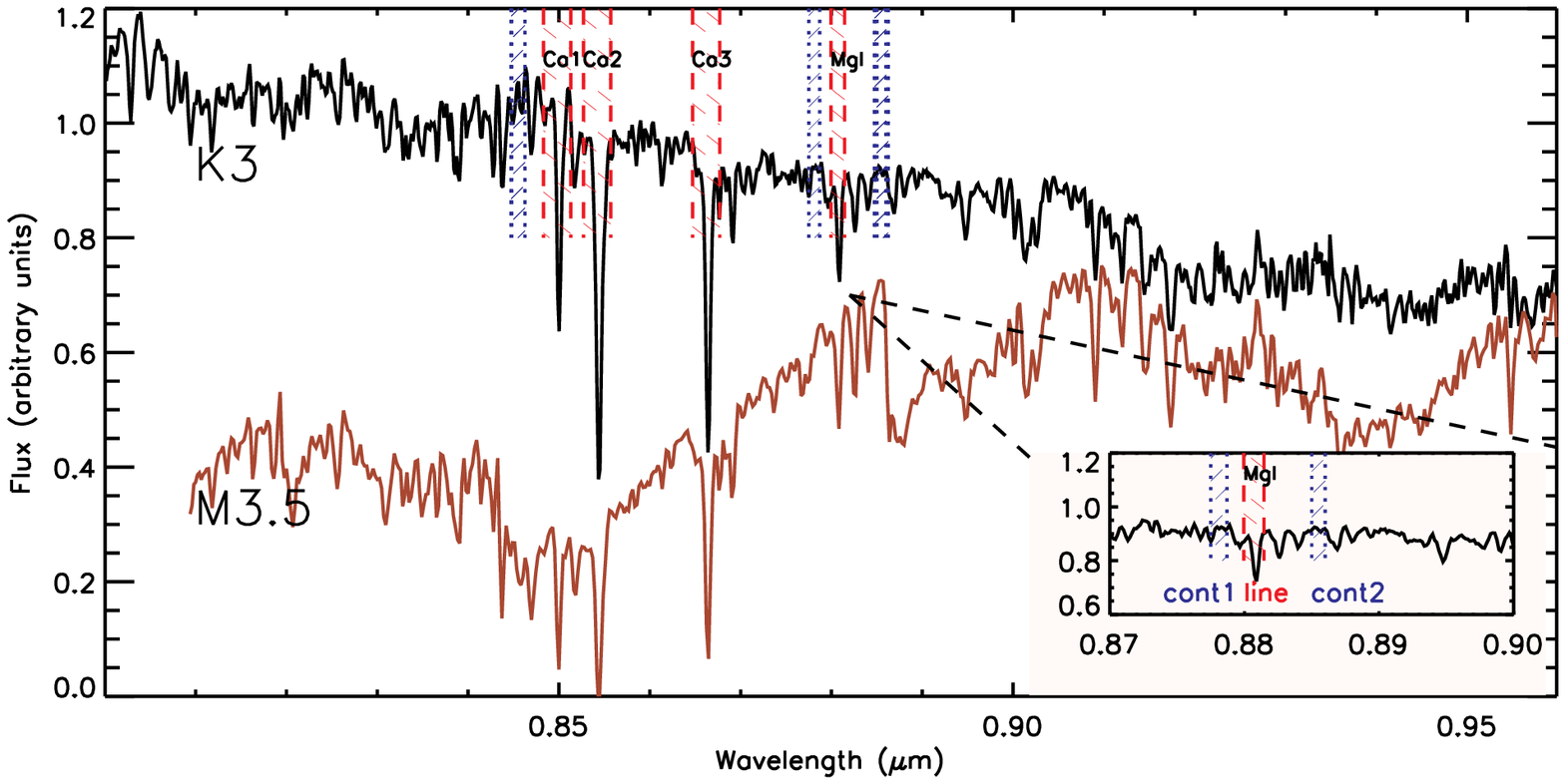}}
\resizebox{0.8\hsize}{!}{\includegraphics[angle=0]{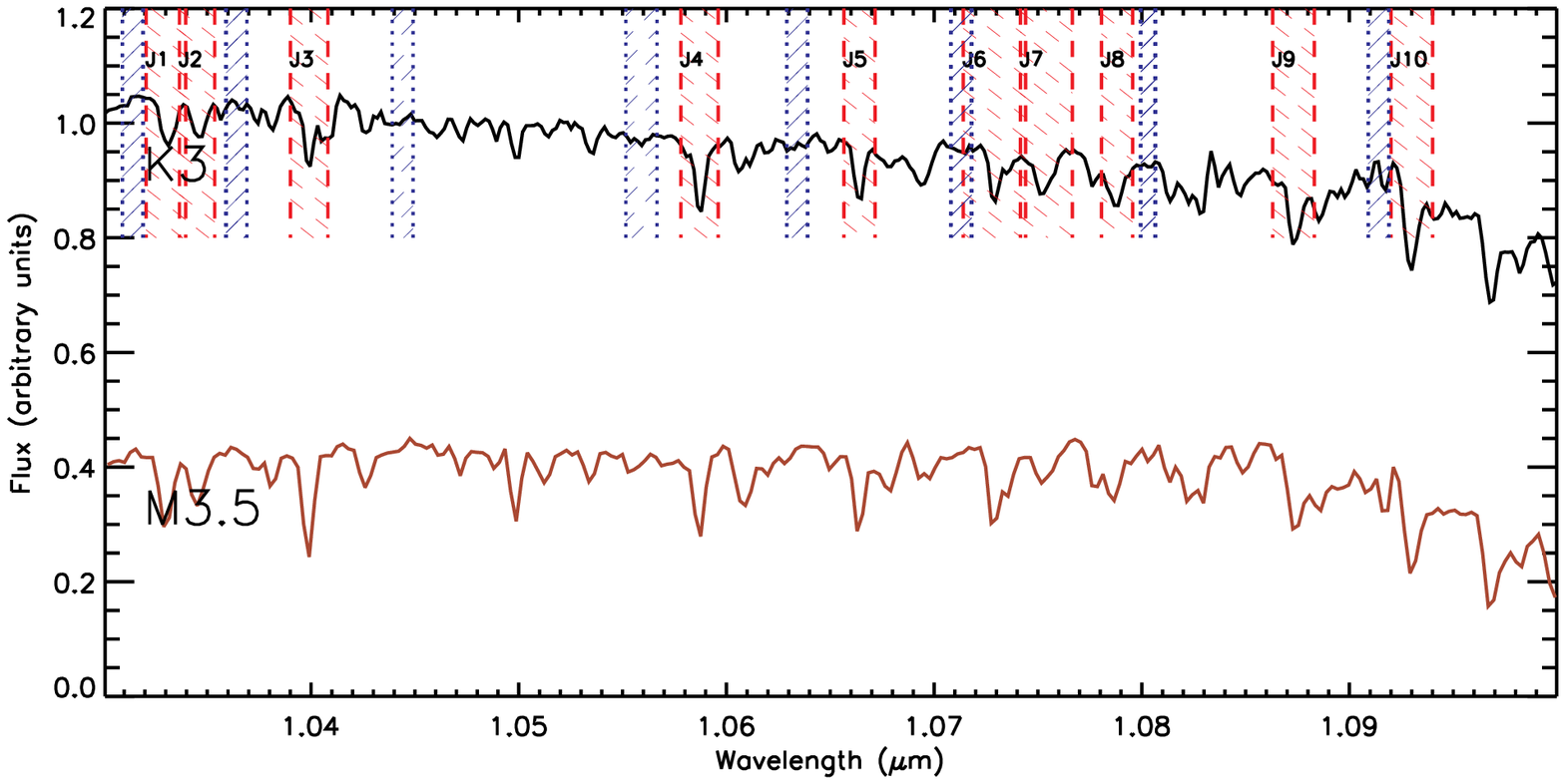}}
\resizebox{0.8\hsize}{!}{\includegraphics[angle=0]{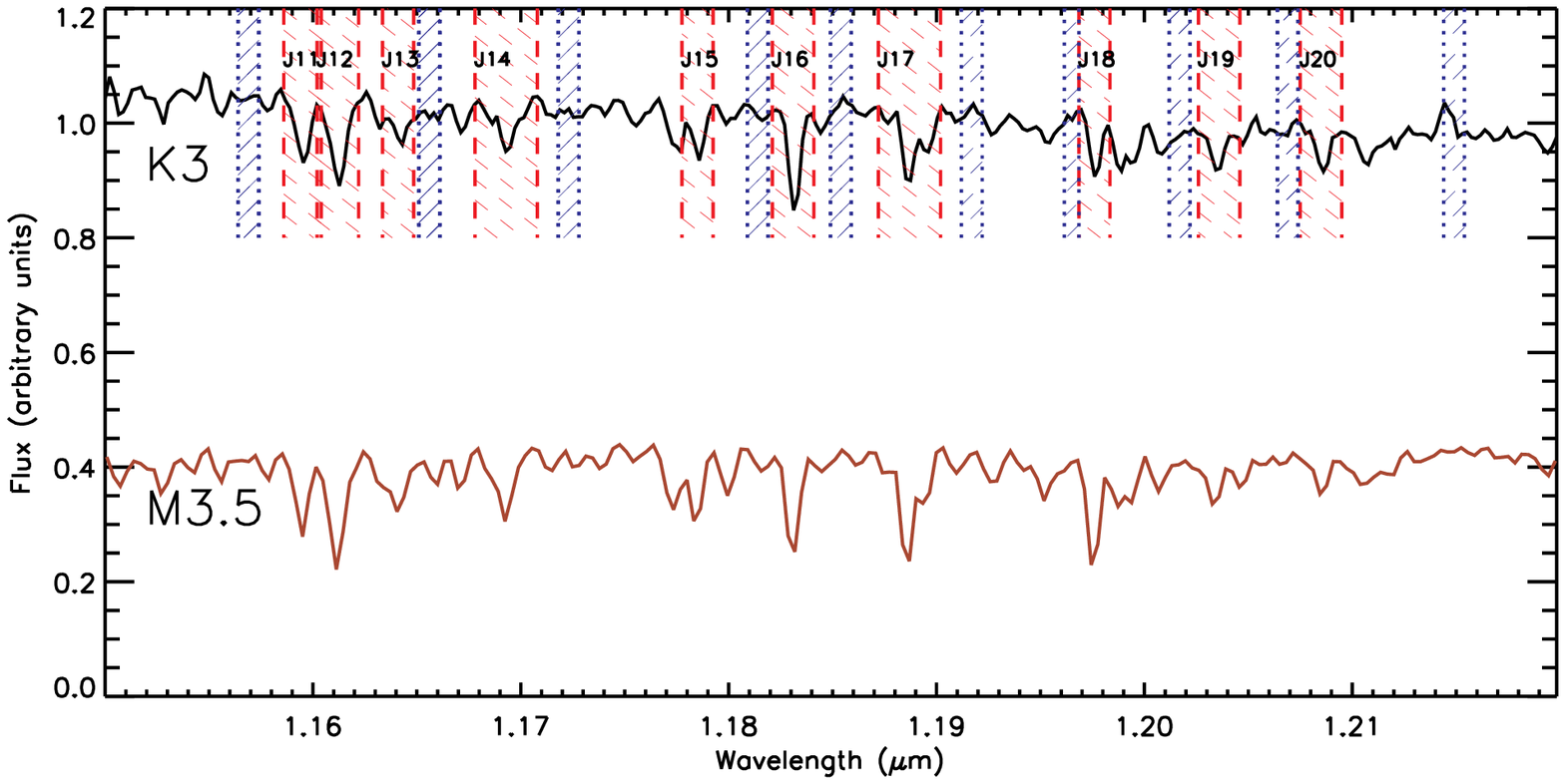}}
\caption{ \label{exampleindex}  Each panel displays a region of the stellar 
spectrum of   2MASS J19461557+1036475 ($\gamma$ Aqlil{\bf ae}) (top curve in black), which is a K3 II star 
in the Morgan \& Keenan system \citep{keenan89}. Below it, 
in brown  the spectrum of 
 2MASS J02220690+5636148/HD 14469 is shown \citep[M3.5Iab,][]{keenan89}.
On the stellar spectrum (top of the figure), 
red dashed-vertical lines  mark the two edges of each 
index defined in Table  \ref{inddef}, while blue dotted lines mark the adjacent continua. 
} 
\end{center}
\end{figure*}

\addtocounter{figure}{-1}
\begin{figure*}
\begin{center}
\resizebox{0.8\hsize}{!}{\includegraphics[angle=0]{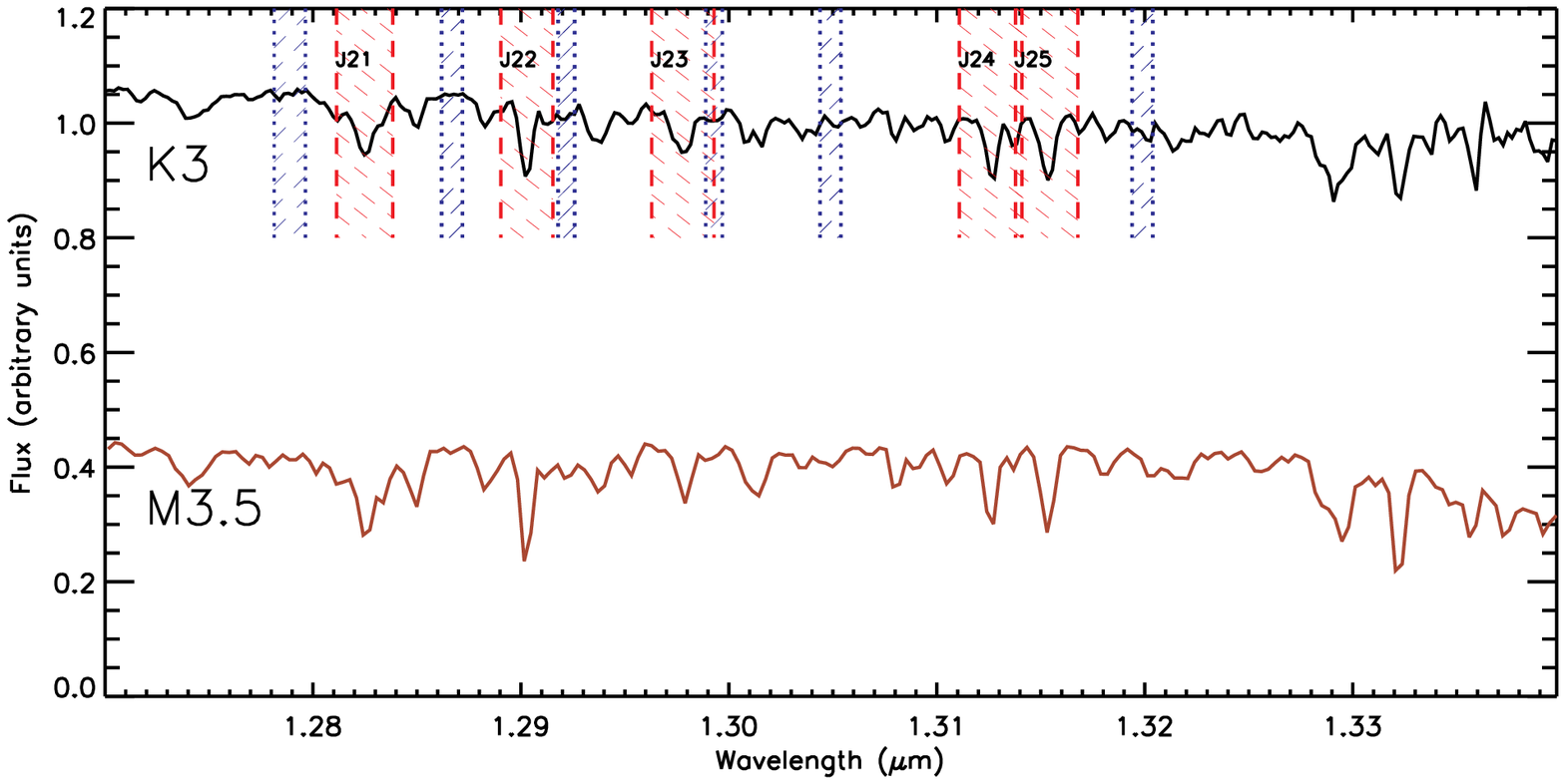}}
\resizebox{0.8\hsize}{!}{\includegraphics[angle=0]{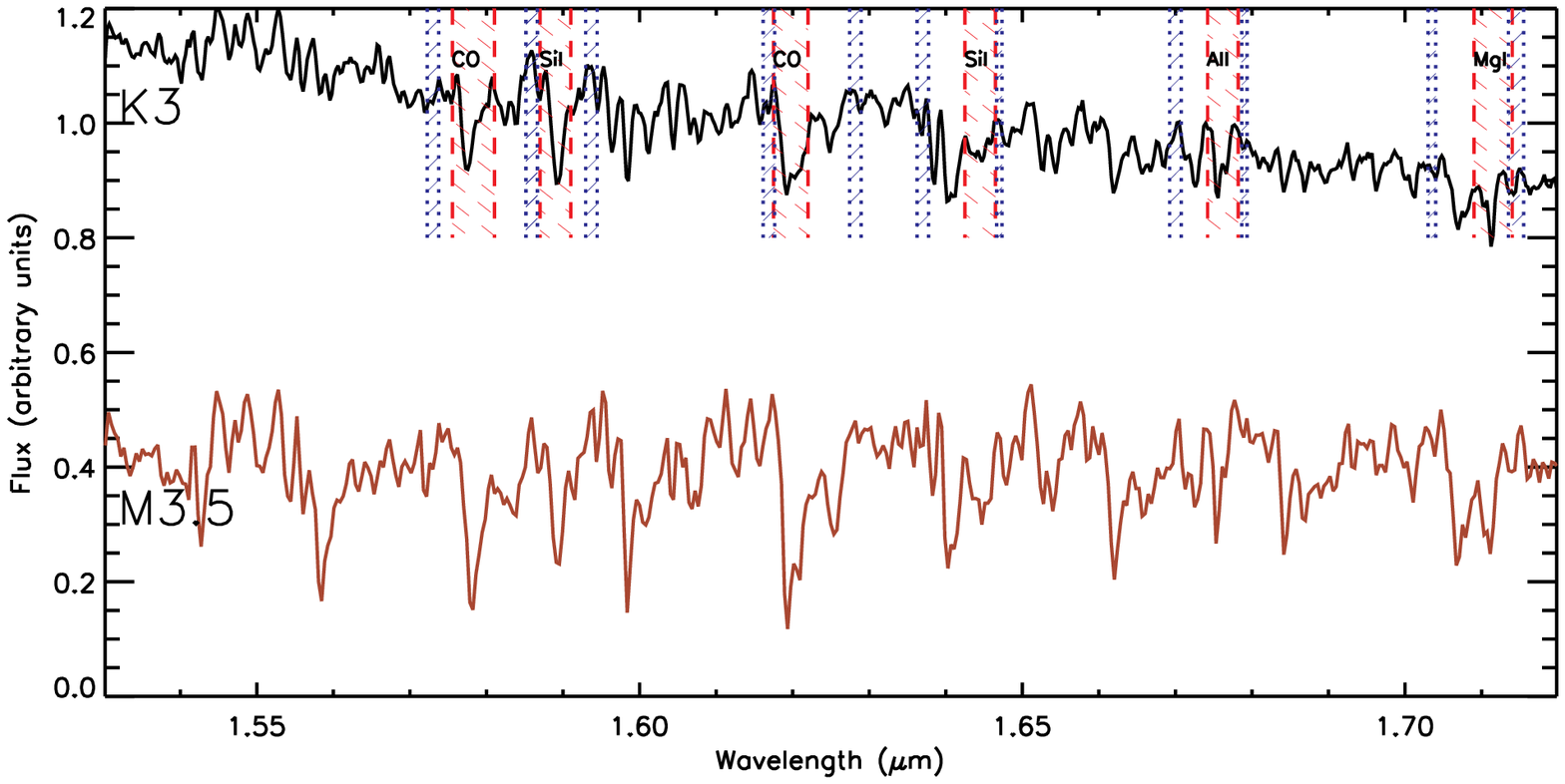}}
\resizebox{0.8\hsize}{!}{\includegraphics[angle=0]{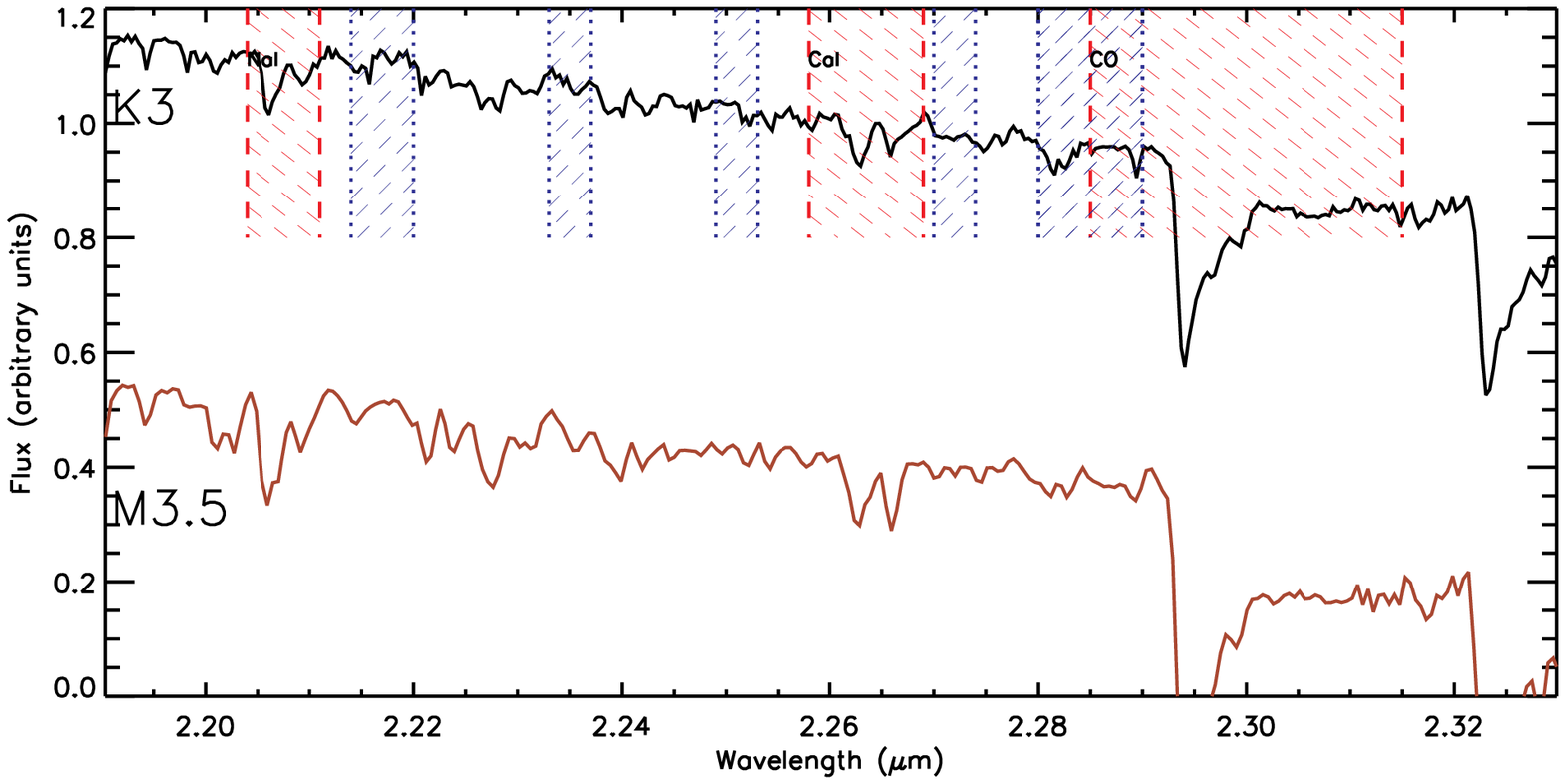}}
\caption{ Continuation of Fig. \ref{exampleindex}  } 
\end{center}
\end{figure*}

\subsection{Spectra from the IRTF libraries}

The IRTF spectral  library  of cold stars  
consists of data taken with the  old SpeX instrument, 
covering from 0.8 \um\ to 5.0 \um\ at a resolution R=2000 
\citep{rayner09}. 
22 stars of class I  are included in the IRTF library of \citet{rayner09} and
also listed in our Table \ref{aliaskrsg}. 
They consist of 21  RSGs, plus   2MASS J17143885+1423253/HD 156014 
\citep[M5 Ib-II, a likely Asymptotic Giant Branch  star, AGB,][]{moravveji13}.
The number of early K-type I stars included in the library 
is scarce, with two K0 I stars and three K2 I stars.\\
A comparison sample of 43 stars of class III 
from the same library is used in the plots. 
The sample includes normal giants and 
AGB stars.

\citet{villaume17} have created an extended IRTF spectral library 
using the new SpeX detector. The library contains 38 cold stars listed 
as class III and two of class I; 
 2MASS J06121911+2254305/HD 042543 is an M1-M2 I star and 
 2MASS J06300229+0755159/HD 045829 is a K2 I star.

\subsection{Contaminants}

The spectra of AGBs can appear similar to those of RSGs,
despite their lower initial masses ($< 8-9$ \Msun).
In the MK system of \citet{keenan89}, AGB stars 
are reported with classes III, II, Ib, and even Iab, 
which depends on initial masses and 
number of thermal pulses \citep{messineo19,moravveji13}.
On the basis of their surface abundances, 
AGBs   divide in C-rich (C/O $> 1$), O-rich (C/O $<1$), and S-types.
S-type stars have a mixed chemistry, in most of the cases 
indicating an intrinsic transition from O- to C-rich
of AGB stars with masses from 1 to 5 \Msun; S-type stars are
rich in s-process elements such as Zr \citep{shetye20}.\\
At infrared light, spectroscopically, C-rich stars are 
easily identified by their strong CN and CO bands;
O-rich AGBs and S-type stars may be mistaken for RSGs depending
on spectral coverage, for example when covering only the $K$-band.
Throughout the manuscript, in each of the presented 
figures, we mark also the distributions of quantities
measured in the spectra of
O-rich AGB stars and S-type stars 
from the IRTF library.  
This allows us to locate a few new diagrams 
that are useful for quantitative decontamination
and that open new direction for theoretical investigations.

\section{Definitions of  spectral indices}
\label{index}

\begin{table*}
\caption{\label{inddef} Definitions of used indices}
\begin{tabular}{llrlrlrllll}
\hline
 ID       & center  & width  & continuum1 & width1 & continuum2 & width2 & Comments \\
          &  [\um]  &  [\AA] & [\um]  &  [\AA]     &  [\um]  &  [\AA]    &\\
\hline   
                  CO  &2.3000  &300  &2.2850  &100  &2.2850  &100  &\citet{messineo17}   \\
                 CaI  &2.2635  &110  &2.2510  & 40  &2.2720  & 40  &as~in~\citet{messineo17}   \\
                 NaI  &2.2075  & 70  &2.2170  & 60  &2.2350  & 40  &as~in~\citet{messineo17}~and~in~\citet{ivanov04}   \\
                 MgI  &1.7115  & 50  &1.7035  & 10  &1.7145  & 20  &as~in~\citet{messineo17}~and~center~as~in~\citet{ivanov04}   \\
                 AlI  &1.6762  & 40  &1.6700  & 15  &1.6790  &  7  &\citet{messineo17}   \\
                 SiI  &1.6445  & 40  &1.6370  & 15  &1.6470  &  7  &\citet{messineo17}   \\
                  CO  &1.6198  & 45  &1.6169  & 15  &1.6282  & 15  &as~in~\citet{messineo17}~and~center~as~in~\citet{origlia93}   \\
                 SiI  &1.5890  & 40  &1.5859  & 15  &1.5937  & 15  &as~in~\citet{messineo17},~center~as~in~\citet{ivanov04}   \\
                  CO  &1.5783  & 55  &1.5730  & 15  &1.5859  & 15  &\citet{origlia93}   \\
\hline
                 J25  &1.3153  & 30  &1.3049  & 10  &1.3199  & 10  &Al~at~1.315438~\um~+Fe+Si   \\
                 J24  &1.3126  & 30  &1.3049  & 10  &1.3199  & 10  &Al~at~1.312703~\um~+Ca   \\
                 J23  &1.2978  & 30  &1.2922  &  8  &1.2993  &  8  &Mn~at~1.2979500~\um   \\
                 J22  &1.2903  & 25  &1.2867  & 10  &1.2922  &  8  &Mn~at~1.290336~\um~   \\
                 J21  &1.2825  & 27  &1.2789  & 15  &1.2867  & 10  &Ti~at~1.281499~\um,~Ca~at~1.281955~\um   \\
                 J20  &1.2085  & 20  &1.2069  & 10  &1.2149  & 10  &Mg~at~1.2086~\um,~Si~at~1.2085~\um,~CN   \\
                 J19  &1.2036  & 20  &1.2017  & 10  &1.2069  & 10  &Si~at~1.2034809~\um,~Mg~1.2086,~CN   \\
                 J18  &1.1976  & 15  &1.1965  &  7  &1.2017  & 10  &Fe~at~1.1973771~\&~1.1976325~\um,~Ti~at~1.1977134~\um   \\
                 J17  &1.1887  & 30  &1.1854  & 10  &1.1917  & 10  &Fe~at~1.1886096~and~1.1887336~\um,~Ti~at~1.1896127~\um,~CN   \\
                 J16  &1.1831  & 20  &1.1814  & 10  &1.1854  & 10  &Mg~at~1.183141~\um~   \\
                 J15  &1.1785  & 15  &1.1723  & 10  &1.1814  & 10  &Ti,~Fe,~CN   \\
                 J14  &1.1693  & 30  &1.1656  & 10  &1.1723  & 10  &Fe~at~1.1693174~\um   \\
                 J13  &1.1641  & 15  &1.1569  & 10  &1.1656  & 10  &Fe,~Si,~Ti,~CN   \\
                 J12  &1.1612  & 20  &1.1569  & 10  &1.1656  & 10  &Fe~at~1.161075~\um,~Cr~at~1.161374~\um,~Si~at~1.161428~\um,~CN   \\
                 J11  &1.1595  & 18  &1.1569  & 10  &1.1656  & 10  &Fe~at~1.1596762~\um,~CN   \\
                 J10  &1.0930  & 20  &1.0803  &  7  &1.0914  & 10  &CN   \\
                  J9  &1.0873  & 20  &1.0803  &  7  &1.0914  & 10  &Fe~at~1.0866494~\um~and~Si~at~1.0871763~and~1.0872520~\um   \\
                  J8  &1.0788  & 15  &1.0713  & 10  &1.0803  &  7  &Si~at~1.0787510~and~1.0789814~\um~and~Fe~at~1.0786009~\um   \\
                  J7  &1.0754  & 25  &1.0713  & 10  &1.0803  &  7  &Si~at~1.0752330~\um,~Fe~1.0755950~\um   \\
                  J6  &1.0729  & 30  &1.0713  & 10  &1.0803  &  7  &Ti~at~1.0729326~and~1.0735811,~Si~at~1.0730351,~Fe~at~1.0728129~\um   \\
                  J5  &1.0664  & 15  &1.0634  & 10  &1.0713  & 10  &Si~at~1.0663891~\um~and~Ti~at~1.0664551~\um   \\
                  J4  &1.0587  & 18  &1.0559  & 15  &1.0634  & 10  &Fe~at~1.0580~\um,~Ti~at~1.05875~\um,~and~Si~1.0588~\um,~VO   \\
                  J3  &1.0399  & 18  &1.0364  & 10  &1.0444  & 10  &Fe~at~1.03986~\um~and~Ti~at~1.03996~\um~   \\
                  J2  &1.0345  & 17  &1.0314  & 10  &1.0364  & 10  &Fe~at~1.0343722~\um~and~Ca~at~1.0346644~\um   \\
                  J1  &1.0330  & 19  &1.0314  & 10  &1.0364  & 10  &Sr~at~1.0330141~\um   \\
                 MgI  &0.8807  & 15  &0.8781  & 12  &0.8855  & 10  &as~in~\citet{diaz89}   \\
\hline
                 Ca3  &0.8662  & 30  &0.8455  & 15  &0.8855  & 15  &as~in~\citet{diaz89}   \\
                 Ca2  &0.8542  & 30  &0.8455  & 15  &0.8855  & 15  &as~in~\citet{diaz89}   \\
                 Ca1  &0.8498  & 30  &0.8455  & 15  &0.8855  & 15  &as~in~\citet{diaz89}   \\
\hline
\end{tabular}
\begin{list}{}
\item {\bf Notes.} The lines listed in the comments are from 
high-resolution  spectra of \citet{hinkle95}.  
\item The FeI-1.1611, FeI-1.641,  FeI-1.1887, FeI-1.976 lines 
of \citet{davies10}  are included in the J12, J13, J17, and J18 
indices, respectively.
\item Our indices contain only two of the best
four  iron lines for temperature determination
of \citet{taniguchi20}. Fe I at  1.0395794 \um\ (pair 8) 
is contained in the J3 index, 
and the Fe I line  at 1.0754753 \um\ (pair 9) in J7.
\item These are the correspondences between 
the lines of \citet{morelli20} and those in this work:  
FeTi (Morelli et al.) $\approx$ J3 (this work), 
CN $\approx$ J9, 
FeCr $\approx$ J12
Mg I $\approx$ J16,
Al $\approx$ J24+J25,
SiMg $\approx$ J20,  
Si I $\approx$ Si I (1.59 \um), CO 1.62 $\approx$ CO 1.62, 
Al  $\approx$ Al (1.67 \um), Mg I $\approx$ Mg I (1.71 \um)+CO.
Generally, we have used closer continua.
The COMg index around 1.71 \um\ of \citet{morelli20}  includes
both the CO band head and Mg I line at 1.71 \um.
We have measured the Mg I line only.
 
\end{list}
\end{table*}

We determine spectral class using apparent single line profiles 
in both K- and M- type stars.
For $HK$-band indices, we adopt the 
indices  defined by 
\citet{messineo17}, \citet{origlia93}, and \citet{ivanov04}.
For $YJ$ bands, we use  25 lines, seven of which 
are included in the work of  \citet{morelli20}.
The adopted lines and continua  
are displayed  in Fig. \ref{exampleindex} along
with the spectra of a K2 RSG and  an M3.5 RSG.

The definitions of the measured spectral indices 
(line bands, and continua), 
are given in Table \ref{inddef}. 
\ew$= bin \times \sum{_j (1- \frac{F_\lambda[j]}{F_{continuum}[j]})}$, 
where $F_\lambda$ is the observed flux density in the line region, 
$F_{continuum}$ is the estimated flux density of the continuum 
in the line region,  and bin is the wavelength size of one bin. 
Continuum estimates are done with 
a linear interpolation of nearby spectral regions.
Generally, the two continuum regions are located 
on each side of the line, when it is possible. 
For line J10 (CN head band), the two continua  
are both at wavelengths shorter than that of the line, 
so to avoid the continuum depression due to CN absorption.
For the Na I line, the two continua  
are both at wavelengths longer 
than that of the line to avoid possible continuum depression 
due to water absorption. 
For the CO head-band at 2.29 \um, a constant continuum is 
assumed,  as in \citet{messineo17} and \citet{figer06}. 
Positive values are  for absorption lines.

Measurements were performed on spectra  corrected for  Doppler shift. 
A linear dispersion in wavelengths was applied.
For each star, Doppler shifts  were measured
in $K$-band and also in $H$-band and applied to the wavelength solution.
The differences between the  velocities obtained from the
$H$-band and $K$-band have a sigma
of 7 \kms.  For 20 stars  velocity measurements are available in the 
Gaia DR2 release, there is a good agreement with our
measurements with a  sigma of 23 \kms. 
The   velocities, 
\Vlsr, are listed in Table \ref{aliaskrsg}.

The following describes the spectral indices, and
measurements are in Appendix  Table C.\\

\begin{figure}
\begin{center}
\resizebox{0.9\hsize}{!}{\includegraphics[angle=0]{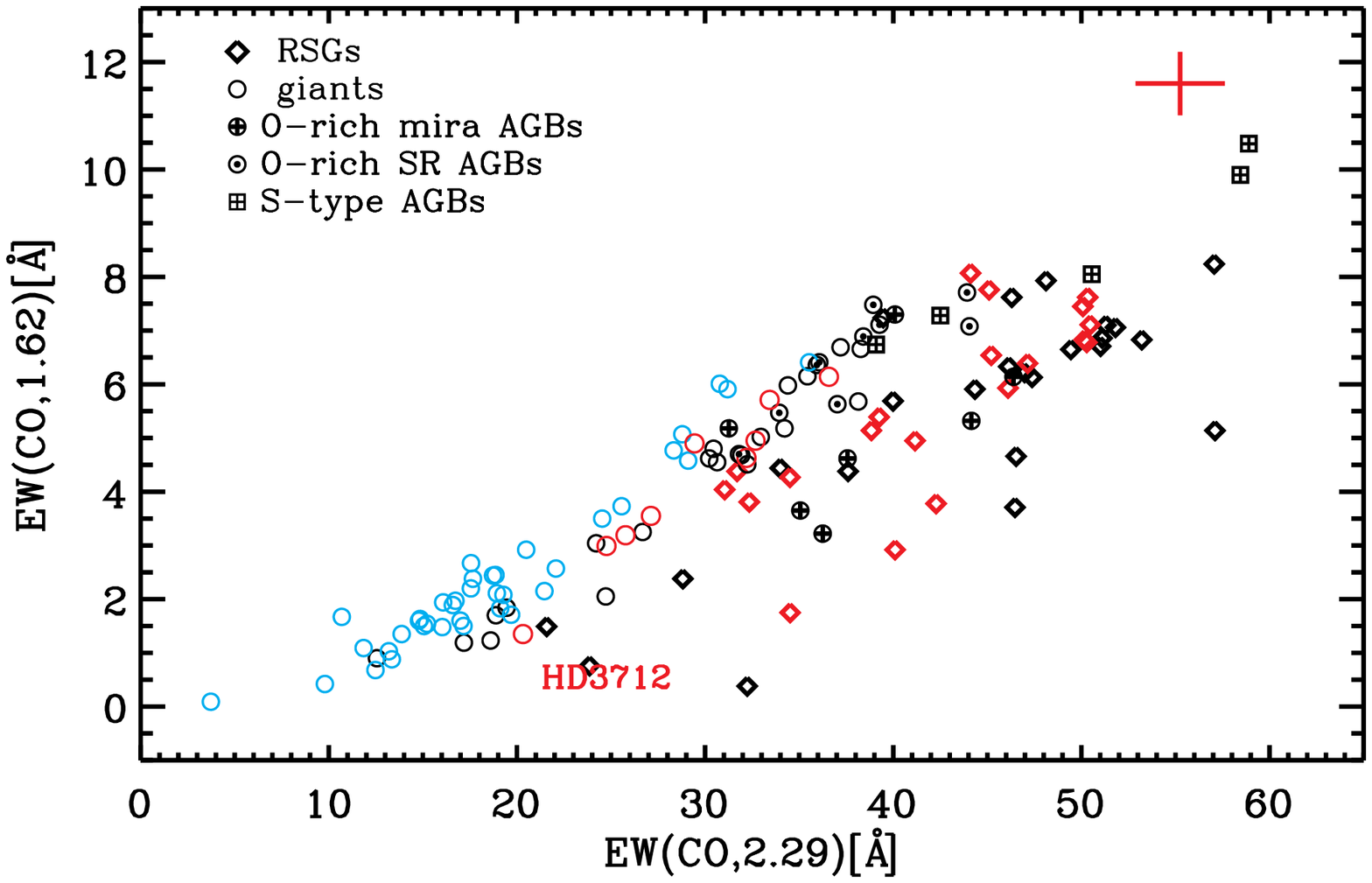}}
\end{center}
\begin{center}
\resizebox{0.9\hsize}{!}{\includegraphics[angle=0]{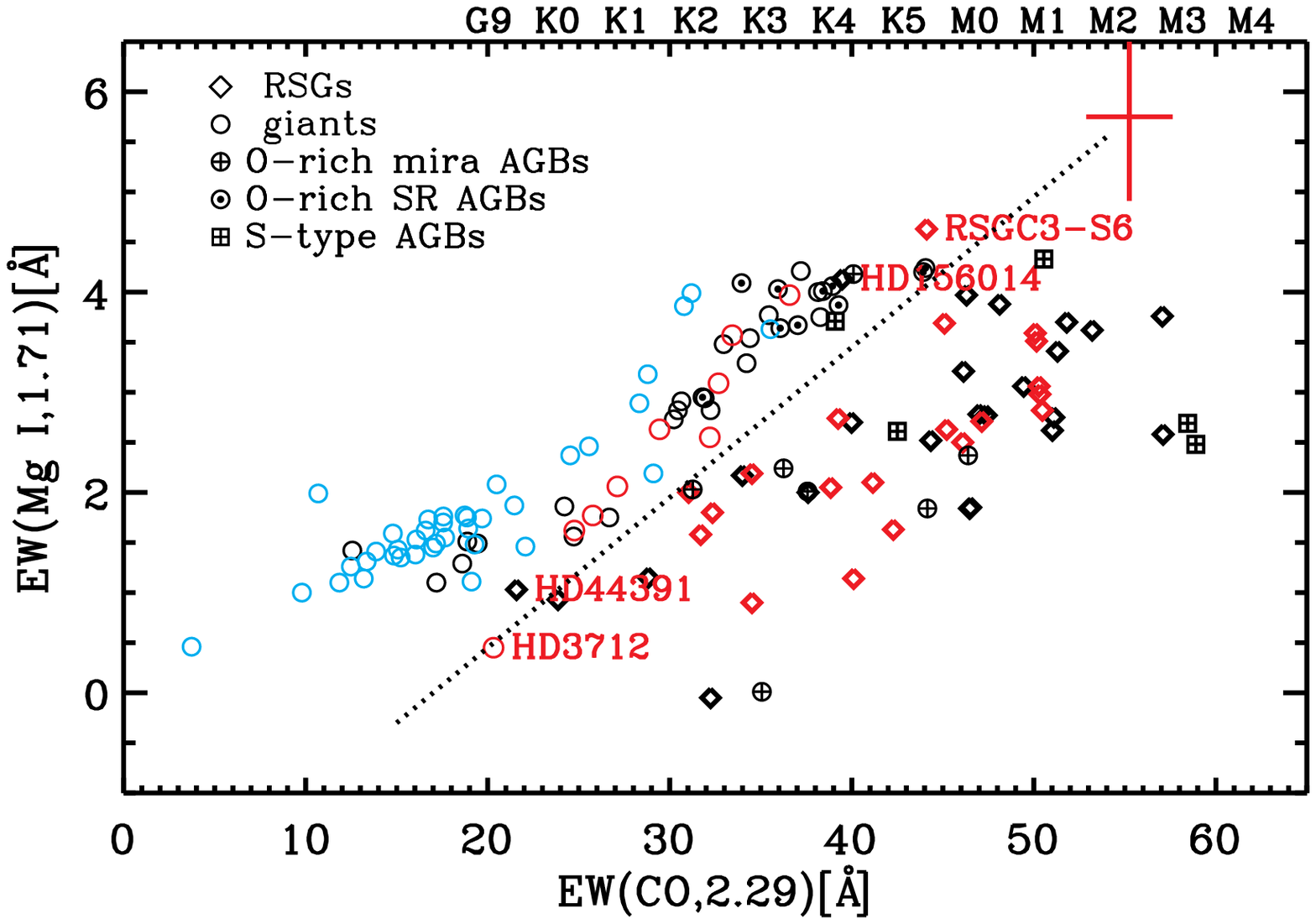}}
\end{center}
\begin{center}
\resizebox{0.9\hsize}{!}{\includegraphics[angle=0]{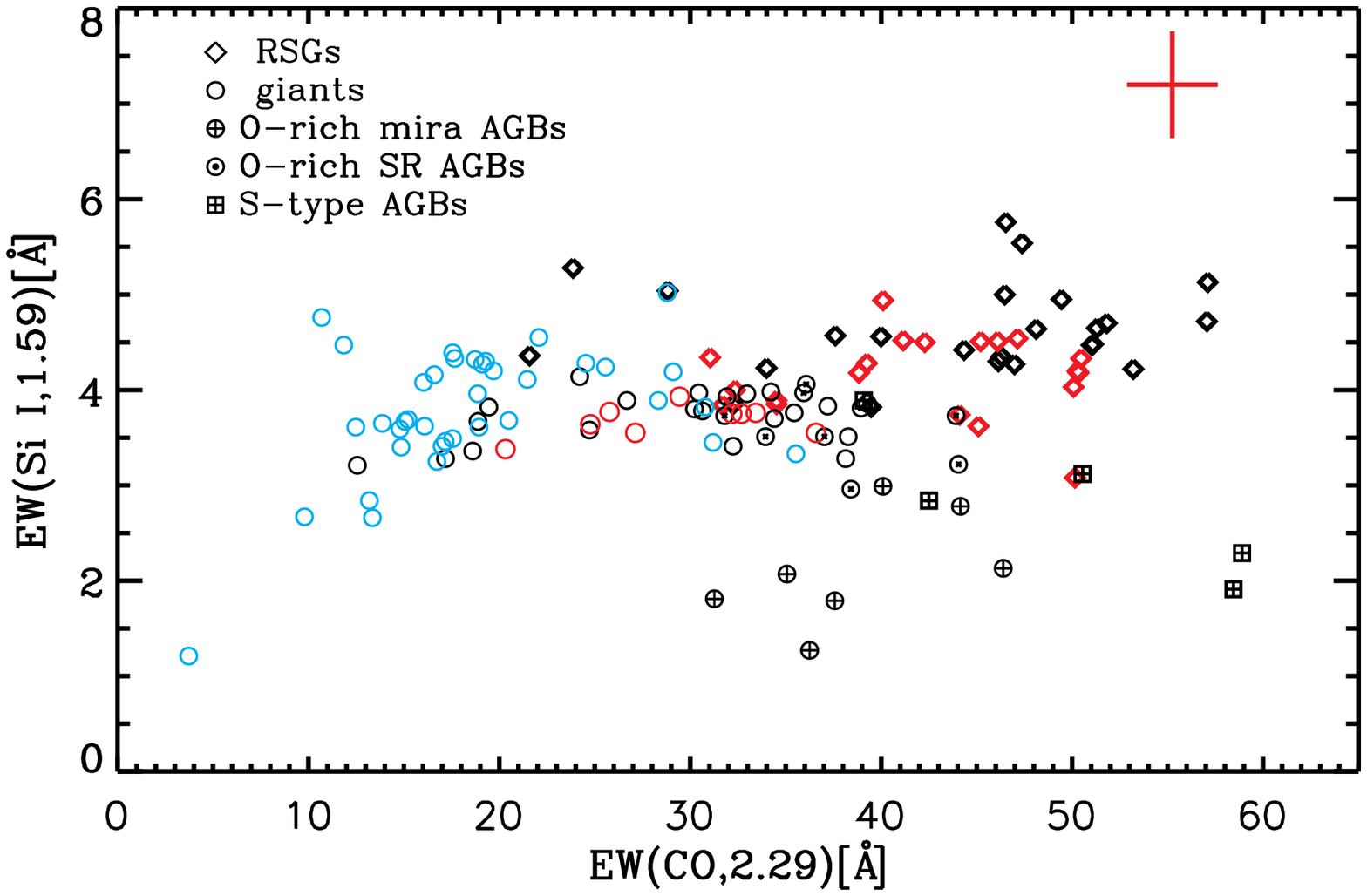}}
\end{center}
\caption{ \label{CO4} {\it Top panel:} 
 The \ewcousd\   vs. the \ewcodue\  values.
Data from the IRTF library of \citet[][black]{rayner09}  
and extended IRTF library of \citet[][cyan]{villaume17}. 
Known O-rich Mira AGBs are marked with an encircled plus and
 O-rich SR AGBs with  dotted circles; 
 S-type AGBs with  squared plus signs.  
 The median of errors  in both axes
are plotted with red segments (length$=2\times$ error) 
at the top-left corner.
Newly observed, but already known, RSGs from Table \ref{aliaskrsg} 
are marked with red diamonds  (2MASS J18345133$-$0713162  
and  2MASS J18345840$-$0714247  from   Table \ref{aliaskrsg}
with \indwater$> 6.6\%$ are not plotted). 
Giants from Table \ref{aliasgiants} are indicated
with red circles. 
\label{MgI5} {\it Middle panel:} 
The \ewmg\ values vs. the \ewcodue\ values. 
Symbols are as in the top panel. 
The dotted curve is as in \citet{messineo17} and  separate giants
and RSGs. The locations of  2MASS J17143885+1423253/HD 156014 (AGB) and 
2MASS J18451939$-$0324483/RSGC3-S6 (RSG), 
2MASS J06224788+2759118/HD 44391 (K0 Ib),
and 2MASS J00403044+5632145/HD 3712 (K0 IIIa) are indicated with labels.
{\it Bottom panel:}  The \ewsiucn\ values vs. the  \ewcodue values. 
} 
\end{figure}

{\it CO band heads}\\
It is well established that the temperature of late-type stars 
can be estimated at infrared wavelengths by using the CO 
band head at 2.29 \um. Giants and RSGs, because of their
different ranges of gravity, 
fall in two different CO versus \Teff\ relations. 
Since RSGs span a larger range of \ewcodue,   
late-type RSGs with types later than an M0
can be unambiguously identified at infrared wavelengths. 
Giants and RSGs overlap in this measurement for 
types earlier than M0.
The \ewco\ values at 1.62 \um\ are plotted versus the \ewco\ 
values at 2.29 \um\  in Fig. \ref{CO4}; 
the \ewco\  at 1.62 \um\ is another
indicator of  temperature.\\

{\it CO band heads and $H_2O$ index}\\
Large amplitude variable (LAV or Mira) AGBs 
stars do not follow  the CO relations found for giants, 
or RSGs \citep[see, e.g.][]{blum03}. 
Usually, at infrared wavelengths, astronomers  distinguish  
RSG stars  and O-rich Mira AGBs by measuring the depth of 
the H$_2$O absorption at the edges of the $H$-band 
\citep[see, e.g.][]{blum03, comeron04}.  
O-rich Mira AGB stars have generally strong continuum absorption 
due to water in their envelopes.

Measurements of water vapour absorptions were performed
as described in \citet[][]{blum03} and \citet{messineo17}.
\\

{\it Ca I at 2.26 \um}\\
The Ca I and the Na I lines are the two strongest atomic lines in the
$K$-band.
The \ewca\ and \ewna\ well correlate with temperature and with the \ew\ 
of the CO band head at 2.29 \um\  \citep[e.g.,][]{ramirez97}.
At the resolution of the SpeX library, the Ca I line is a 
good temperature indicator   \citep{cesetti13,rayner09}.\\

{\it Na I at 2.21 \um}\\
is a temperature indicator \citep[e.g.][]{ramirez97,cesetti13,rayner09,park18}.\\

{\it Mg I at 1.71 \um}\\
The Mg I line at 1.71 \um\  is prominent \citep{origlia93} 
and correlates with temperature for evolved late-type stars 
\citep{rayner09}. 
\citet{messineo17} inspected 
various correlations between this Mg I line and other lines. 
The \ewmg\ value, in combination with the \ewco\
value,  is a powerful diagnostic for luminosity class. 
Indeed, a given value of \ewco,  the \ewmg\  value 
 appears weaker in the spectra of RSGs than in those of 
 normal giants, see Fig. \ref{MgI5}. 
 Sec. \ref{magnesiumsec} outlines the importance 
 of this line for classifying evolved cool stars. \\

{\it Si I  line at 1.59 \um}\\
The  Si I  line at 1.59 \um\ was reported as one of the most prominent 
metal lines in $H$-band by \citet{origlia93}.
In the \ewsiucn\ versus \ewcodue\ plot, 
this Si I line appears stronger in the spectra of RSGs 
than in those of giants;   when the \ew\ of the CO band heads  
at 2.29 \um\ are larger than $\approx 42$ \AA, 
i.e., outside the range of red giants,  RSGs can be distinguished from
Mira AGBs by their larger \ewsiucn\ 
($\ga 3.25$ \AA, see  Fig.  \ref{CO4}).
Its strength does not appear to depend on the stellar 
temperature of the RSGs and it 
breaks the degeneracy between giants and supergiants.
This line may serve, therefore, as an empirical gravity indicator, 
to locate the youngest population of evolved late-type stars, the RSGs.
Recently, \citet{thorsbro20} has outlined the importance
to study  Si I lines to determine the history of Galactic 
$\alpha$-elements, which are found  to be enhanced in the metal rich 
population of the central nuclear Disk.\\

{\it Si I  line at 1.64 \um}\\
The  Si I  line at 1.64 \um\ appears as the second  Si I line 
in strength  present in the $H$-band models
of \citet{origlia93}, but it is much fainter than the line at 1.59 \um.
We defined an index for it in \citet{messineo17}.
The \ewsiusq\ is here found to  correlate  with the stellar \Teff\ values.\\

{\it Al I }\\
\citet{origlia93} also identified the Al I line at 1.67 \um.\\
In the work of \citet{park18}, it does not appear to be
a temperature indicator for bright giants. 
We did not find correlations with
luminosities.\\

{\it  $Y$ and $J$-bands indices}\\
The  $Y$-band (0.95-1.13 \um) and $J$-band (1.13-1.35 \um)   
contain prominent 
Ti I, Si I, Al I, Mg I, Mn I, Cr I, Sr I, and Fe I lines, and a variety 
of molecular bands 
\citep[e.g.][]{joyce98,hinkle95,rayner09,davies10,morelli20}.
The 25 lines are listed   in Table \ref{inddef}, and along the text
are shortly named as J1,J2, ...J25 as most of the lines 
considered are   blends of several atomic lines.
Among them, ten  lines have been already analysed in the work of 
\citet{davies10} and \citet{morelli20} --
J3 (FeTi), J9 (CN), J12 (FeCr), J13 (Fe+Si+Ti+CN), J16 (Mg I), 
J17 (Fe+Ti+CN), J18 (Fe+Ti),  J20 (SiMg), and J24+J25 (Al).                              
Their centerings and bandwidths are similar to those of \citet{morelli20}, 
but generally, we have adopted closer and narrower continua.
Indeed, the  indexes of \citet{morelli20} were defined with the goal of 
studying integrated stellar populations, and
the continua had to be placed away from H and He lines.

The dominant atomic elements or molecules contributing to these lines 
in the early K-type star Arcturus  were identified by \citet{hinkle95}
and are annotated in the comments of Table \ref{inddef}. 
 The J9 (CN) is a good marker of CN enhancement, 
while J12 (FeCr) is a marker of gravity \citep{morelli20}.
The J12, J13, J17, and J18  lines are useful for  
metallicity determination
\citep{davies10} (see Sect. \ref{metsec}). 
We anticipate that 
only eight (out of 25) lines were found 
to be extremely useful for spectroscopically confirming
RSGs: 3  for CN enhancement (J8, J9, J10) and 
5 for luminosity estimates
(J12, J17, J18, J21, and J22) (see Sect. \ref{secgravity}). \\

{\it Ca II triplet (CaT)}\\
The  SpeX detector in the chosen configuration covers  the region 
of the Ca II triplet (CaT).
Three prominent lines were detected at the expected wavelengths; 
8498 (Ca1), 8542 (Ca2), 8666 (Ca3) \AA. 
This well-studied region was also chosen for the Gaia spectrograph, 
as it allows us to classify  spectral types and to 
assign luminosity classes.
The CaT feature is useful to classify stars earlier than M3, 
as the spectra
of stars with later types have strong TiO bands at this wavelength 
\citep[see also][]{lopezcorredoira99}. \\

As shown in \citet{diaz89},  the \ewCaT\ defined as 
the sum of the \ew s of the Ca2 and Ca3 lines is $> 9$ \AA\
in RSGs for spectral types earlier than M2.5.
\ewCaT\ is larger than 9 \AA\ in 19 stars out of the 24 class I stars 
of the IRTF libraries \citep{rayner09,villaume17}.
The remaining  five  stars have spectral types later than M2.5 
and  2MASS J17143885+1423253/HD 156014 (M5 Ib-II) is an AGB \citep{moravveji13}.
In our observations, the  \ewCaT\ could be measured only in 33\%
of the 72 observed spectra, with 15 \ewCaT\ $> 9$ \AA\ 
for 13 previously known RSGs  and two candidate RSGs.\\

{\it Mg I at 0.88 \um}\\
This line is a temperature indicator, discovered by \citet{diaz89}.\\

\subsection{Homogeneity of the spectral library}
\label{omo}

Following \citet{messineo17},
we measured the \ew\ measurements after rebinning and smoothing the IRTF data.
After rebinning, the \ew s were smaller  by a variable factor from a few 
percent up to 35\% depending on the line and its signal-to-noise.
This partly explains  the large scatter among \ew s of giants and RSGs observed 
when using different libraries. 
Among the lines, we note that the band and continua used for the 
Mg I at 1.71 \um\ yield stable results; Mg I
at 1.71 \um\ is the strongest line  in $H$-band after the CO at 1.62 \um.
It is  recommended to quantitatively compare low-resolution spectra from 
the same instrument to build a reference frame  that allows us to separate 
late-type giants and RSGs.
\citet{cesetti13} and \citet{morelli20} have selected 
sensitive lines across the entire SpeX coverage for
integrated-light stellar population studies with signal-to-noise as low as 30.
Beside the already well studied lines in $H$ (CO, Si, Mg, Al) 
and in $K$-band (Na, Ca, CO) and in $Z$-band (CaT), we
explore  25 other lines in $Y$ and $J$-bands which characterize the
spectra of  RSGs. 
For our EW measurements we required a minimum signal to noise of 50.
Six of these lines correspond to spectral indices 
selected also by \citet{morelli20} (see footnote of \ref{inddef}).

Data with the new SpeX yield \ew s consistent within errors with  those
obtained with the old SpeX. Indeed, we re-observed three giants
from the IRTF library, as shown in Table \ref{giantcal}.

\begin{table}
\caption{\label{giantcal} 
Ratios between the \ew s from the  newly observed spectra 
and those in the IRTF library for the giants  2MASS J23300740+4907592/HD 221246, 
2MASS J20224530+4101338/HD 194193, and 2MASS J13492867+1547523/HD 120477.}
\begin{tabular}{lrrr}
\hline
\hline
       EW(line)     &  Ratio         & (EW(Lib.)-EW(New obs)) & $<$EW(RSG)e$>^*$ \\
                    &                &  [\AA]                 &   [\AA]   \\
\hline

               \ewcodue & 0.98$\pm$ 0.05 & 0.55$\pm$ 1.51  &  2.28    \\
                  \ewna & 1.20$\pm$ 0.08 &$-$0.54$\pm$ 0.20  &  0.42    \\
                  \ewca & 1.02$\pm$ 0.07 &$-$0.04$\pm$ 0.20  &  0.89    \\
               \ewsiucn & 0.97$\pm$ 0.02 & 0.11$\pm$ 0.08  &  0.59    \\
               \ewcousd & 1.00$\pm$ 0.03 & 0.00$\pm$ 0.14  &  0.62    \\
               \ewsiusq & 1.63$\pm$ 0.21 &$-$0.89$\pm$ 0.31  &  0.68    \\
                  \ewmg & 0.94$\pm$ 0.05 & 0.19$\pm$ 0.13  &  0.77    \\
                  \ewal & 0.69$\pm$ 0.12 & 0.38$\pm$ 0.15  &  0.67    \\
                \ewJuno & 1.05$\pm$ 0.19 &$-$0.03$\pm$ 0.09  &  0.08    \\
                \ewJdue & 1.03$\pm$ 0.11 &$-$0.01$\pm$ 0.06  &  0.06    \\
                \ewJtre & 1.01$\pm$ 0.09 &$-$0.00$\pm$ 0.05  &  0.09    \\
            \ewJquattro & 0.94$\pm$ 0.09 & 0.03$\pm$ 0.05  &  0.09    \\
             \ewJcinque & 1.00$\pm$ 0.04 &$-$0.00$\pm$ 0.02  &  0.11    \\
                \ewJsei & 1.01$\pm$ 0.09 &$-$0.01$\pm$ 0.06  &  0.31    \\
              \ewJsette & 0.91$\pm$ 0.09 & 0.04$\pm$ 0.05  &  0.24    \\
               \ewJotto & 0.99$\pm$ 0.07 & 0.01$\pm$ 0.03  &  0.11    \\
               \ewJnove & 1.70$\pm$ 0.91 &$-$0.32$\pm$ 0.39  &  0.70    \\
              \ewJdieci & 2.92$\pm$ 1.95 &$-$0.81$\pm$ 0.65  &  0.76    \\
              \ewJundici & 1.23$\pm$ 0.18 &$-$0.09$\pm$ 0.06  &  0.18    \\
              \ewJdodici & 1.02$\pm$ 0.03 &$-$0.02$\pm$ 0.03  &  0.22    \\
              \ewJutre & 1.14$\pm$ 0.27 &$-$0.04$\pm$ 0.10  &  0.14    \\
              \ewJuquattro & 1.15$\pm$ 0.34 &$-$0.05$\pm$ 0.13  &  0.34    \\
              \ewJucinque & 1.05$\pm$ 0.19 &$-$0.02$\pm$ 0.09  &  0.10    \\
              \ewJusei & 0.99$\pm$ 0.06 & 0.01$\pm$ 0.07  &  0.14    \\
              \ewJusette & 0.92$\pm$ 0.12 & 0.11$\pm$ 0.15  &  0.22    \\
              \ewJuotto & 0.66$\pm$ 0.36 & 0.30$\pm$ 0.31  &  0.07    \\
             \ewJunove & 0.86$\pm$ 0.09 & 0.09$\pm$ 0.06  &  0.10    \\
              \ewJdzero & 0.84$\pm$ 0.15 & 0.08$\pm$ 0.08  &  0.09    \\
              \ewJdu & 0.92$\pm$ 0.16 & 0.13$\pm$ 0.23  &  0.07    \\
              \ewJdd & 1.12$\pm$ 0.14 &$-$0.07$\pm$ 0.08  &  0.33    \\
              \ewJdtre & 3.02$\pm$ 1.89 &$-$0.11$\pm$ 0.18  &  0.30    \\
              \ewJdquattro & 1.03$\pm$ 0.12 &$-$0.01$\pm$ 0.08  &  0.25    \\
              \ewJdcinque & 0.94$\pm$ 0.25 & 0.04$\pm$ 0.14  &  0.25    \\

\hline
\end{tabular}
\begin{list}{}
\item $(^*)$  Average errors on the measured \ew s of RSGs in the IRTF library 
\citep{rayner09}.
\end{list}
\end{table}

\begin{figure}
\begin{center}
\resizebox{0.99\hsize}{!}{\includegraphics[angle=0]{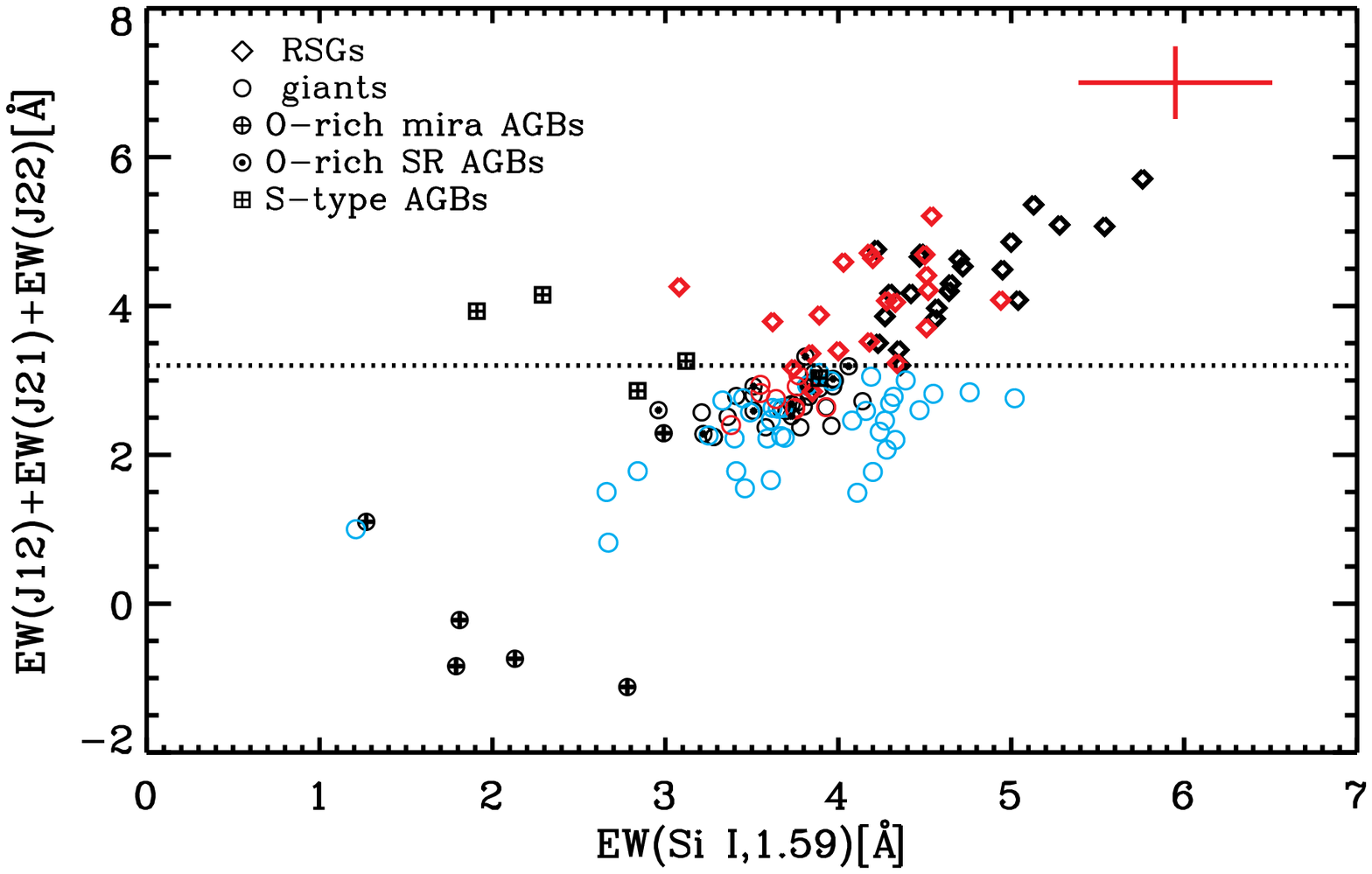}}
\resizebox{0.99\hsize}{!}{\includegraphics[angle=0]{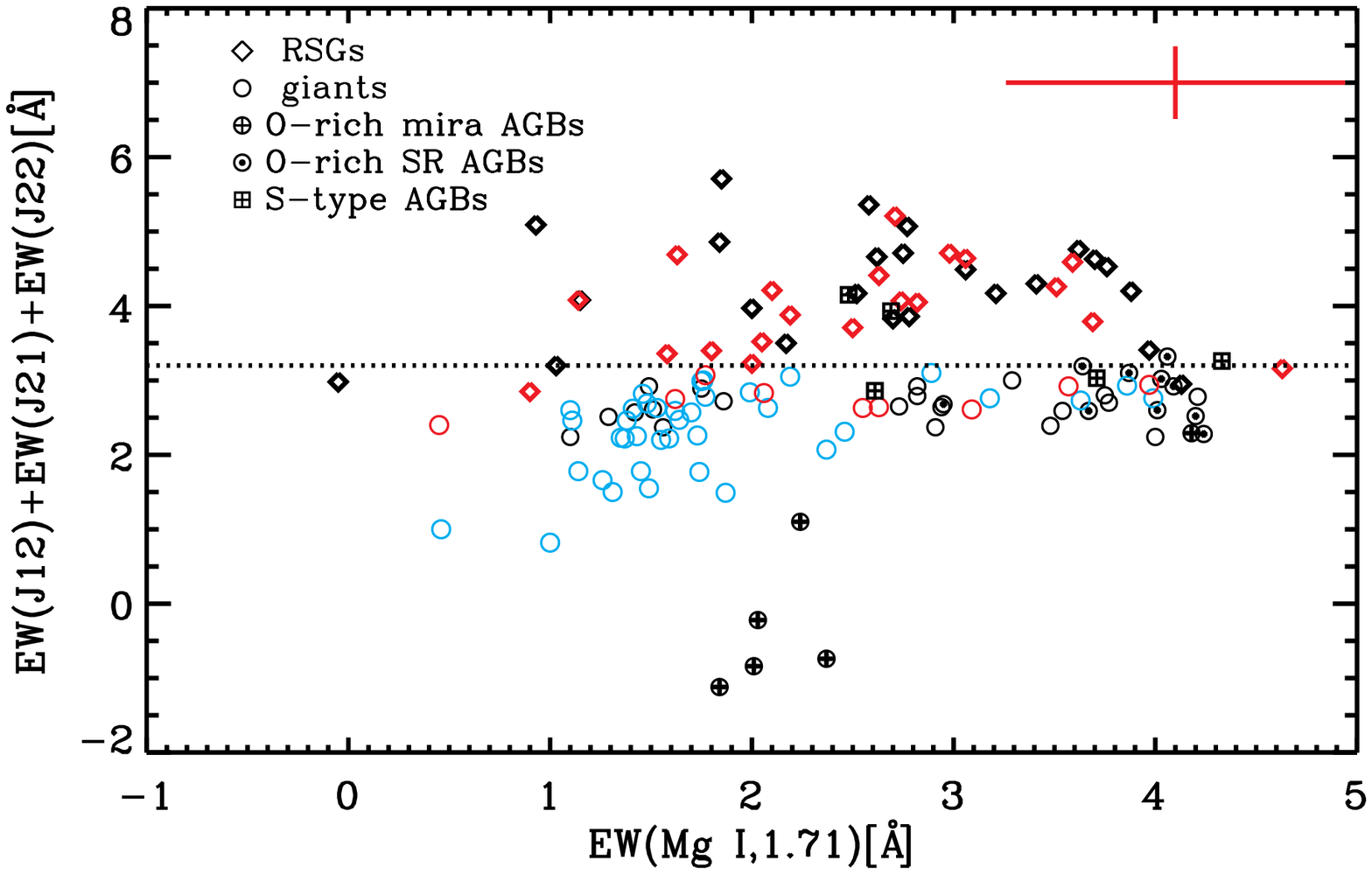}}
\end{center}
\caption{ \label{figalJplus} 
{\it Top panel:} \ewJudJduJdd\ values are plotted vs. the  \ewsiucn\ values.
Stars above   the dashed line at 3.2 \AA\ are RSGs.
{\it Bottom panel:} \ewJudJduJdd\ values are plotted vs. the  \ewmg\ values.
Stars above  the dashed line at 3.2 \AA\ are RSGs.
Symbols and colors are as in Fig. \ref{CO4}.
} 
\end{figure}

\begin{figure}
\begin{center}
\resizebox{0.99\hsize}{!}{\includegraphics[angle=0]{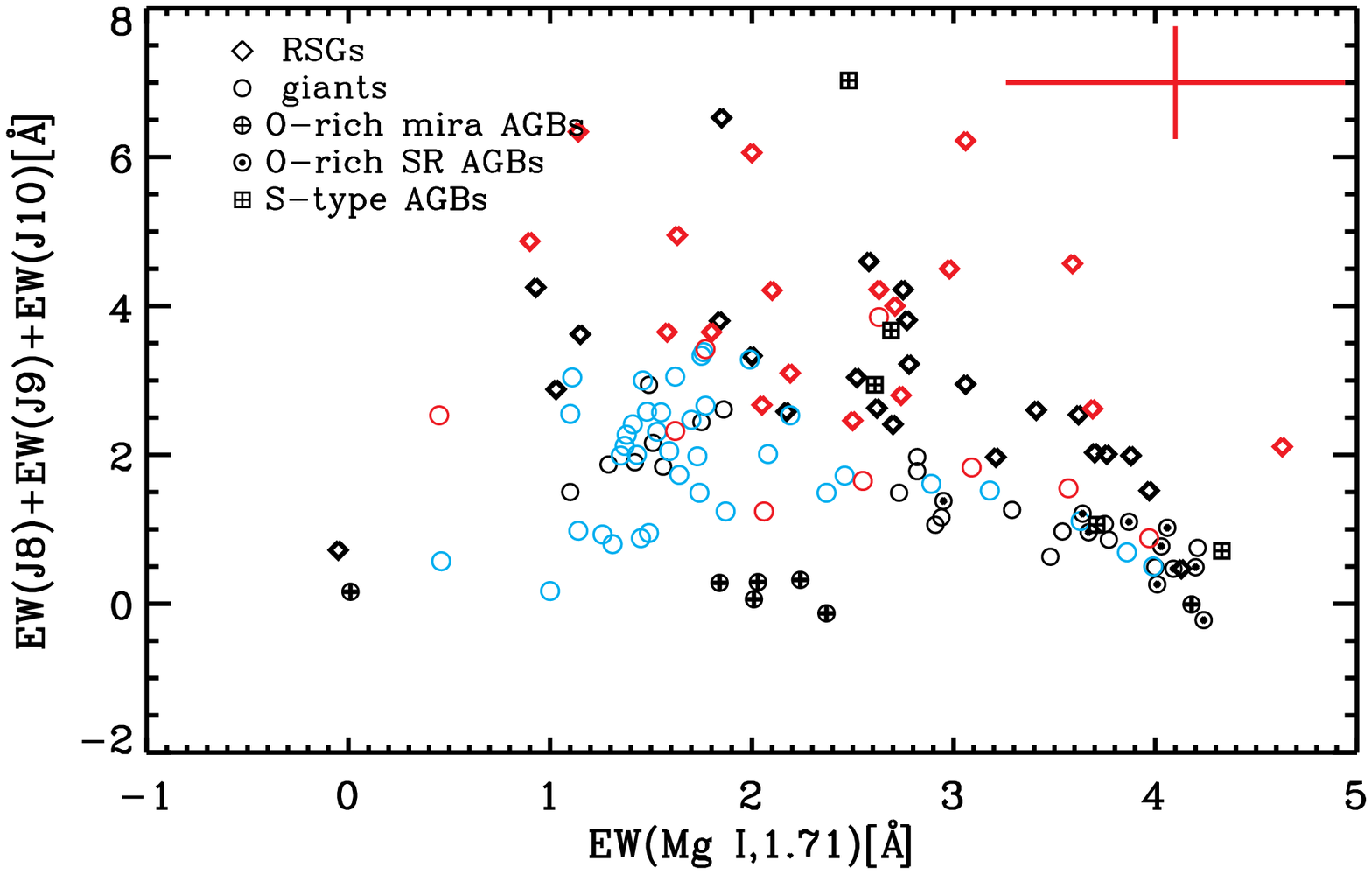}}
\resizebox{0.99\hsize}{!}{\includegraphics[angle=0]{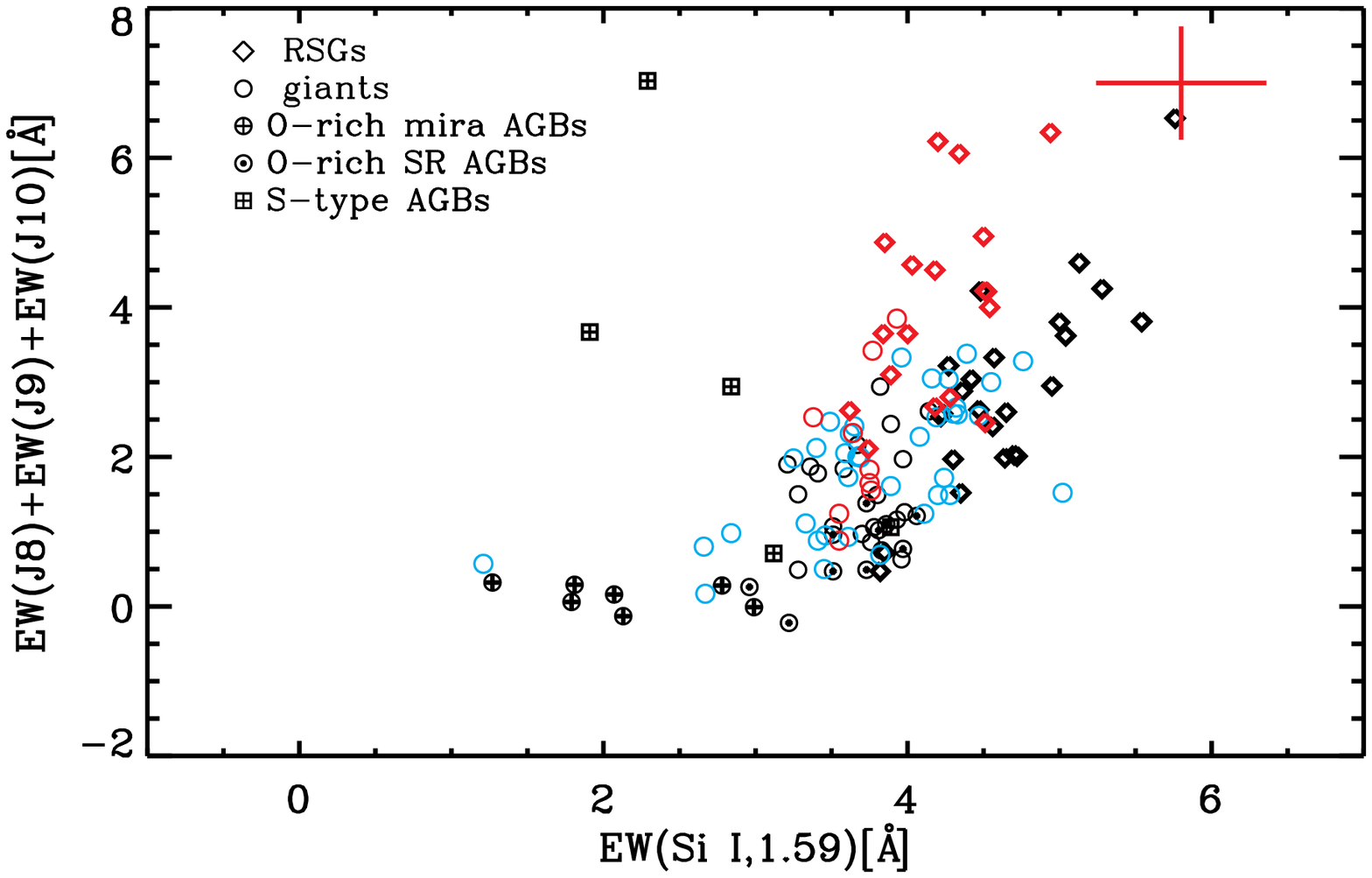}}
\resizebox{0.99\hsize}{!}{\includegraphics[angle=0]{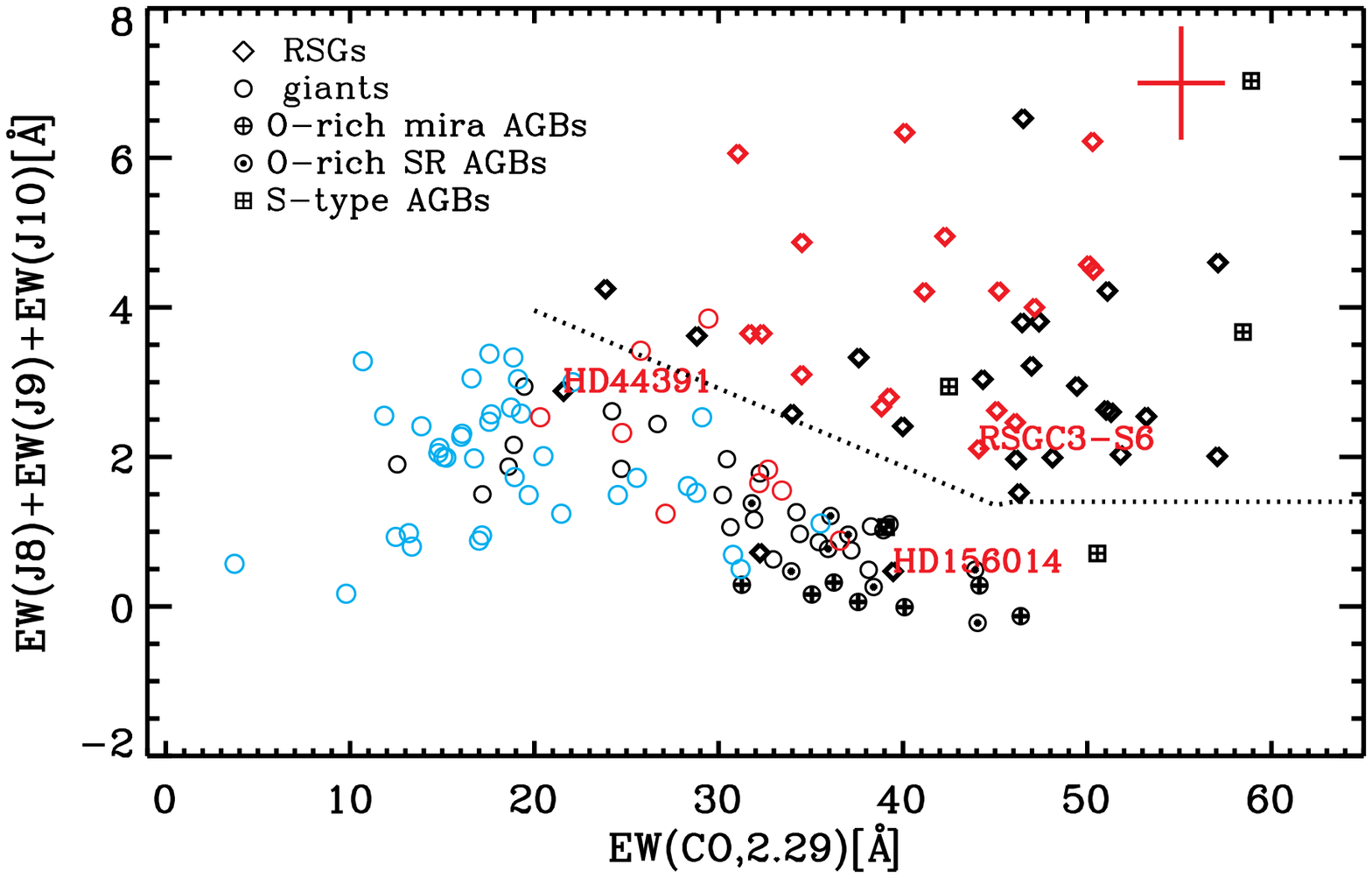}}
\end{center}
\caption{\label{comboCN} 
{\it Top panel:} \ewJoJnJd\ values are plotted vs. the  \ewmg\ values.
{\it Middle panel:} \ewJoJnJd\ values are plotted vs. the  \ewsiucn\ values.
{\it Bottom panel:} \ewJoJnJd\ values are plotted vs. 
the  \ewcodue\ values. Symbols and colors are as in Fig. \ref{CO4}.
Typically, RSGs are located above the dotted line.
}
\end{figure}

\begin{figure}
\begin{center}
\resizebox{0.9\hsize}{!}{\includegraphics[angle=0]{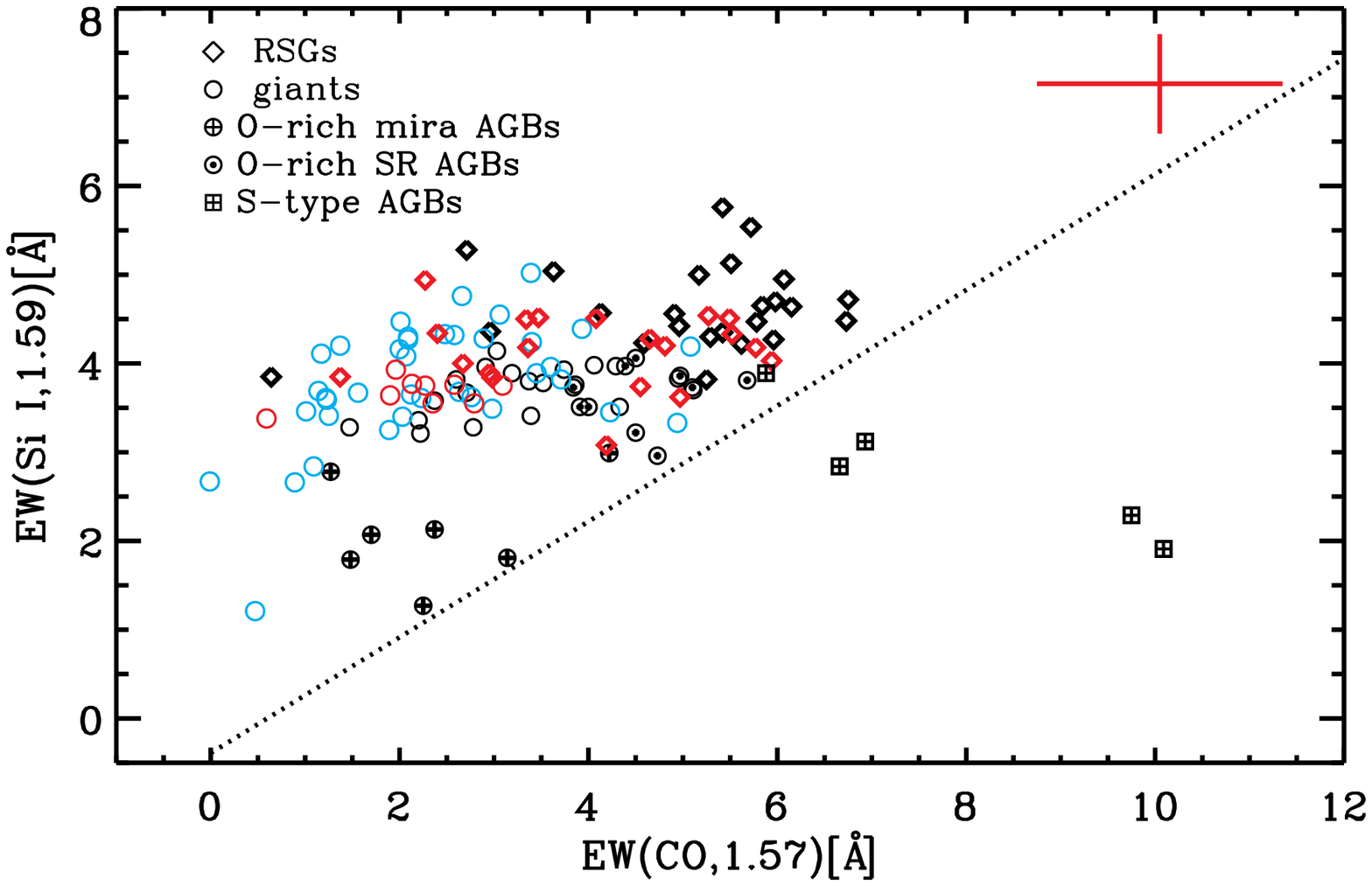}}
\resizebox{0.9\hsize}{!}{\includegraphics[angle=0]{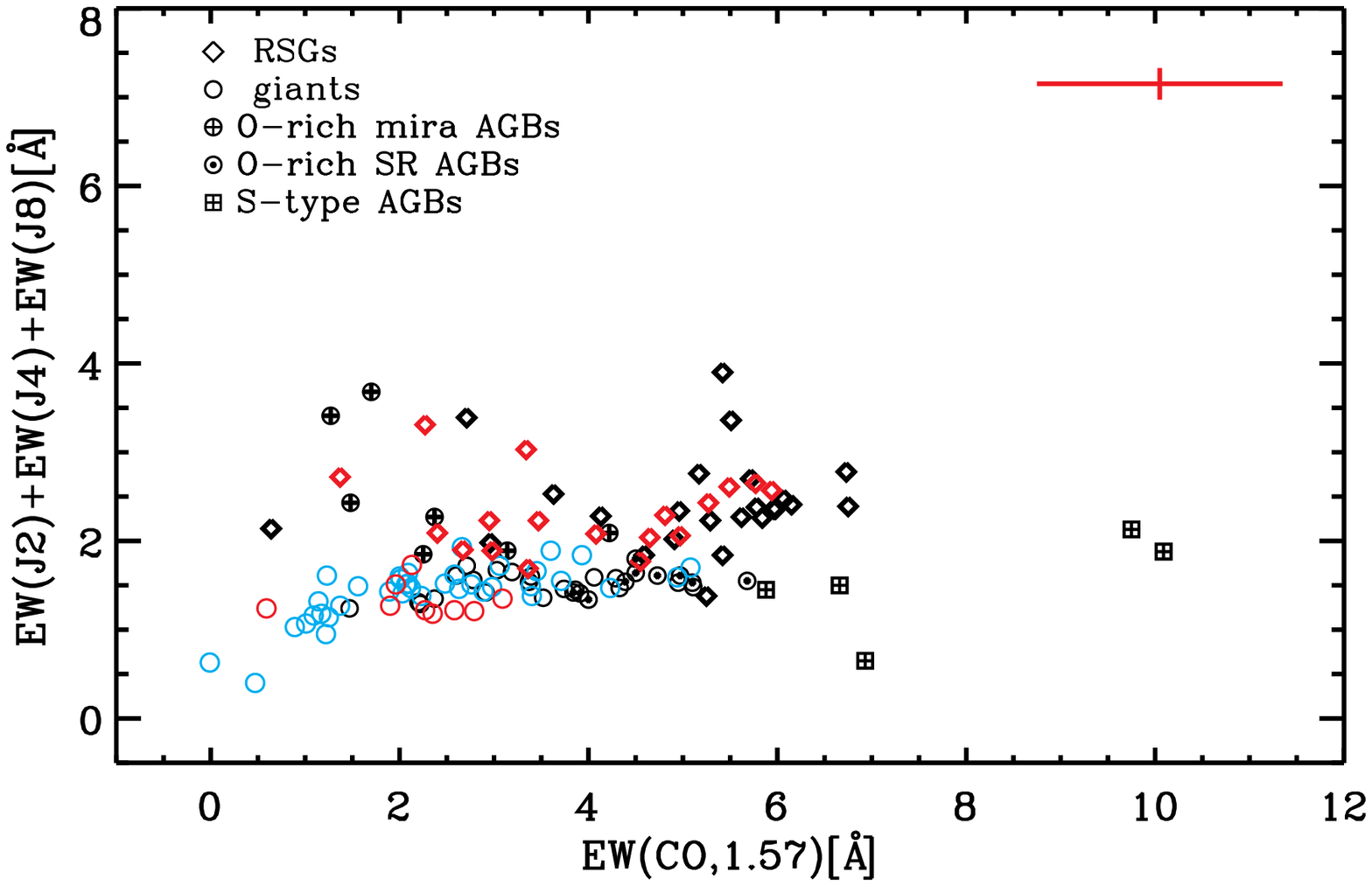}}
\caption{ \label{figStype} 
{\it Top panel:}  The \ewcoucs\ vs. \ewsiucn\ diagram.
Symbols and colors are as in Fig. \ref{CO4}.
Data points lying below the dotted line are typical 
of S-type stars  \citep{rayner09}.
{\it Bottom panel:}  The \ewcoucs\ vs. \ewJdJqJo\ diagram.
} 
\end{center}
\end{figure}

\begin{figure}
\begin{center}
\resizebox{0.9\hsize}{!}{\includegraphics[angle=0]{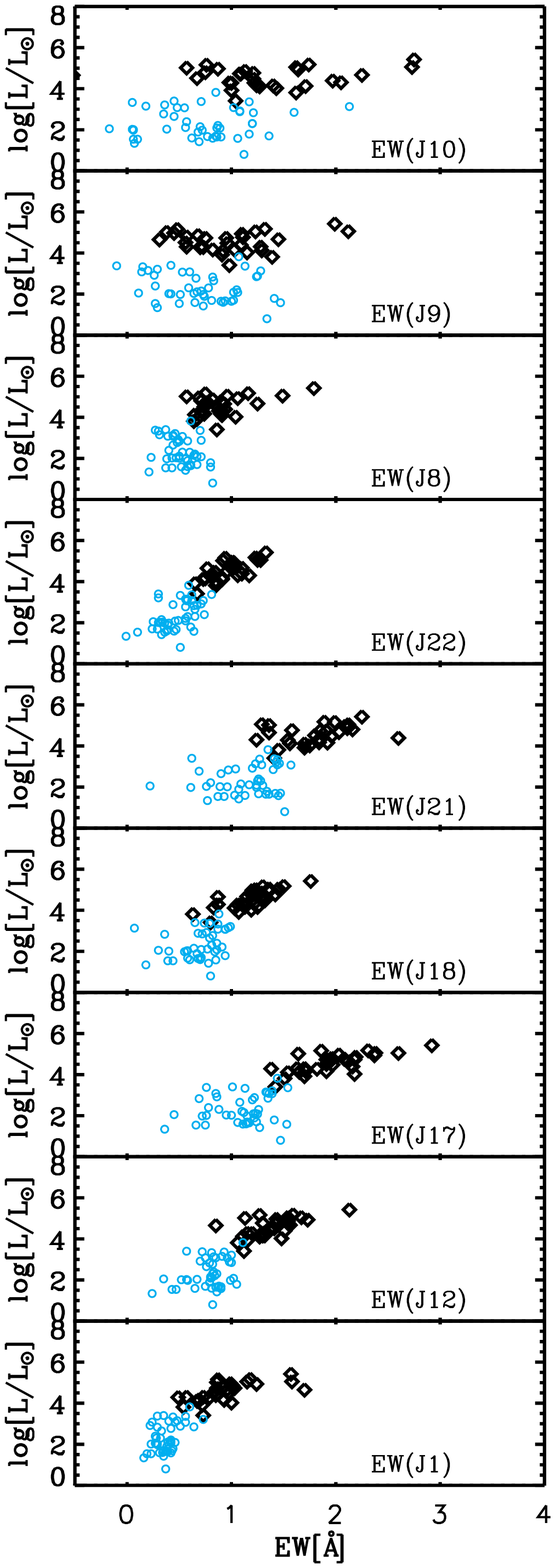}}
\end{center}
\caption{\label{lumsepa} 
{\bf The  log(L/\Lsun) (with Gaia EDR3 distances with fractional errors $< 25$\%)
are plotted vs.
EW values of the J1, J12, J17, J18, J21, J22 lines and J8, J9, and J10. }
 Indices  of RSGs from Table \ref{aliaskrsg} are 
{\bf marked with black diamonds}.
The indices of giants 
with luminosities estimated in \citet{villaume17} 
and the newly observed giants from Table \ref{aliasgiants}
are over-plotted {\bf with cyan open circles}.}
\end{figure}

\begin{figure*}
\begin{center}
\resizebox{0.48\hsize}{!}{\includegraphics[angle=0]{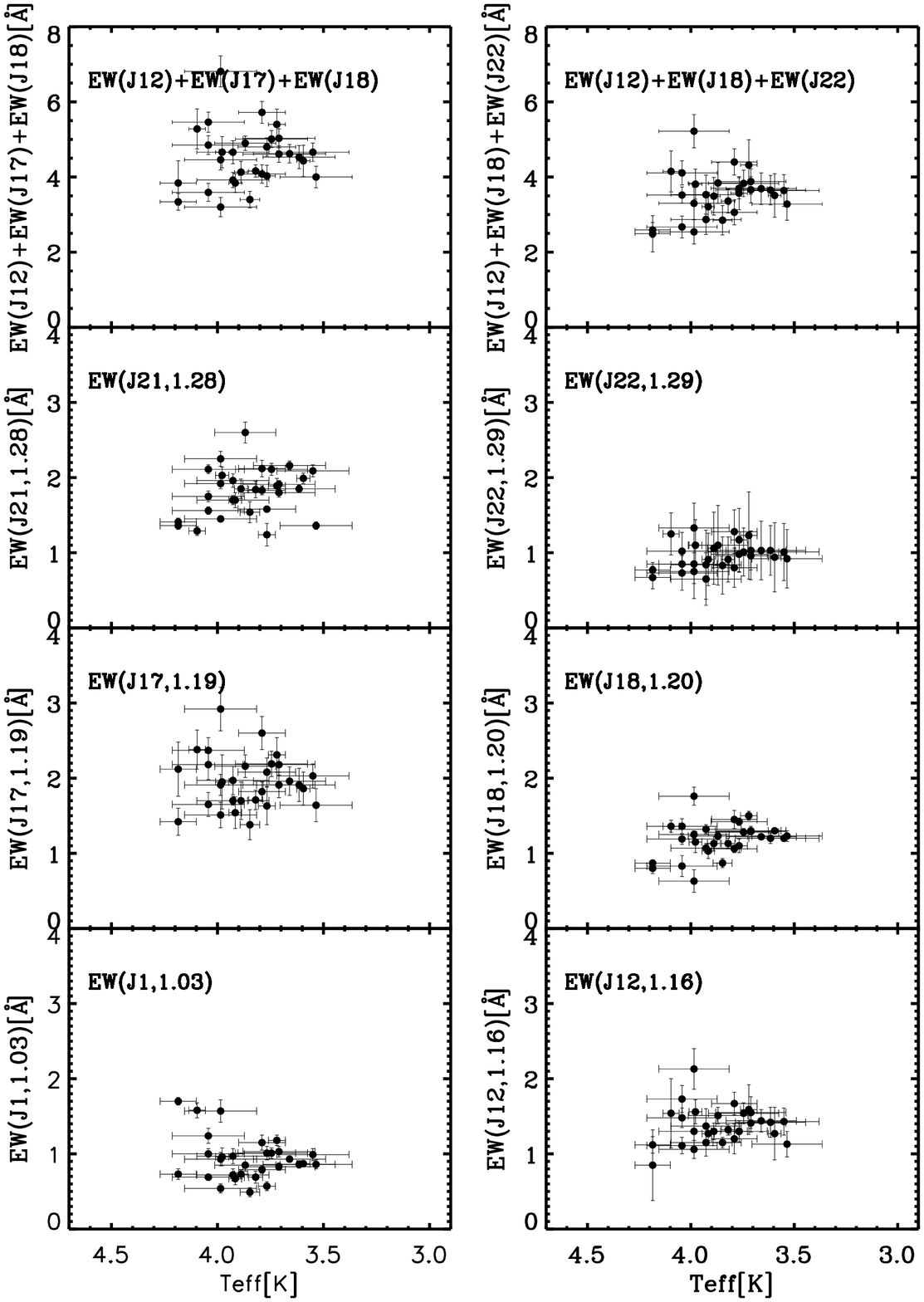}}
\resizebox{0.48\hsize}{!}{\includegraphics[angle=0]{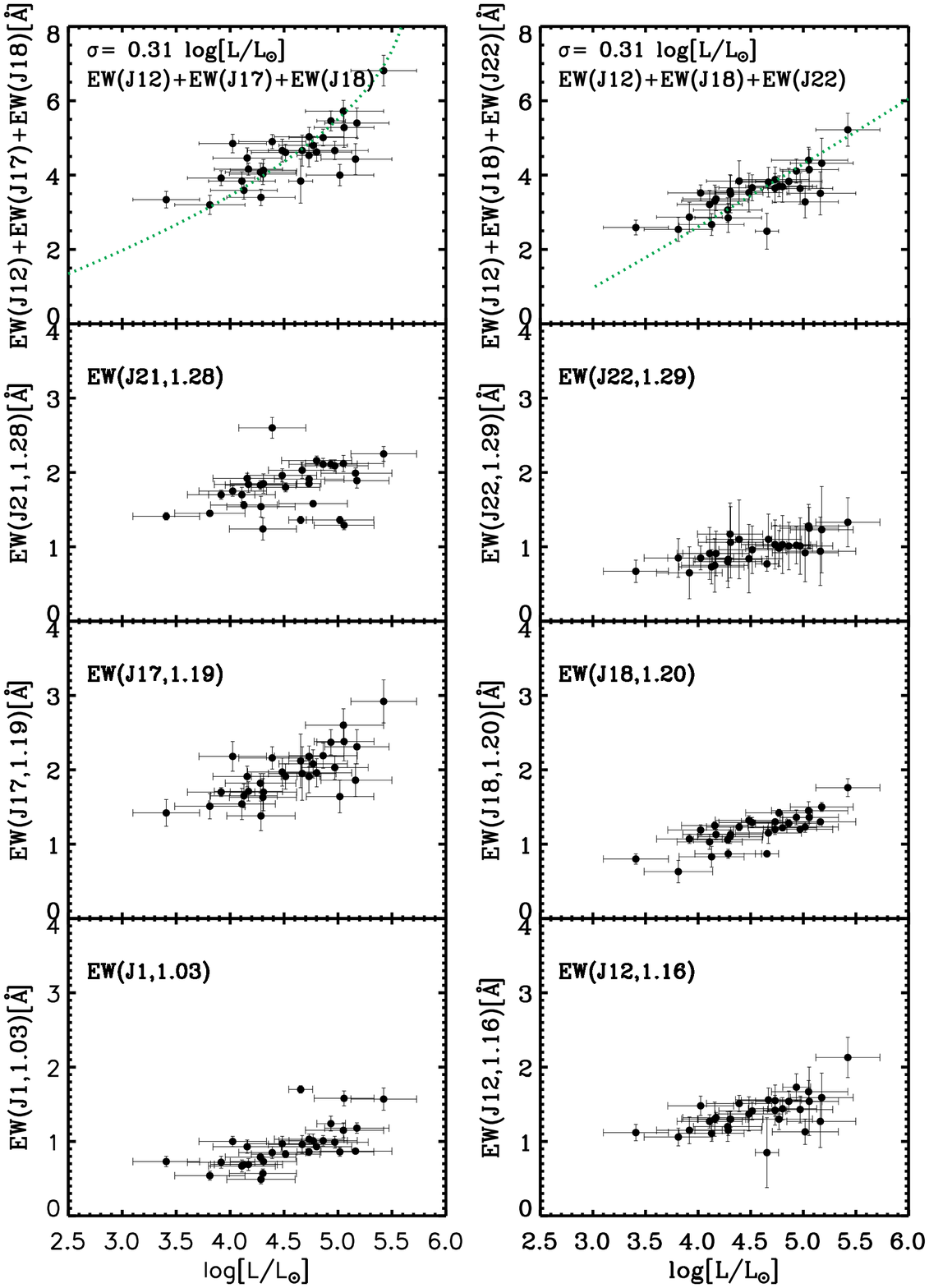}}
\end{center}
\caption{ \label{figEWlum-teff}  EW indicators of luminosity. 
{\it Columns 1 and 2:} 
The EW values of the J1, J12, J17, J18, J21, J22 
lines of known RSGs are plotted vs. the  \Teff\ values.
{\it Columns 3 and 4:}  For a direct comparison,  
their  EWs  are also plotted vs. 
the {\bf log(L/Lsun)} (Gaia EDR3 distances). 
These are the same RSG data points displayed 
in Fig. \ref{lumsepa}. 
Symbols are as in Columns 1 and 2. 
{\bf In the top two panels, EW(J12)+EW(J17)+EW(J18) vs. log(L/\Lsun) 
and  EW(J12)+EW(J18)+EW(J22) vs. log(L/\Lsun) diagrams,}
the two green dotted lines are polynomial fits to the data points
{\bf (used to estimate the 
spectroscopic log(L/\Lsun), see text below).}
}
\end{figure*}

\section{  Luminosities of supergiants}
\label{secgravity}

In this section, we analyze  the correlation between line strength and
 luminosities, with the aim  of locating 
spectral lines suitable for determining luminosity classes,
or even of predicting luminosities from \ew s.
Luminosities and \ew s are  independently obtained.

The luminosities of Galactic RSGs range from 
log(L/L$_\odot$)=3.5 to 5.6 \citep{levesque05, messineo19}.
Evolutionary tracks predict that gravity linearly correlates 
with $L$ and \Teff\ \citep{limongi17,bertelli09}, and an estimate of the stellar 
gravity can be obtained as described in Appendix  A.  

\subsection{Measurements of luminosity and \Teff\ of  RSGs} 
\label{seclum}

We collected a total of 45 spectra of 44 known RSGs observed 
with the SpeX detector on the IRTF telescope.
21  of these spectra  are from the IRTF library 
(i.e.,  all those inherent class I stars without counting  $\alpha$ Her, which is an AGB), 
two from the extended library, and 21 others 
were observed during our runs. 

Their luminosities were estimated  
by using parallactic  distances from Gaia EDR3\footnote{ The quasar median 
 zero point of $-0.017$ mas was applied to the parallaxes. 
The use of the provided ``tentative recipe'' \citep{gaiadr3} for obtaining 
more precise individual corrections as a function of magnitudes, 
colors, and positions would yield a mean correction of $-0.043$ mas with a  $\sigma$ of $0.019$ mas. 
The effect on the  luminosity would be a mean decrease of $-0.048$ with a $\sigma$ of 0.054.
Brightness fluctuations characterize  the surface of RSGs \citep{chiavassa11}.
However, this does not yields systematic parallax errors, but only 
degrades the quality of the fit \citep{messineo19}.
}
\citep{gaiadr3} and infrared magnitudes 
from 1 \um\ to 21 \um, 
as well as with the \BCK\ of \citet{levesque05} 
and \citet{neugent20}, as in  the catalog 
of \citet{messineo19}.
Interstellar extinction \Aks\ was estimated using the 
2MASS $JH$\Ks\ photometry \citep{skrutskie06}, the intrinsic colors 
from the work of \citet{koornneef83}, and the extinction 
power law described 
in \citet{messineo05} \citep[see also,][]{messineo17,messineo19}.
\Teff\ values were estimated using the temperature 
scale of \citet{levesque05} and the adopted spectral types 
\citep[see also,][]{messineo19}.
Estimated luminosities are listed in Table \ref{aliaskrsg}.
As the  spectral types adopted by \citet{messineo19}
were often the average 
of historical records listed in \citet{skiff14}, 
the original references to the spectroscopic works 
 are also provided in Table \ref{aliaskrsg}.

For the comparison sample of spectra of giants from the  
extended IRTF library,  luminosity estimates are given 
in the catalog of \citet{villaume17}.
Their luminosities were estimated with direct integration under the 
stellar energy distributions, which were builded 
with observed flux densities from UV to infrared light and
with the Gaia DR2 parallactic distances. 
For these giants, in the plots 
we use the \Teff\ listed in the catalog of \citet{prugniel11}.
The \Teff\ values of the MILES spectra 
were determined by comparison with 
the template spectra of the ELODIE library using an interpolator.

\begin{table}
\caption{\label{lumind} Infrared luminosity-class indicators 
for evolved late-type stars}.
\begin{tabular}{lrlll}
\hline
\hline
Line ID & Center & Elements & separator     &  \\
        & [\um]  &          &  \ew(RSG)[\AA] &\\
\hline
J1   &1.0331     & Sr at 1.0330141 & $> 0.7$ \\
J12$^a$  &1.1613 & Fe+Cr+Si        & $> 1.0$ \\ 
J17  &1.1888     & Fe+Ti           & $> 1.5$ \\
J18  &1.1977     & Fe+Ti           & $> 1.0$ \\ 
J21  &1.2826     & Ti+Ca           & $> 1.7$ \\
J22  &1.2904     & Mn at 1.290336  & $> 0.8$ \\
\hline
\end{tabular}
\begin{list}{}
\item {\bf Notes:} For Galactic RSGs,
the above infrared lines have \ew s with similar behaviors.
They do not correlate with  \Teff, and neither with \ewmg. 
They do correlate with \ewsiucn\ and with each other (Fig. \ref{figalJplus}), 
and  are larger in RSGs than in giants.
\item $^a$ J12 corresponds to FeCr gravity indicator of \citet{morelli20}. 
\end{list}
\end{table}

\subsection{New indicators of luminosity in  $Y$- and $J$-bands}
\label{secJlum}

The  $Y$- and $J$-bands  indices defined in Table \ref{inddef}  
were plotted 
against the \ewcodue, \ewmg, and \ewsiucn\  values, respectively.
The \ewcodue\ is  very sensitive  to temperature,
as well as the Mg I line at 1.71 \um.
The strengths of the Si I at 1.59 \um\  and  Mg I line at 1.71 \um\ 
are uncorrelated; the Si I line  tends to be  higher in RSGs 
\citep[e.g. plots in][]{origlia93}.
The analysis of these plots  promptly allowed us to locate other  
lines with analogous behaviors.
We found that the strengths of the J1  (1.033 \um), 
J12  (1.161 \um), J17 (1.188 \um), J18  (1.198 \um), 
 J21  (1.282 \um), and J22  (1.290 \um) lines 
remained quite constant with
increasing strength of  Mg I, i.e. 
they were independent of temperature, 
and RSGs were located visibly 
above a certain threshold, 
with giants and AGBs below it, as listed in Table 
\ref{lumind}.
Moreover, they would linearly correlate with the strength 
of the Si I at 1.59 \um.
When plotting these $Y$- and $J$-band indices  against each other,
a linear correlation is  seen.
To minimize the observational errors, combined indices 
were also measured, for example, \ewJudJusJuo\  and  \ewJduJdd\   
(see Fig. \ref{figalJplus}). 
For these two  indices, using the thresholds  given  in Table 
\ref{lumind}, we were able to retrieve  82\% and 82\% 
of the sample of known RSGs, respectively
(or 79\% and 74\% of the K-type RSG sample).
The J12  feature at 1.161 \um\  may contain Fe I, Cr I, and Si I 
atomic lines, as well as strong CN bands, 
as seen in the high-resolution spectrum of 
the K-type giant Arcturus \citep{hinkle95}. 
The J17 and J18 features at 1.189 and 1.198 \um\ 
are made up of blended  Fe I and Ti I lines. 
The J21 feature at 1.283 \um\  is made up of Ti I, Ca I, 
and possible CN  lines.
The J1 (1.033 \um) and J22 (1.290 \um) lines  are  unblended 
lines from Sr I and Mn I, respectively.

In contrast to  the  $Y$- and $J$-bands lines above mentioned, the
 \ewJoJnJd\ values  appear slightly 
anti-correlated with those of CO and Mg I in red giants 
(the J9 and J10 lines  at 1.087 and  1.093 \um\  are CN dominated).
The diagrams of  \ewJoJnJd\ values 
versus CO or Mg I \ew s are  useful indicators of luminosity classes,
as shown in Fig. \ref{comboCN}. 
The area above the dotted curve encloses 95\% of the known RSGs
(90\% of the known K-type RSGs). 
Unfortunately, S-type stars occupy  the same area as RSGs.
However, in the \ewcoucs\ vs. \ewJdJqJo\
S-type stars have fainter  \ewJdJqJo\ and larger \ewcoucs\
than those in RSGs (see Fig. \ref{figStype}).
The J4 blend contains the Si I line at 1.0588 \um\ (J4) and
the J8 the Si I line at 1.078 \um.

These  indices  enable us to separate RSGs  from giants using only
the  $Y$- and $J$-bands. Figs. \ref{lumsepa}, 
and \ref{figEWlum-teff}
{show the relation between  these indices and luminosity. }
The \ew s of the J1, J12, 
J17, J18, J21, and J22 lines  mildly 
correlate  with the stellar luminosities, 
 but not with the \Teff\ values
(Fig. \ref{figEWlum-teff}).

\begin{figure}
\begin{center}
\resizebox{0.48\hsize}{!}{\includegraphics[angle=0]{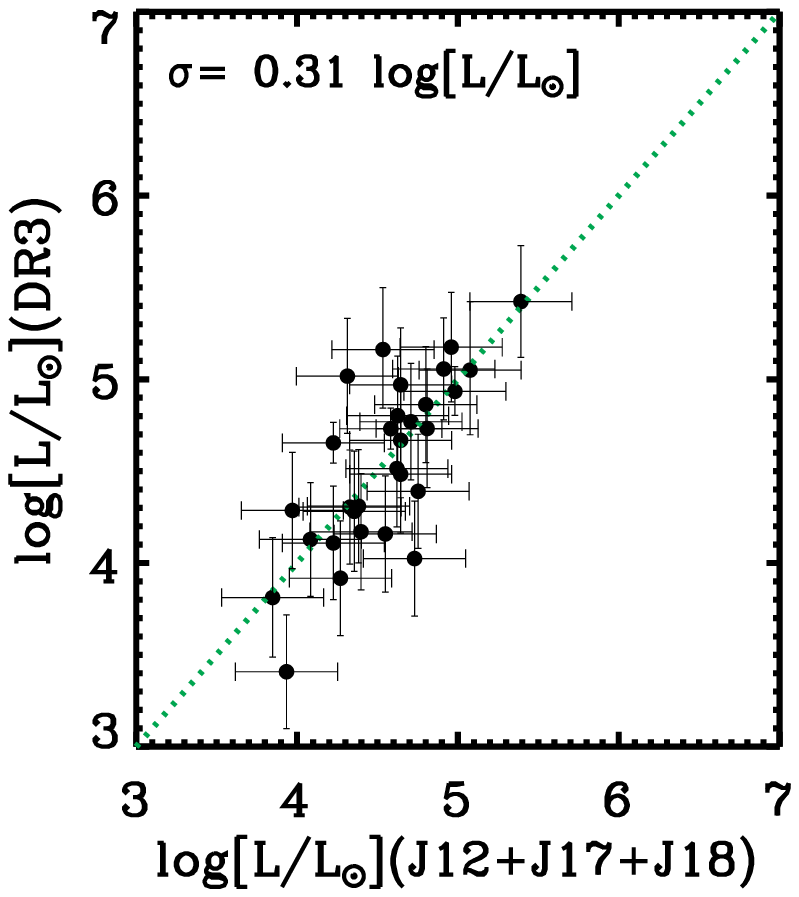}}
\resizebox{0.48\hsize}{!}{\includegraphics[angle=0]{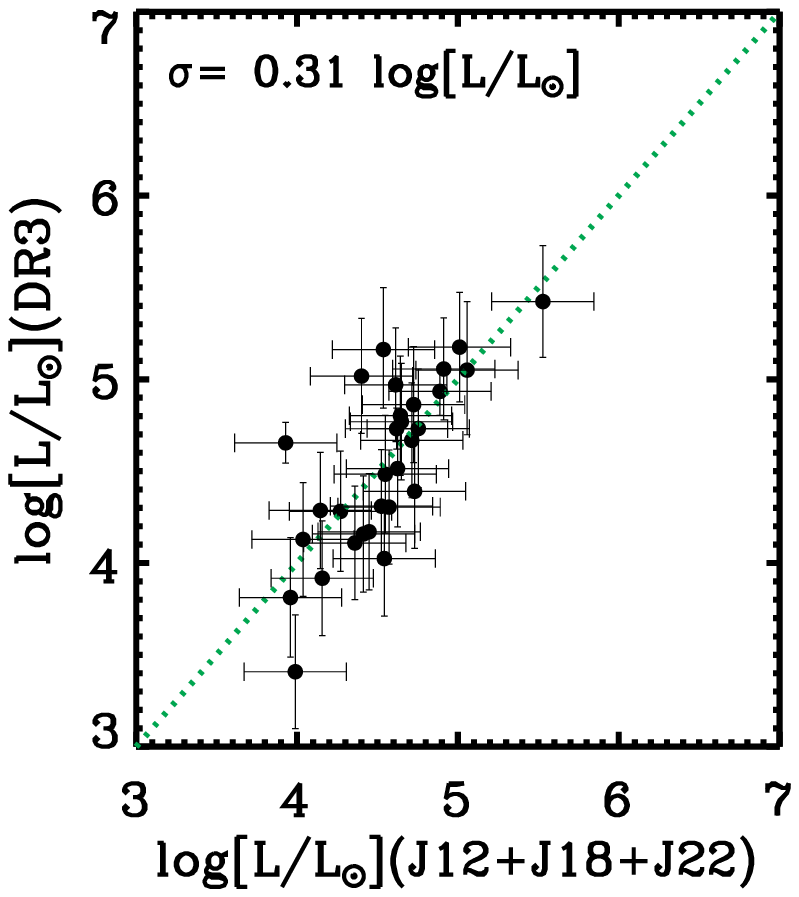}}
\end{center}
\caption{\label{speclum} 
Examples of spectroscopic luminosities. Luminosities inferred from the J12, J17, J18  
lines of RSGs {\it (left panel)} and from the J12, J18, J22  lines of RSGs {\it (right panel)}
are compared with the photometric luminosities (based on the Gaia EDR3 distances). 
Identity is depicted by the green dashed line.}
\end{figure}

A direct estimation of luminosities from these spectral indices 
is possible.
For example, \ewJudJusJuo\ yields luminosities in agreement
with those from photometry with a sigma=0.31 Log(L/\Lsun),
as elucidated in Fig.\ \ref{speclum}.

We remark that all findings of this work are based on empirical
correlations of observational \ew s at infrared wavelengths.
These tools are stable and independent of any modeling.

\section{Magnesium: the \ewcodue\ versus \ewmg\ diagram}
\label{magnesiumsec}
The Mg I at 1.71 \um\ is a key infrared line that allows 
us to separate giants  from RSGs.

In the \ewcodue\ versus \ewmg\  diagram, 
giants are located above RSGs  \citep{messineo17}, 
as shown in  Fig.\ \ref{CO4}. A linear curve is drawn to roughly mark
the two sectors of the diagrams (that occupied by giants, and that
of RSGs).  More precisely, 95\% of  the known RSGs  listed 
in Table \ref{aliaskrsg} are located to the right of this curve.
This diagram separates well RSGs (with spectral types 
from K0 to M5) and normal giants, 
independently of their spectral types.
At the two extremes, three data points deviate from the trend:
the  2MASS J06224788+2759118/HD 44391 (K0 Ib), 
which is nevertheless identified by its 
strong CaT lines and CN head bands, and
 2MASS J18451939$-$0324483/RSGC3-S6 (M4.5 I) and 
 2MASS J17143885+1423253/HD 156014 (M5 Ib-II),  which
remain located on the giant sequence of the diagram. 

C-rich and S-type AGB stars, as well as the O-rich Mira AGB stars,
 are distributed similarly to the RSGs.

\begin{figure*}[h]
\begin{center}
\resizebox{0.47\hsize}{!}{\includegraphics[angle=0]{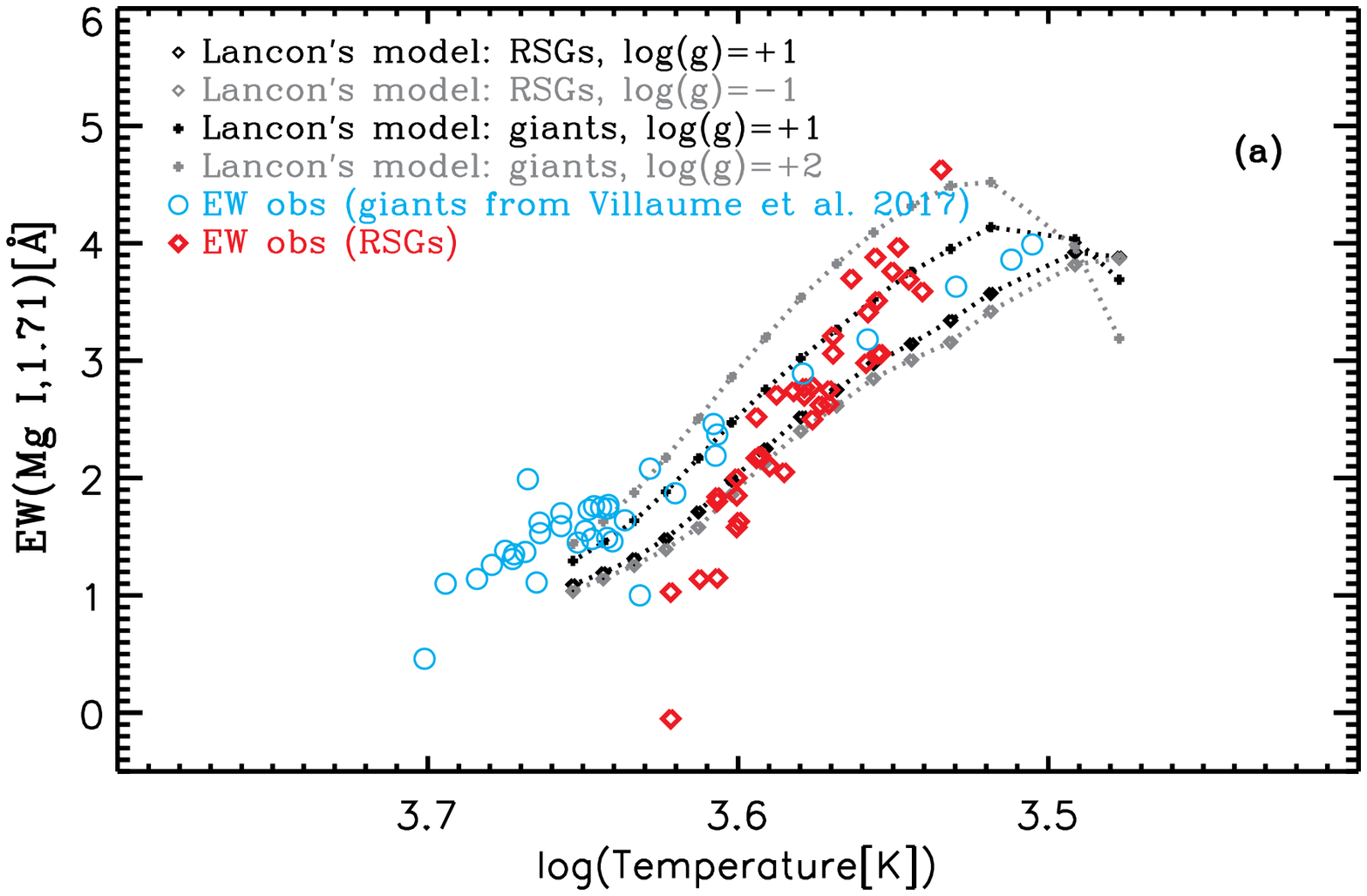}}
\resizebox{0.47\hsize}{!}{\includegraphics[angle=0]{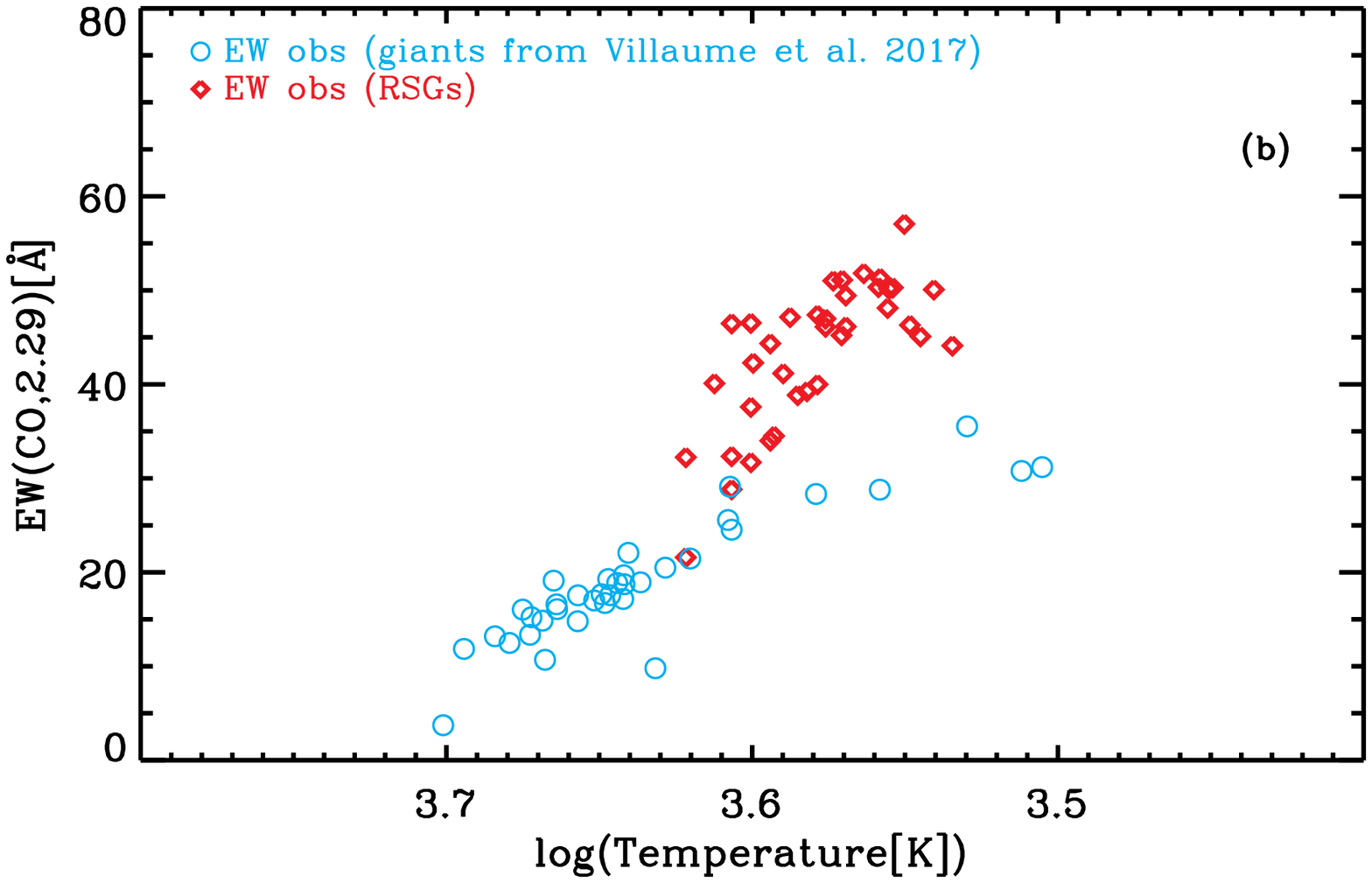}}
\resizebox{0.47\hsize}{!}{\includegraphics[angle=0]{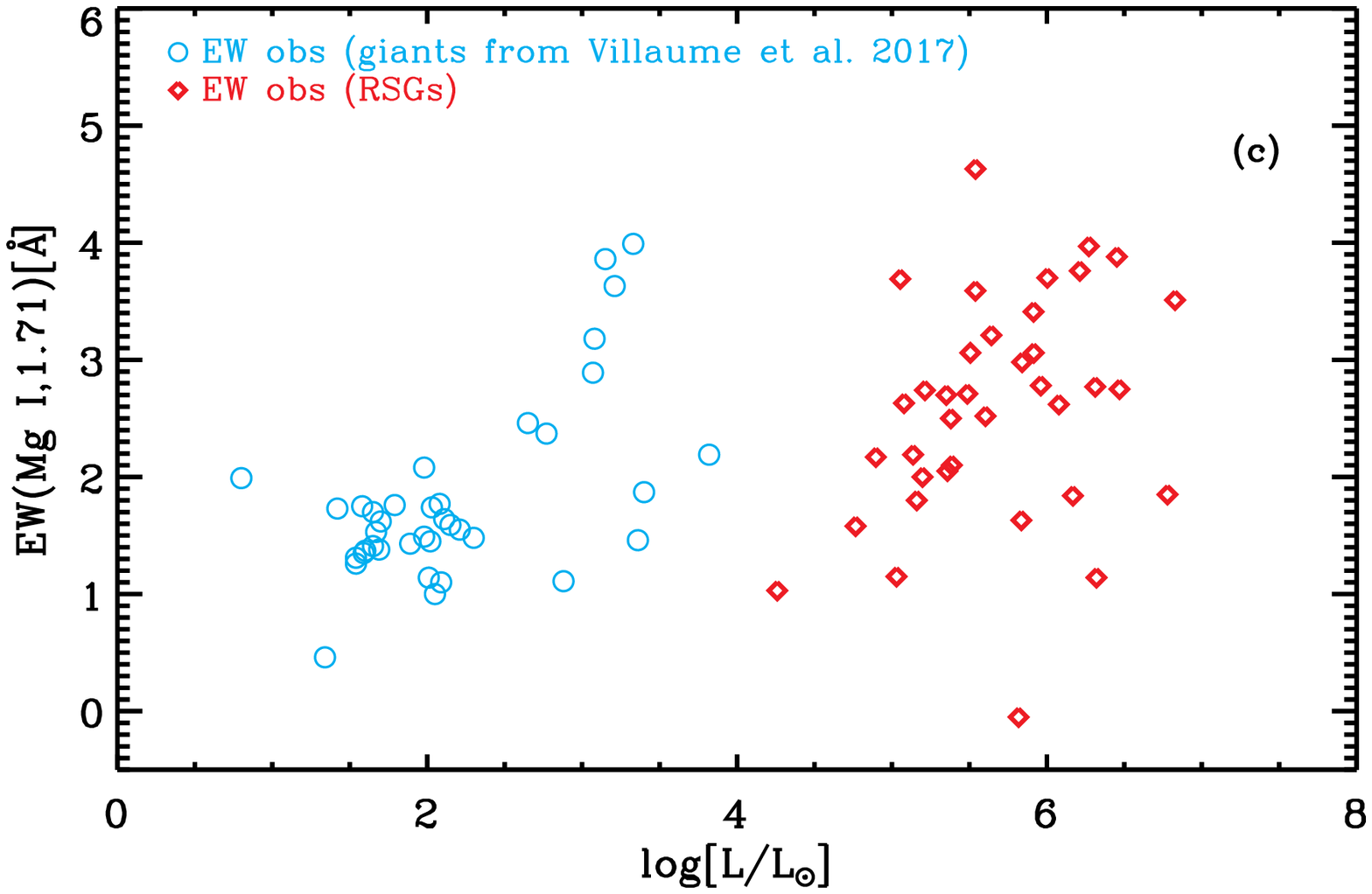}}
\resizebox{0.47\hsize}{!}{\includegraphics[angle=0]{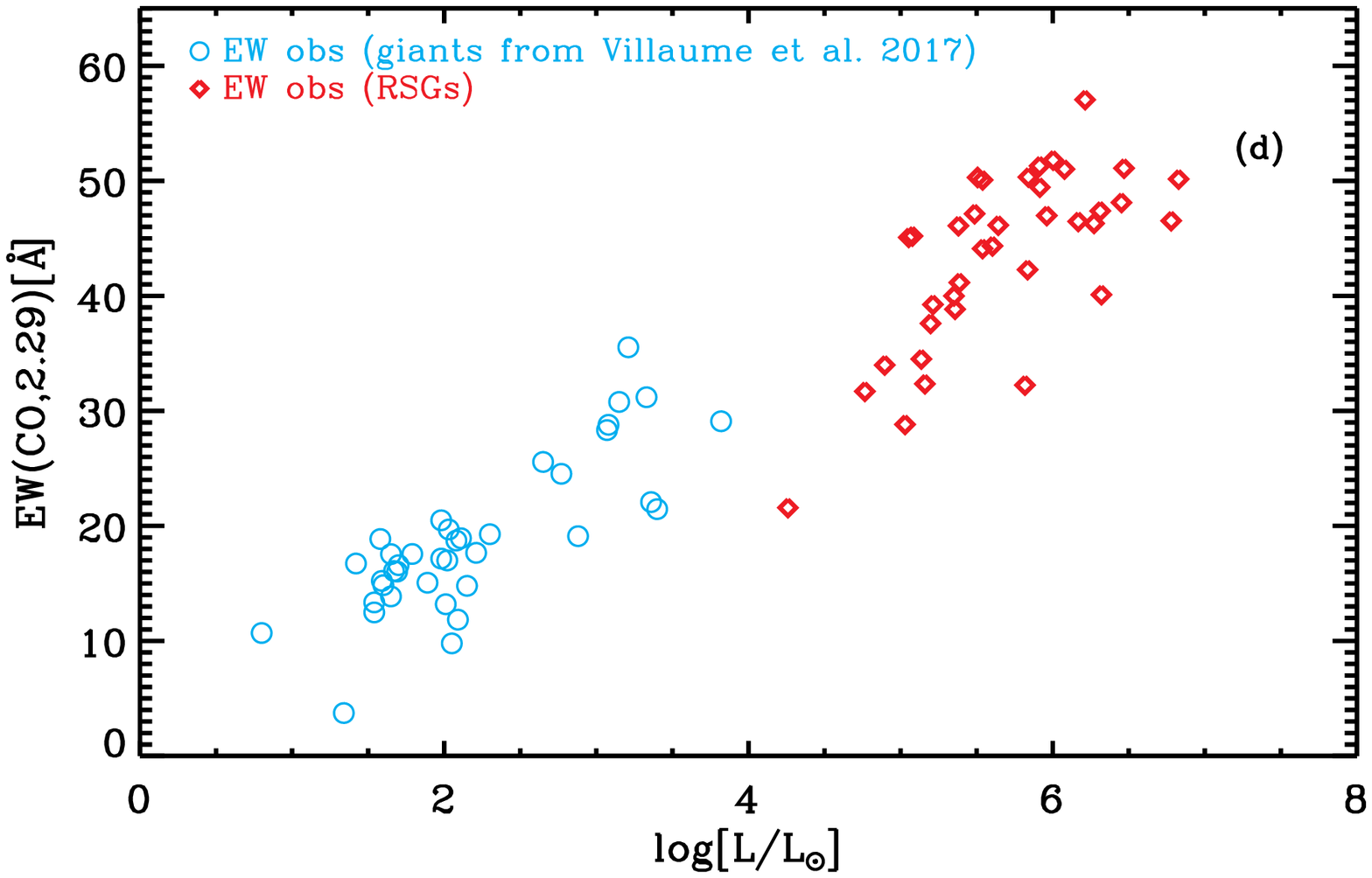}}
\resizebox{0.49\hsize}{!}{\includegraphics[angle=0]{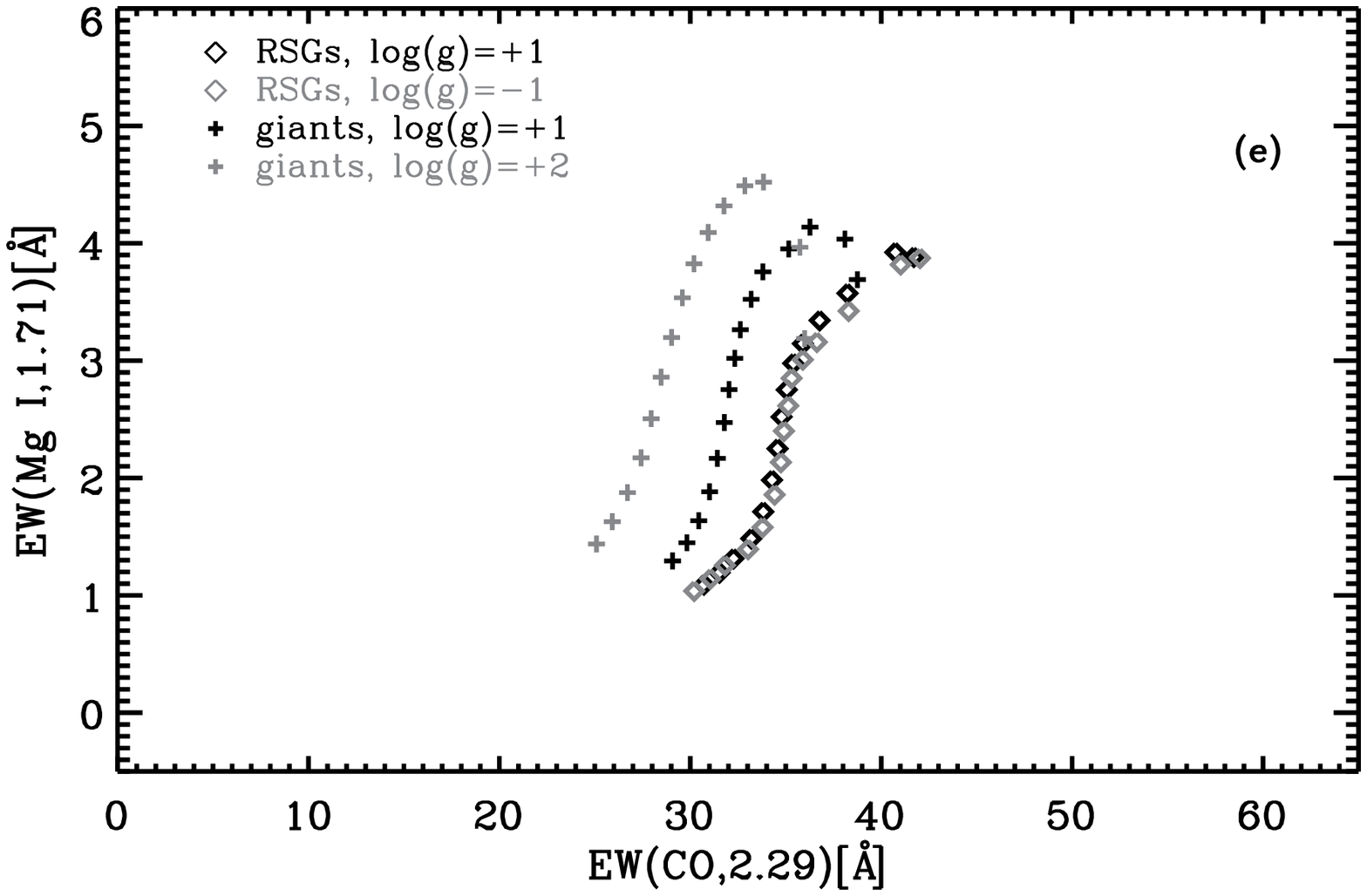}}
\resizebox{0.47\hsize}{!}{\includegraphics[angle=0]{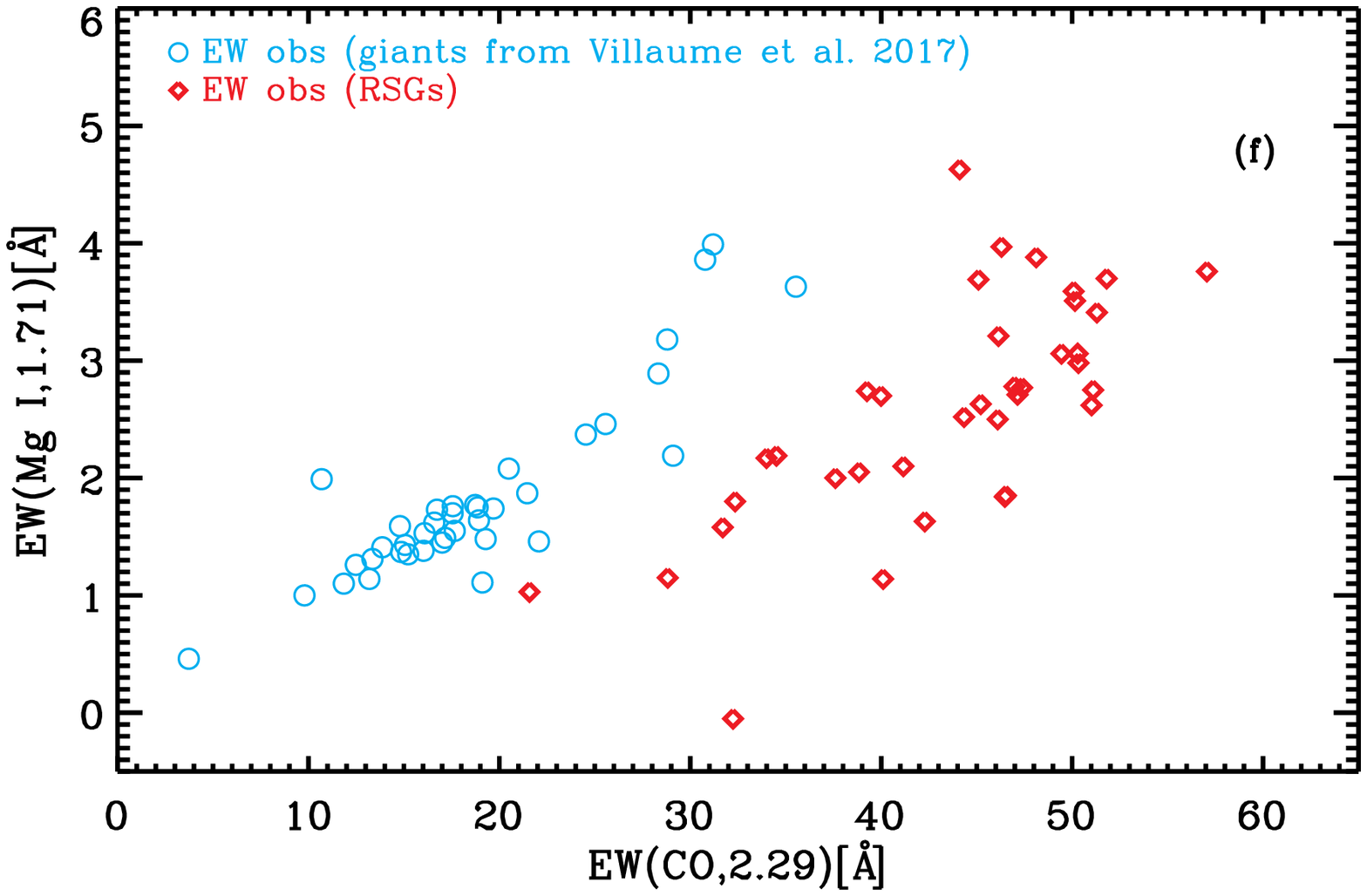}}
\caption{ \label{figlancon} 
{\it Panel (a):}  The \ewmg\ values vs. stellar \Teff\ estimated from 
the models of \citet{lancon07}. Connected gray crosses  mark the locus of giants 
with log(g)=+2, black crosses  the locus of giants with log(g)=+1. 
Connected gray diamonds  mark the locus of RSGs with log(g)=$-1$, black diamonds  
the locus of RSGs with log(g)=+1. 
 The observed \ewmg\ and \Teff\ of giants  
from the work of \citet{villaume17} are overplotted with cyan circles;
those of known RSGs (IRTF libraries and newly observed) with red diamonds; 
\Teff\ values are taken from  Table \ref{aliaskrsg}.
{\it Panel (b):} \ew s of the CO band head at 2.29  
\um\ vs. stellar \Teff. 
Symbols are as in panel (a).
{\it Panel (c):} \ewmg\ values vs. luminosities.  
Symbols are as in panel (a).
{\it Panel (d):}  The \ewcodue\ values vs. the  stellar luminosities. 
Symbols are as in panel (a).
{\it Panel (e):}  The theoretical \ewmg\ values vs. the \ewcodue\ 
values estimated from the models of \citet{lancon07}.
{\it Panel (f):}  The observed \ewmg\ values  vs. the \ewcodue\ values;
symbols are as in panel (a).
} 
\end{center}
\end{figure*}


For stars of 1 and 15 \Msun, a grid of synthetic infrared spectra 
with solar metallicity covering temperatures 
from  2900 K to 4500 K and log(g $[$cm s$^{-2}]$) from $-1$ to 2   
has been published by  \citet{lancon07}.
Galactic RSGs have a typical log(g $[$cm s$^{-2}]$) value ranging from $-0.5$ to 1 
\citep{davies10}  (see Appendix  A).
Giants cover a broader range from +3  (near 5000 K) 
to $-0.2$ (near 3000 K) \citep{bessell98}.
In the panel (a) of Fig.\  \ref{figlancon}, we used the 
Lancon et al. synthetic spectra to reproduce 
the theoretical distribution of \ewmg\  values 
with temperatures.
The Mg I absorption line at 1.71 \um\ strengthens  
with decreasing  temperature.  
By lowering the stellar  surface gravity, this line 
notably weakens (panel (e) of Fig.\  \ref{figlancon}).
Indeed, as shown in panel (f),  observed Galactic  giants  
have a stronger Mg I line than RSGs;  
The \ewmg\ values  of giants sit on a different sequence above that of RSGs. 
The  Mg I line at 1.71 \um\ correlates only with the stellar temperature, 
as shown in panels (c) and (d) of Fig. \ref{figlancon},  unlike the CO lines.
These behaviors of the EWs are 
confirmed with the high-resolution spectra of \citet{park18}.

When plotting the Mg I versus the CO absorption strengths 
measured from   the models  of \citet{lancon07} 
(panel (e) of Fig. \ref{figlancon}), a 
luminosity separation is visible, with the giants data points lying on 
the top of those of supergiants. 
The diagram of \ewmg\ values versus \ewcodue\ values 
 provides an observational tool to separate normal giants and supergiants,
because it removes the ambiguity  in the relation of \ewcodue\  
and  \Teff\  diagram.

\begin{figure*}
\begin{center}
\resizebox{0.48\hsize}{!}{\includegraphics[angle=0]{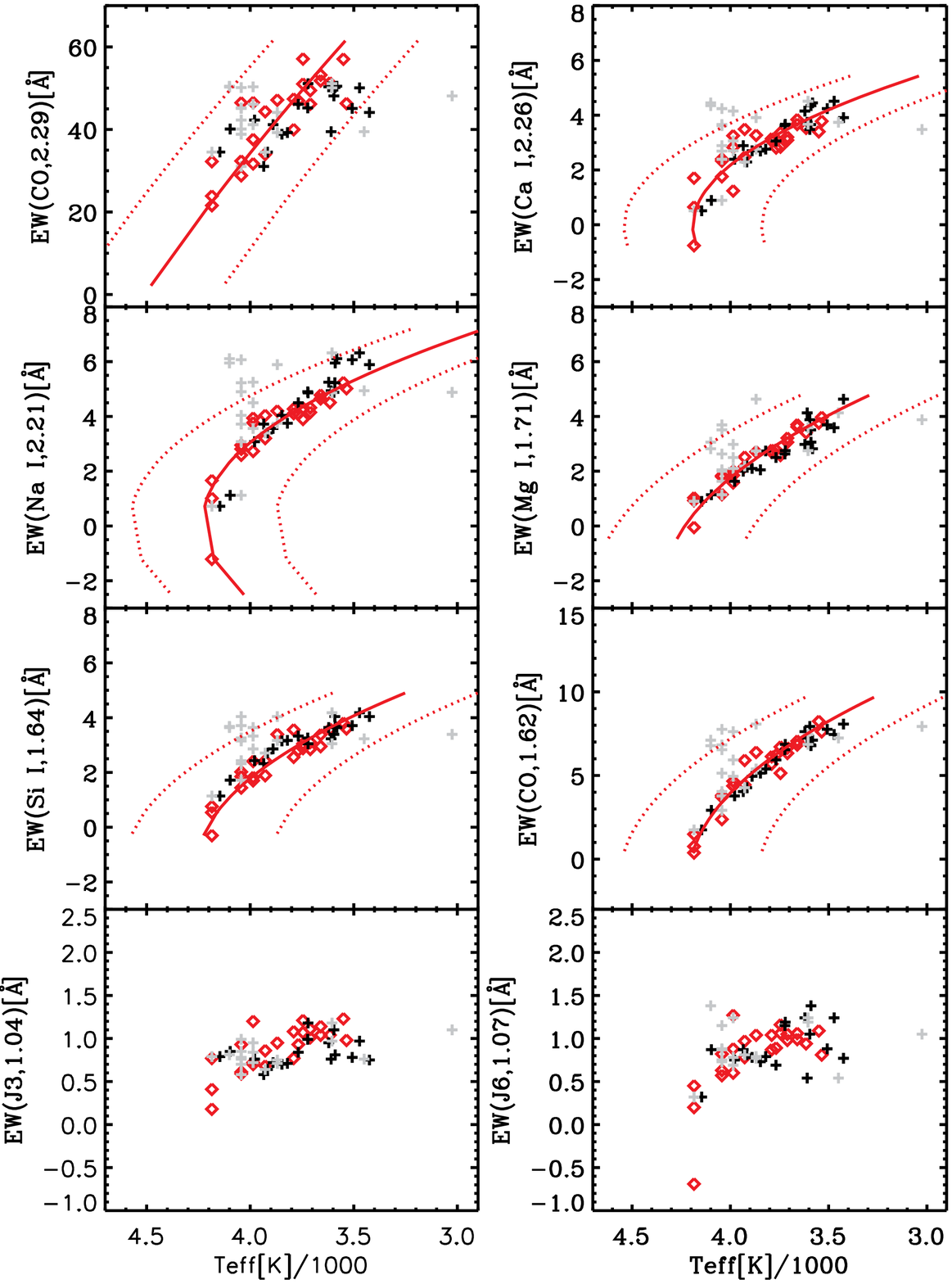}}
\resizebox{0.48\hsize}{!}{\includegraphics[angle=0]{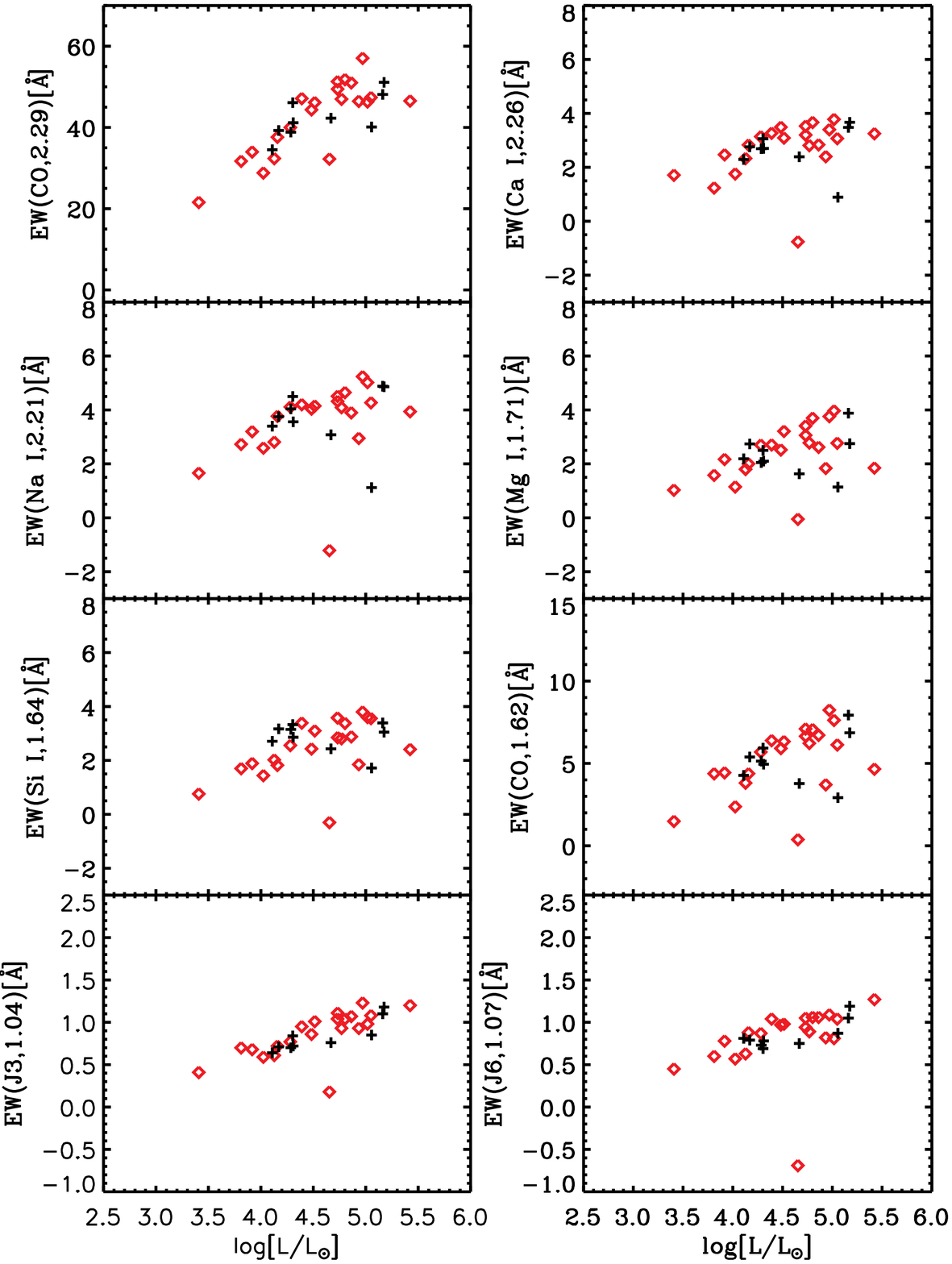}}
\caption{ \label{measureTeff}  
 EW indicators of temperature.  
{\it Columns 1 and 2:} 
In each panel,  red diamonds mark  the \ew s of  reference RSGs 
with optical spectral type vs. the \Teff,  
as estimated by \citet{messineo19} with the  scale 
of \citet{levesque05}.
Most of these stars come from the sample of  \citet{keenan89}, 
as listed in Table \ref{aliaskrsg}.
From top-left to bottom right, \ew s from CO at 2.29 \um, 
Ca at 2.26 \um, Na at 2.21 \um,
Mg at 1.71, Si at 1.64 \um, CO at 1.62 \um, J3 at 1.04 \um, 
and J6 at 1.07 \um, respectively. 
Continuous red lines show  2nd order polynomial fits to the 
EWs of reference RSGs.
Dashed red lines show the 2nd order polynomial fits shifted by $\pm350$ K.
Crosses marks the \ew s of  other known RSGs, mostly with 
infrared spectral types (CaT or CO),
in gray when using the literature \Teff, and in 
black when using the \Teff\ estimated in this work,  \tcombi.
{\it Columns 3 and 4:} In each panel,  
the same \ew s of   RSGs are plotted  vs. the stellar luminosities
(Gaia EDR3 distances). Symbols are as in Columns 1 and 2.
} 
\end{center}
\end{figure*}

\section{ Temperatures of supergiants}
\label{sectemperature}

The \Teff\ values of known RSGs range 
from about 4500 K to 3500 K.
At optical wavelengths, spectral types of RSGs
and temperatures (\ttio) are usually defined on the basis of the strengths of 
their  TiO bands.
In recent literature, this temperature scale 
has been  intensely discussed  and even questioned. 

Infrared estimates of spectral types have been 
carried out with the \ew s of the CO band head at 2.29 \um\ 
\citep[e.g.,][]{figer06}.
At shorter wavelengths, 
the TiO band heads at 0.88 \um\ is often used 
\citep{dorda16b,negueruela16}. 
 The infrared spectral types are more uncertain that 
those from optical studies.

We found that the \ewcodue, \ewca,  \ewna, \ewmg, 
\ewsiusq, \ewcousd,  \ewJtre, 
and \ewJsei\ values  correlate  with 
the \ttio\  temperatures
as shown in Fig.\  \ref{measureTeff}.
We neglected  any dependence on Z, because the  Z range of Galactic
RSGs is small anyway, and we used  2nd order polynomial fittings  
 to   estimate the  temperatures of RSGs. 
 We used 21 spectra of RSGs from the two IRTF libraries 
(we excluded  the three outliers 
 2MASS J20285059+3958543 (RW Cyg), 
2MASS J17143885+1423253 (HD 156014), and 
 2MASS J22543171+6049388 (MY Cep), 
plus  2MASS J19461557+1036475 ($\gamma$ Aql), 
 2MASS J21045763+4708491 (HD 200945), 
and  2MASS J20112810+2201354 (BD+21 4089), 
 which were observed during our runs.  
The results are listed in Tables \ref{aliaskrsg}, \ref{aliasnewrsg}, 
and \ref{aliasgiants}. The fits were done with temperatures 
from 4180 K (K0) to 3535 K (M5). 
We used extrapolations for  temperatures  lower than that 
(i.e, larger \ew s) 
 resulting in M5-M6 types (see Table \ref{aliaskrsg}).

\begin{figure*}
\begin{center}
\resizebox{0.33\hsize}{!}{\includegraphics[angle=0]{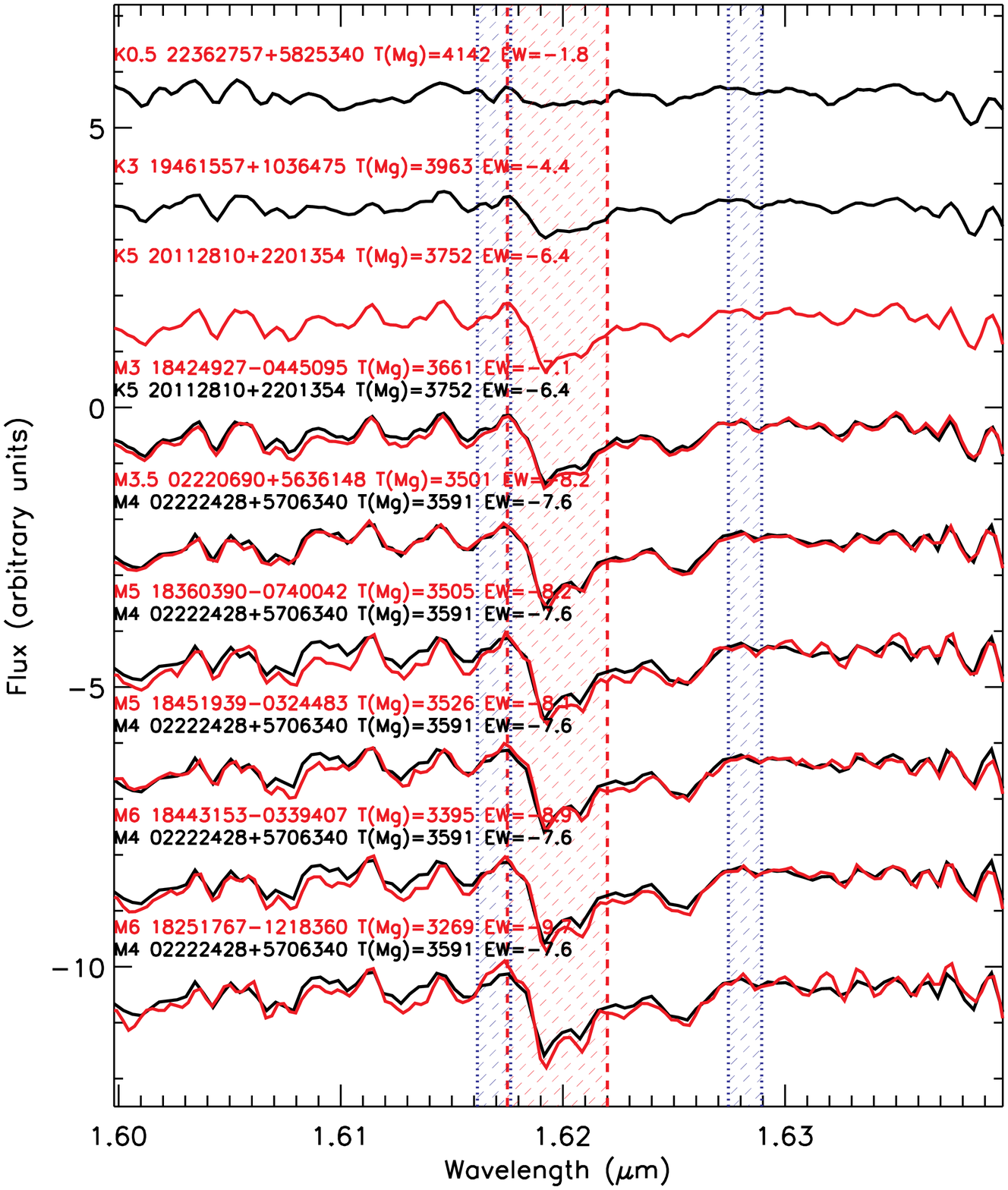}}
\resizebox{0.33\hsize}{!}{\includegraphics[angle=0]{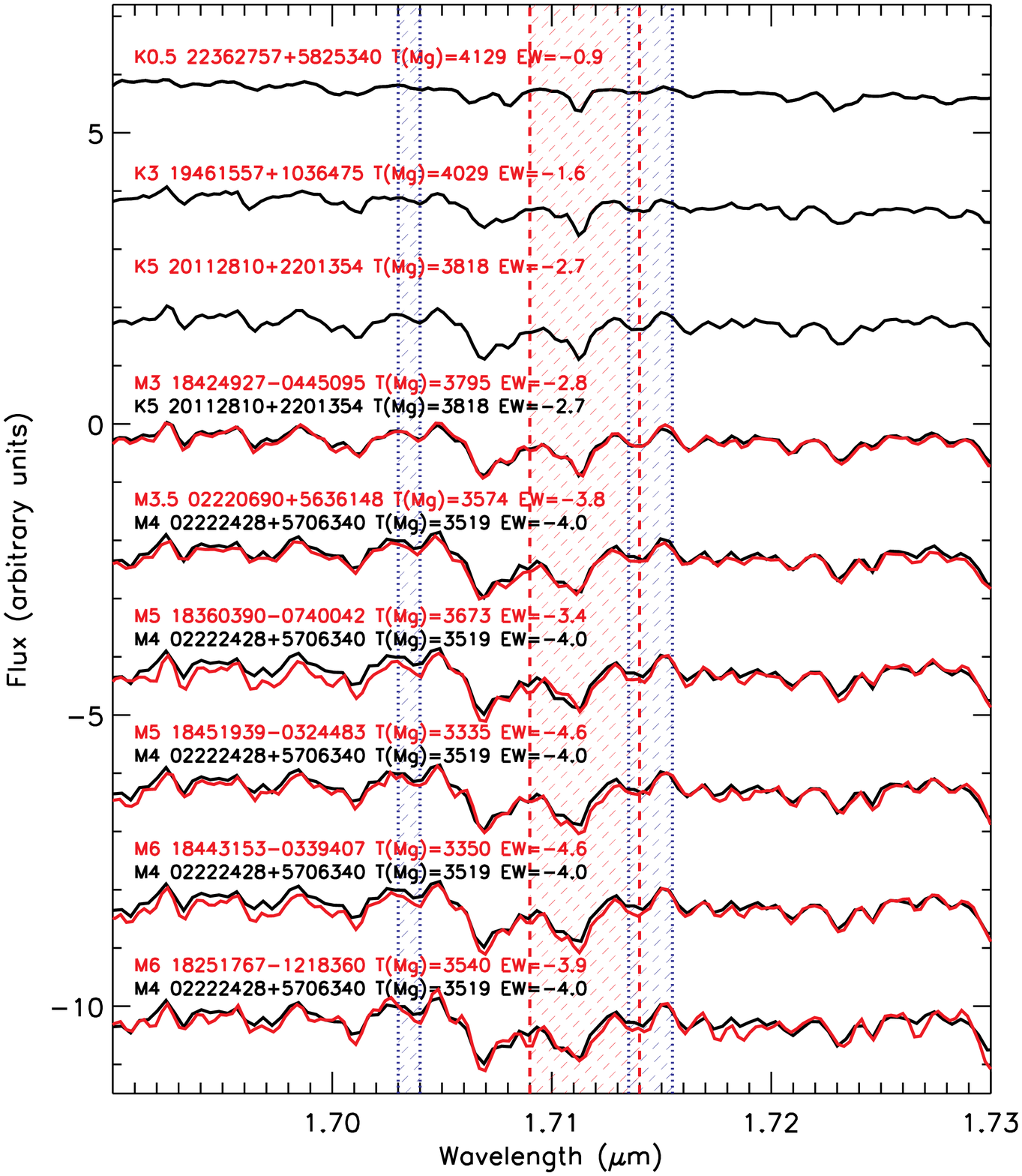}}
\resizebox{0.33\hsize}{!}{\includegraphics[angle=0]{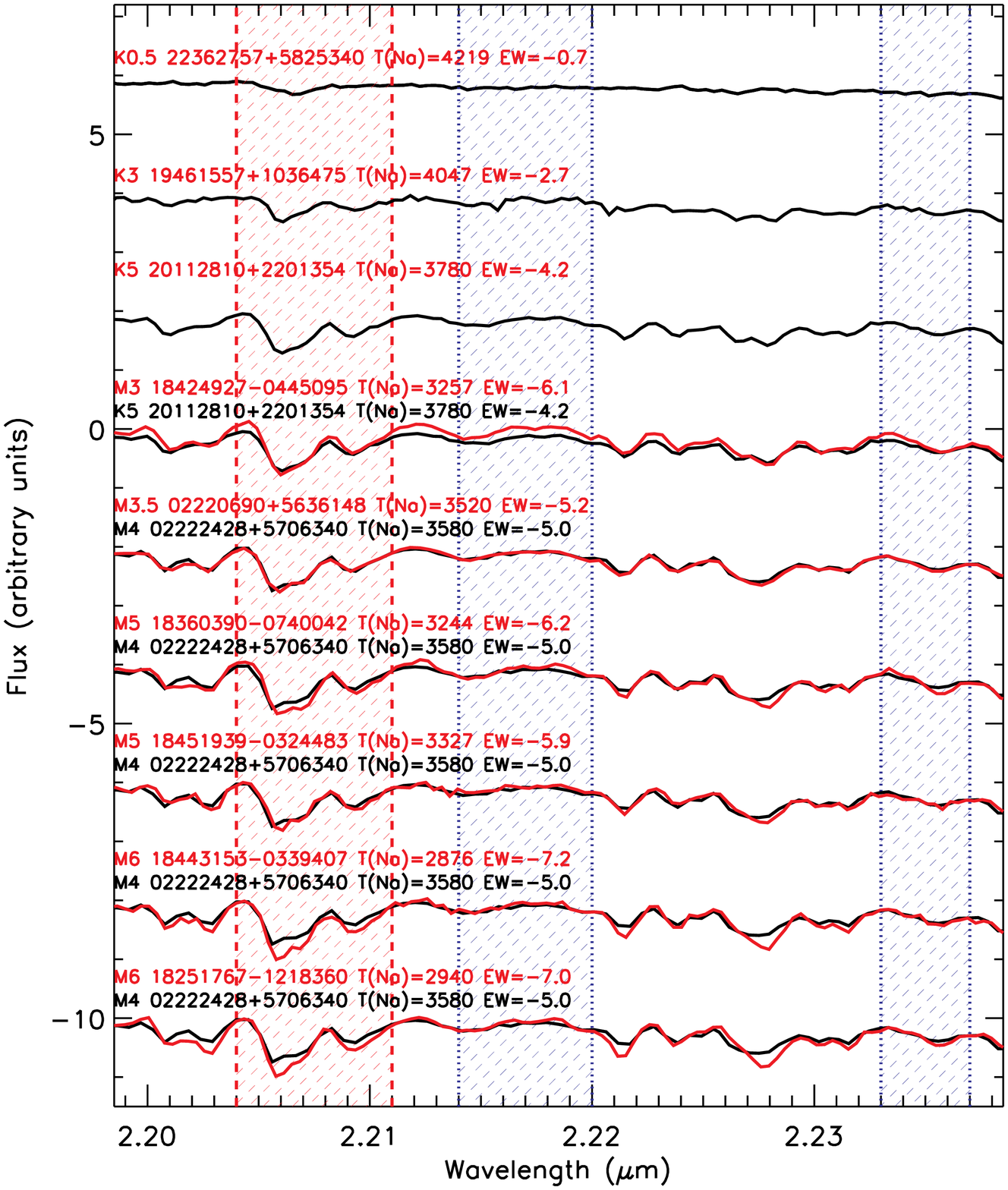}}
\end{center}
\caption{  \label{sequencetemp} Examples of spectral features
due to CO at 1.62 \um\ (left panel), 
Mg at 1.71 \um\ (central panel) , Na at 2.2 \um\ (right panel). } 
\end{figure*}

The \ewcodue,  \ewJtre, and \ewJsei\ show a dependence 
on both luminosity and \Teff;
the  \ewca, \ewna, 
\ewmg, \ewsiusq, and the \ewcousd\  values 
 do not show a clear  correlation with  the luminosities; 
so the latter \ew s are  better indicators of temperature.
 The high-resolution data of \citet{park18} 
show  that the \ewca, \ewna, and \ewmg\ of 
cold supergiants 
 do not correlate  with  gravity,
and increase  with increasing temperatures.\\
The average temperatures and spectral types 
 were inferred by averaging  the
 \ewna, \ewmg, \ewsiusq, and  \ewcousd\
of  our targets with those of reference RSGs, 
 and are indicated as
\tcombi\ and \spcombi, respectively. 
The average \ew s of RSGs per spectral type are listed in Table 
\ref{table.averagetemp}, and  three spectral features 
per spectral type are shown in Fig.  \ref{sequencetemp}.

\begin{figure*}
\begin{center}
\resizebox{0.7\hsize}{!}{\includegraphics[angle=0]{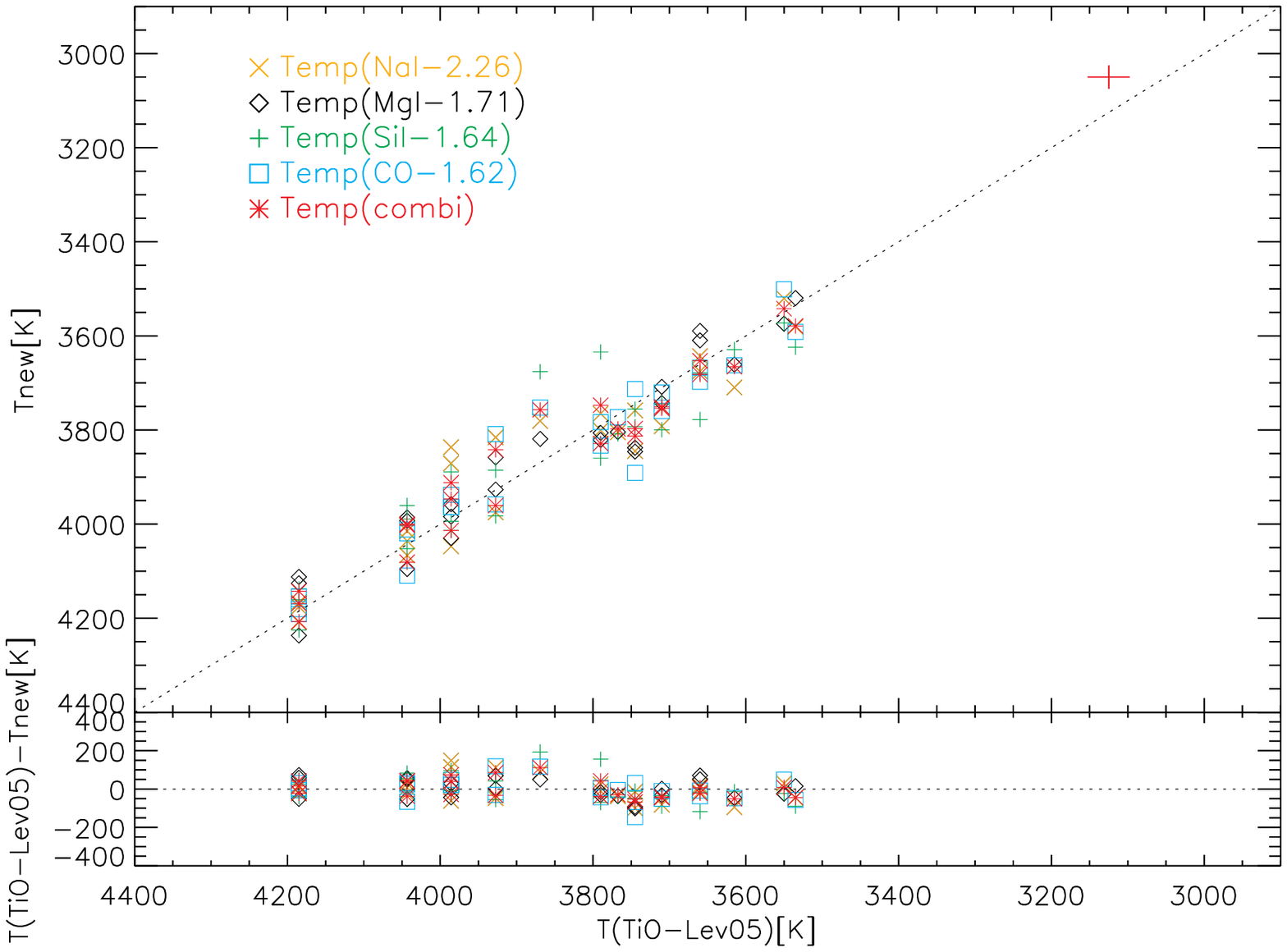}}
\caption{ \label{DeltaTeff}  
{\it  In the top panel}, the infrared temperatures estimated with 
a 2nd order polynomial fitting of the \ew s
are plotted vs. the \ttio\ temperatures (inferred from the
optical spectral types and the temperature scale of \citet{levesque05}).
Orange crosses mark temperature estimates from Na I at 2.26 \um, 
black diamonds those from Mg I at 1.71 \um,
green pluses those from Si I at 1.64 \um, 
cyan squares those from CO at 1.62 \um,
and red asterisks refer to an average temperature (\tcombi) 
from NaI, MgI, SiI, and CO at 1.62 \um.
The identity line is marked with a dashed line. 
The vertical segment of the red cross shows the $\sigma$ 
of the  \tcombi$-$T(TiO) differences, 
while the horizontal segment is the average difference 
between the temperatures of two 
spectral types, e.g. K4 and K5.   
{\it In the lower panel}, the temperature differences are plotted 
vs. the \ttio\ temperatures. 
} 
\end{center}
\end{figure*}

For the reference stars, the differences between 
the inferred temperatures and 
those from the optical classification of 
\citet{keenan89} and \citet{levesque05} 
are shown in Fig. \ref{DeltaTeff}. 
The estimated temperatures 
agree with those based on TiO \citep{levesque05} 
with a $\sigma$ of about 100 K.
In conclusion, by using infrared \ew s 
it is possible to accurately reproduce temperatures (within 150 K)
which are consistent with those optically made.
We followed the calibration of \citet{levesque05}.

Note that  the Levesque scale may be offset by 50 to  150 K.
In the recent work of \citet{tabernero18},
the relation between spectral types and stellar 
temperatures (based on TiO) is confirmed  (R=10,000), 
but the authors also confirmed a warmer temperature scale 
as suggested by \citet[R=3000,][]{gazak14} and 
\citet[R = 4000--8000,][]{davies15}.
The latter authors used a  modeling of atomic lines in the $J$-band
to infer  metallicity and stellar temperatures.
 The scale of Tabernero et al.
linearly correlates with that of \citet{davies15}, 
but is 168 K cooler ($\sigma$=72 K). 
The scale of Levesque et al. is about $220\pm100$ K 
(LTE models) cooler than that given in \citet{gazak14} -- 
$270 \pm 130$ K with non-LTE models.
These offset trends (hotter temperatures) are confirmed by the  RSG 
parameters inferred by  \citet[][]{arentsen19} with spectra at R=10,000, 
though statistics is still poor with only 1 or 2
stars per spectral type (Table \ref{cal.table}).
The temperatures inferred  with 
iron lines in $Y$-band by \citet[R=28,000,][]{taniguchi20} 
agree better with those of \citet{levesque05}
with offsets from $-91$ to +37 K.\\
Future consistent modeling of  atomic lines  from the 
same element in multiple 
regions of the stellar spectrum  (and for several elements)
 may help to solve the issue 
about the absolute scale,  as well as improve the modeling itself.

\begin{table*}
\caption{ \label{cal.table} Mean temperatures  of RSGs per spectral type.}
\begin{tabular}{llrrlrrr}
\hline
\hline
Sp      &   T(Lev)$^a$   & T(Xsho)$^b$      & T(Xsho)-T(Lev) &  T(WIN)$^c$        &T(WIN)-T(Lev)&T(IGRI)$^d$\\
        &  [K]        & [K]         &  [K]           &     [K]         &   [K]   &[K]\\
\hline

    K0   &   4185   &   4581 $\pm$   56   &  396  &     $..$      &   $..$    & 4573  \\
    K1   &   4100   &   4251 $\pm$   46   &  197  &     $..$      &   $..$    &   \\
  K1.5   &   4057   &     $..$      &   $..$    &   4073 $\pm$   31   &   15  &   \\
    K2   &   4015   &   4125 $\pm$   44   &  110  &     $..$      &   $..$    &   \\
    K3   &   3956   &   4024 $\pm$   43   &   47  &   3940 $\pm$   29   &  $-$16  &   \\
  K4.5   &   3869   &     $..$      &   $..$    &   3829 $\pm$   61   &  $-$39  & 4031  \\
    K5   &   3840   &   3977 $\pm$   53   &  137  &   3749 $\pm$   49   &  $-$91  &   \\
    M0   &   3790   &   3868 $\pm$   32   &   78  &     $..$      &   $..$    &   \\
    M1   &   3745   &   3780 $\pm$   62   &   35  &   3676 $\pm$  117   &  $-$69  &   \\
  M1.5   &   3710   &   3747 $\pm$   51   &   37  &   3747 $\pm$  136   &   37  &   \\
    M2   &   3660   &   3588 $\pm$   59   &  -72  &   3651 $\pm$   31   &   -9  &   \\
    M3   &   3605   &   3768 $\pm$   62   &  163  &     $..$      &   $..$    &   \\
  M4.5   &   3506   &   3460 $\pm$   57   &  $-$46  &     $..$      &   $..$    &   \\

 \hline
\end{tabular}
\begin{list}{}
\item 
$^a$=Average temperatures per spectral type \citet{levesque05}.\\
$^b$=Average temperatures per spectral type made with  the XShooter spectral catalog of \citet{arentsen19}.\\
$^c$=Average temperatures per spectral type made with  the WINERED catalog of \citet{taniguchi20}.\\
$^d$=Average temperatures per spectral type made with  the IGRINS catalog of \citet{park18}.\\
\end{list}
\end{table*}

For 14 RSGs from the IRTF library, temperatures
and metallicities were estimated with a modeling of the $J$-band
by \citet{davies10}. But,  at the SpeX resolution, 
temperatures were not well resolved and their estimates 
were $153\pm167$ K cooler than those of Levesque.

\begin{table*}
\caption{\label{table.averagetemp}  Average \ew s per 
spectral type of the infrared indicators 
of temperature.}
\begin{tabular}{lrrrrrrrrrrrrrr}
\hline
Sp &   \Teff &      \ewcodue    & $\sigma$  & \ewca  &   $\sigma$ &  \ewna &  $\sigma$ &   \ewmg  &$\sigma$   & \ewsiusq& $\sigma$ &  \ewcousd &$\sigma$ & \\
   &   [K]   &     [\AA]        & [\AA]       &  [\AA]    &    [\AA]     &   [\AA]   &  [\AA]      &  [\AA]      & [\AA]       &  [\AA]      & [\AA]    &   [\AA]  & [\AA] \\
\hline
   K0   &  4185   &    25.9   &     5.6   &    0.5   &     1.2   &    0.5   &     1.5   &    0.6   &     0.6   &    0.3   &     0.6   &    0.9   &     0.6   &   \\
   K2   &  4043   &    35.9   &     9.3   &    2.2   &     0.4   &    2.8   &     0.2   &    1.6   &     0.4   &    1.8   &     0.3   &    3.3   &     0.8   &   \\
   K4   &  3927   &    39.2   &     7.3   &    3.0   &     0.7   &    3.6   &     0.6   &    2.3   &     0.2   &    2.2   &     0.4   &    5.2   &     1.0   &   \\
   M0   &  3790   &    43.7   &     5.2   &    3.1   &     0.0   &    4.2   &     0.1   &    2.7   &     0.0   &    3.1   &     0.7   &    5.9   &     0.3   &   \\
   M1   &  3745   &    54.1   &     4.3   &    3.1   &     0.4   &    4.1   &     0.3   &    2.6   &     0.0   &    3.0   &     0.1   &    5.9   &     1.1   &   \\
 M1.5   &  3710   &    47.8   &     2.3   &    3.1   &     0.1   &    4.2   &     0.1   &    3.1   &     0.1   &    3.0   &     0.2   &    6.5   &     0.2   &   \\
   M2   &  3660   &    52.5   &     1.0   &    3.8   &     0.1   &    4.7   &     0.1   &    3.7   &     0.1   &    3.2   &     0.3   &    6.9   &     0.2   &   \\
 M3.5   &  3550   &    57.1   &    $..$   &    3.4   &   $..$   &    5.2   &   $..$   &    3.8   &   $..$  &    3.8   &   $..$   &    8.2   &   $..$   &   \\
   M4   &  3535   &    46.3   &    $..$   &    3.8   &   $..$   &    5.0   &   $..$   &    4.0   &   $..$   &    3.6   &   $..$   &    7.6   &   $..$   &   \\
\hline
\end{tabular}
\end{table*}

\subsection{Template temperature sequence for RSGs}

We built a template sequence sorting our spectral atlas by increasing \tcombi\ 
(see Tables \ref{aliaskrsg},  \ref{aliasnewrsg}, and \ref{aliasgiants}) 
and fortunately got a nice sequence of molecular absorption,
as described in Sec.\ \ref{molecules}.  This sequence
is useful for spectral typing in heavily obscured regions and for stellar
population modeling work.

\begin{figure*}
\begin{center}
\resizebox{0.33\hsize}{!}{\includegraphics[angle=0]{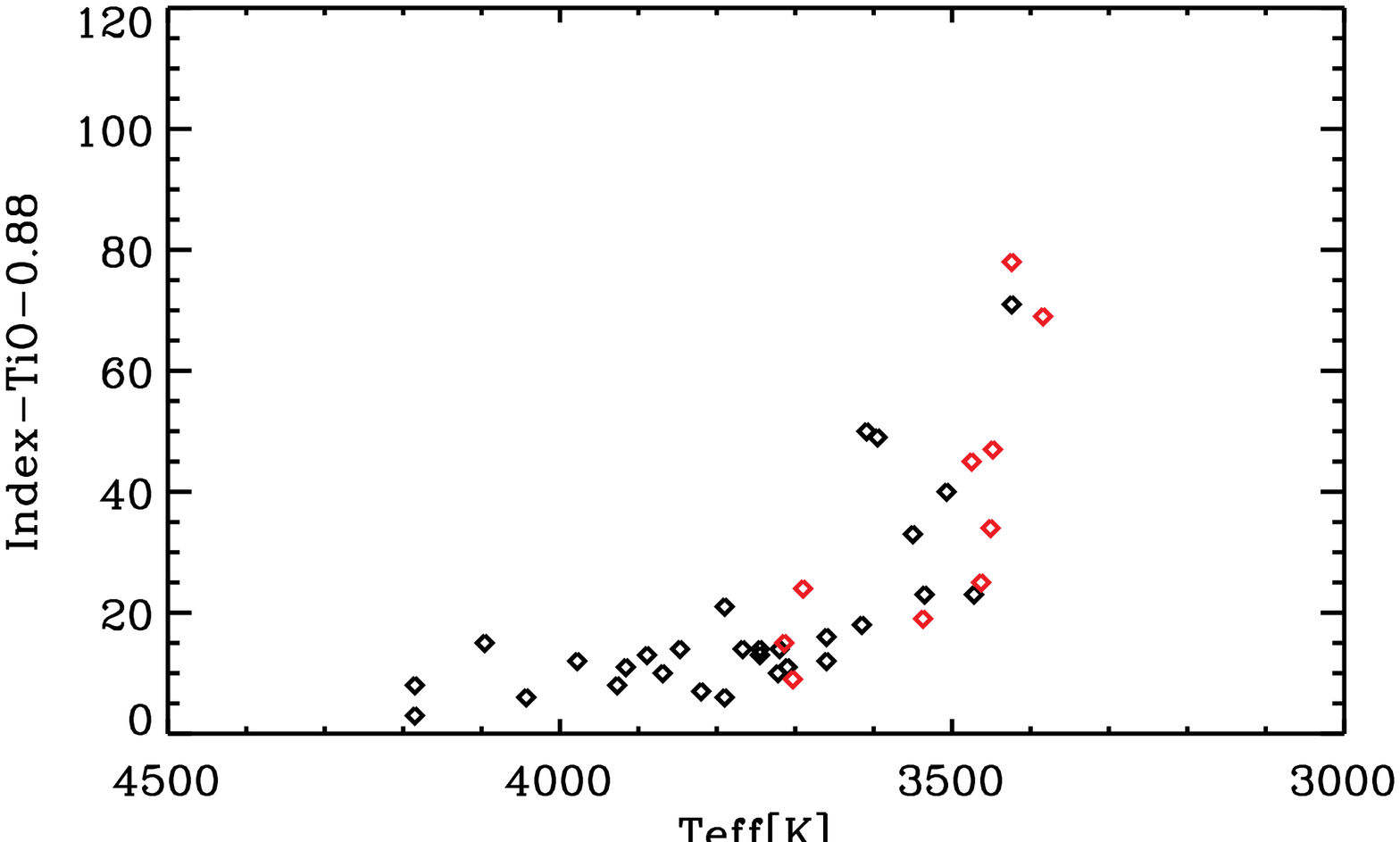}}
\resizebox{0.33\hsize}{!}{\includegraphics[angle=0]{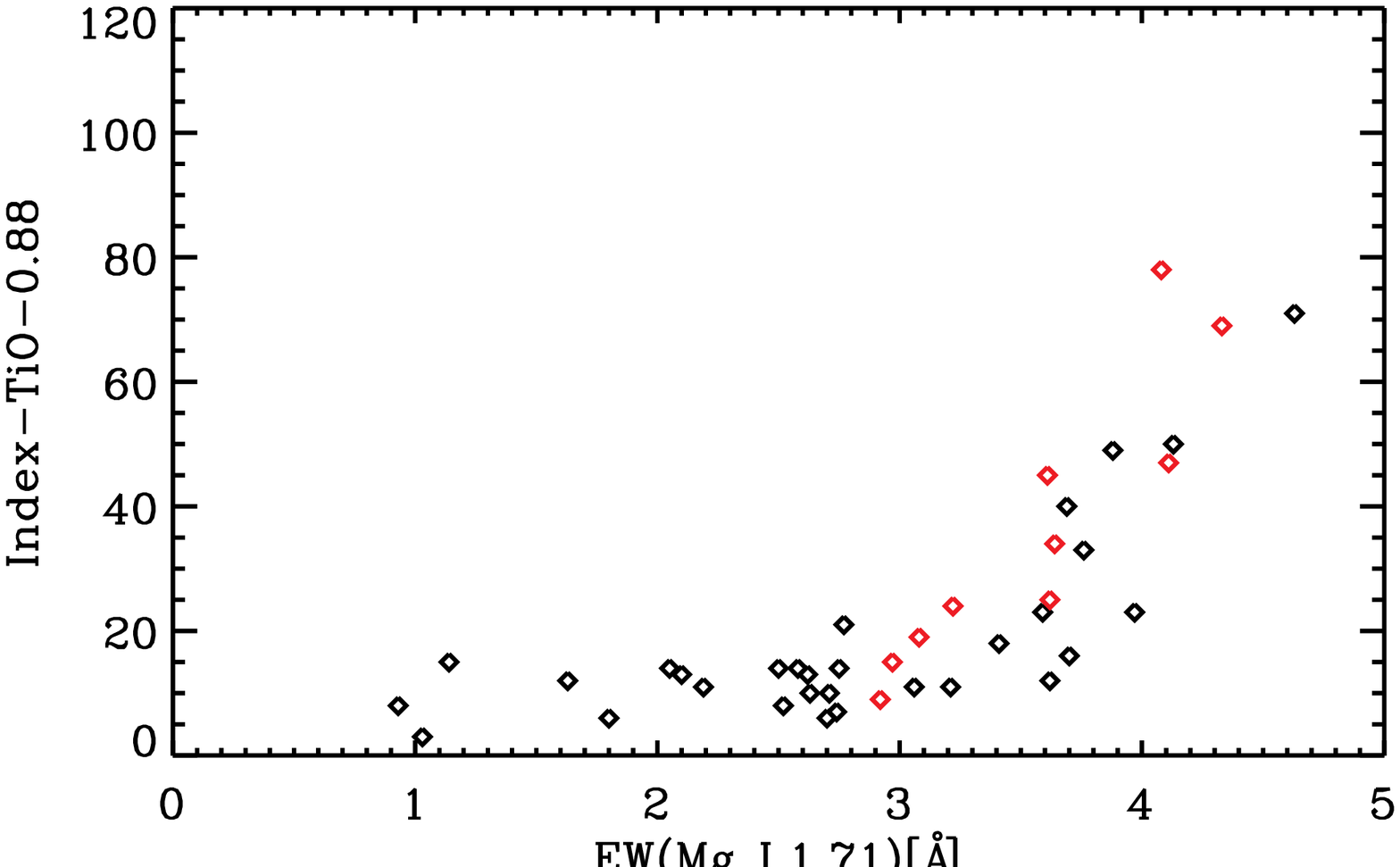}}
\resizebox{0.33\hsize}{!}{\includegraphics[angle=0]{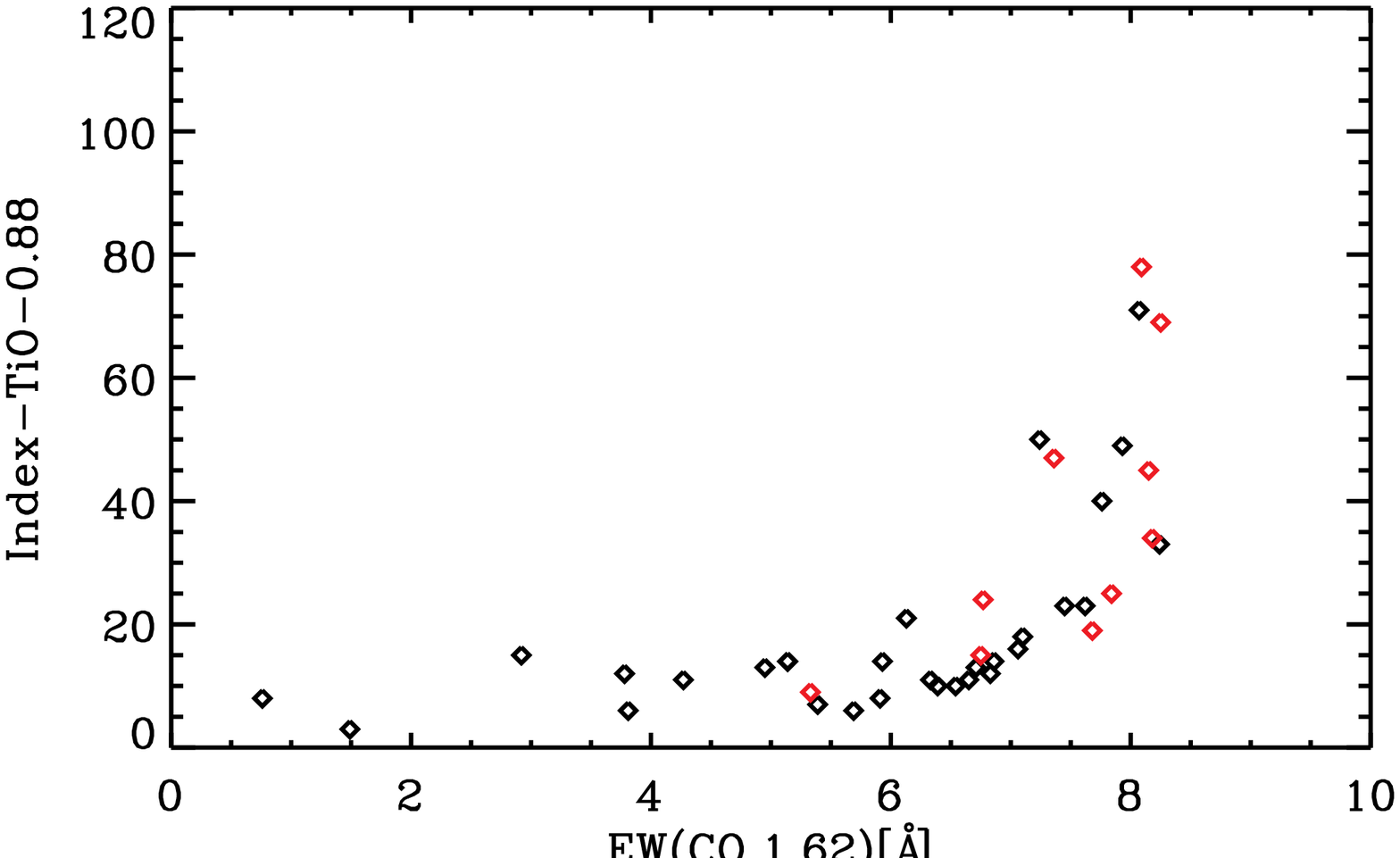}}
\end{center}
\caption{  \label{TiOtcombi} {\it Left panel:} 
Index for the strengths of the TiO 
band heads at 0.88 \um\ vs. \tcombi\  
(the \Teff\  estimated from the SpeX
spectra with $H$- and $K$-band lines).
{\it Middle panel:} 
Index for the strengths of the TiO 
band heads at 0.88 \um\ vs.  \ewmg.
{\it Right panel:} 
Index for the strengths of the TiO 
band heads at 0.88 \um\ vs.  \ewcousd.
}
\end{figure*}

We have estimated spectroscopic  stellar temperatures at infrared 
wavelengths (using only the $H$ and \Ks\ part of the spectrum)
 using the  calibration of optical spectral-types 
based on TiO bands \citep{levesque05}.
In Fig.\ \ref{TiOtcombi},  the strengths of the TiO band heads at 0.88 \um\
appear to correlate with \tcombi, and 
with the \ewmg\ and \ewcousd\ measured  in the SpeX spectra.
The TiO strength is estimated by fitting a linear continuum and measuring 
the depth of the TiO absorption from 0.8857 \um\ to 0.9099 \um. 

\begin{figure*}
\begin{center}
\resizebox{0.7\hsize}{!}{\includegraphics[angle=0]{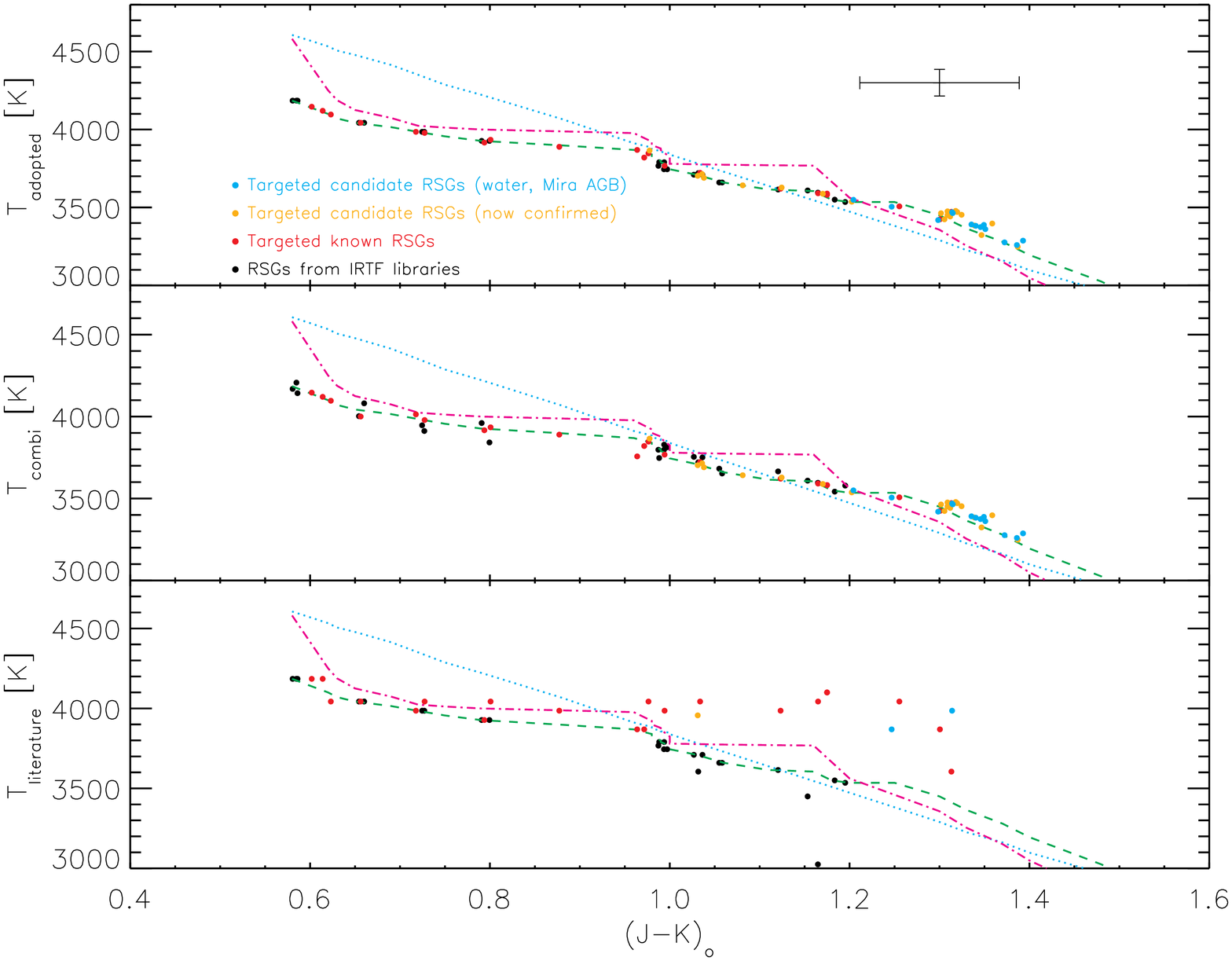}}
\end{center}
\caption{\label{JKovsTall}  $(J-$\Ks$)_o$ colors vs. stellar temperatures 
\citep[dereddening was done with the intrinsic colors of][]{koornneef83}.
{\it In the bottom panel}, we used the \Teff\  estimated from the literature 
\citep[\Teff\ are  estimated with collected spectral types and the temperature scale of][]{levesque05}.
{\it In the middle panel} the \tcombi\ estimated with the SpeX spectra. 
{\it In the top panel}, we use the literature values 
for well-known RSGs optically analyzed, e.g., by \citet{keenan89} and \citet{levesque05},
and our \tcombi\ for other stars (mostly infrared classified stars), as listed in Table 1.
The average 2MASS $J-$\Ks\  error and temperature error is marked on the top-right corner.
The data points are colored as described in the legend; black circles indicate data points
of RSGs with spectra in the SpeX libraries; red circles mark  known RSGs observed during our runs,
and orange and cyan circles candidate RSGs  
 with and without water absorption, respectively.
The dashed green curve is the relation between spectral-types and  $J-K$ colors of 
\citet{koornneef83} converted to temperatures with the scale of \citet{levesque05}.
The long-dashed magenta curve shows the relation between the $J-K$ colors of 
\citet{koornneef83} and the average temperatures per spectral 
type of \citet{arentsen19} that are listed in Table  \ref{cal.table}.
The cyan dotted line is the relation between temperatures and $J-K$ colors
predicted with atmosphere models by \citet{levesque05} and \citet{neugent20}.
}
\end{figure*}

In Fig. \ref{JKovsTall}, we plot the temperatures versus 
the  $(J-$\Ks$)_o$ colors, 
and it appears that water absorption dominates 
the   colors redder than $1.3$ mag.
For comparison, we overplot the empirical relation between   
the $(J-$\Ks$)_o$ colors and the spectral types by \citet{koornneef83}
by using the temperature scale of \citet{levesque05}, as well as 
by using the temperatures inferred from the work of \citet{arentsen19}.
For K-type stars, the theoretical temperatures  obtained with models 
of atmospheres by \citet{neugent20}
remain higher, up to about 400 K, than that estimated with 
the empirical relation of \citet{koornneef83};  for M-types,
the models predict temperatures up to 150 K cooler.

\section{Metallicity of RSGs}
\label{metsec}

 RSG stars are a young population of  the Galactic disk, which has
high metallicity.
Metallicity measurements for only 25 of the 
known RSGs included in this work are reported in {\bf the} literature
\citep{davies10, gazak15,arentsen19,prugniel11,stevens17,
anderson12,pastel16,ivanov04,davies09}.
Their values range from $-0.2$ dex to +0.35 dex (see Appendix  A).
16 out of these 25  measurements have been made in $J$-band 
\citep{davies10,gazak15}.
\citet{davies10} used some  spectra of RSGs from the IRTF library 
 to study atomic lines in $J$-band, from 1.15 \um\ to 1.23 \um.
The authors located several iron-dominated lines 
that {\bf allowed} the measurement 
of a global metallicity with an accuracy of 0.1 dex 
with synthetic spectra.
In this paper, we only present the new spectra and empirical quantities.
Metallicity determinations with synthetic spectra will be presented
elsewhere.

\begin{figure}
\begin{center}
\resizebox{0.9\hsize}{!}{\includegraphics[angle=0]{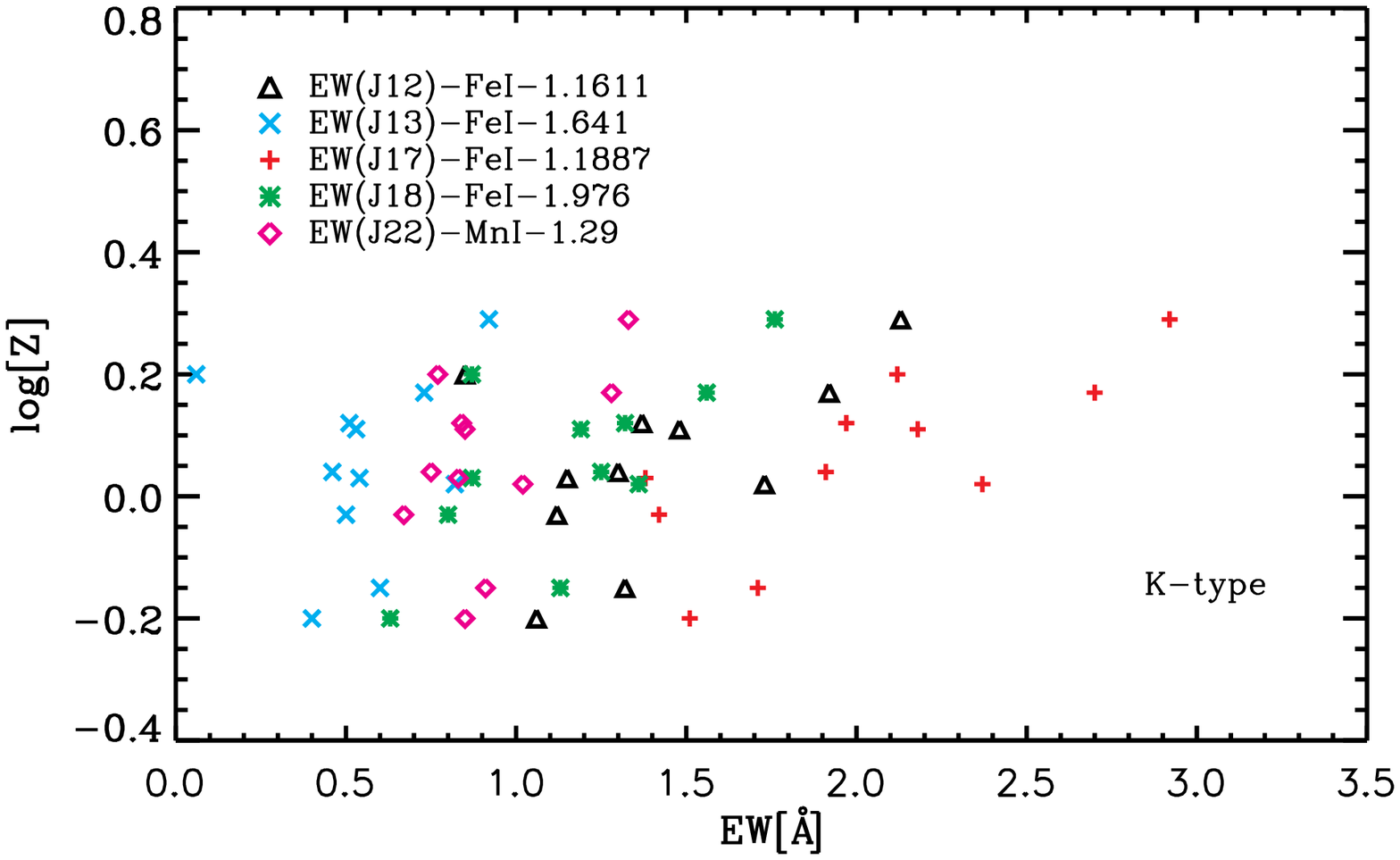}}
\resizebox{0.9\hsize}{!}{\includegraphics[angle=0]{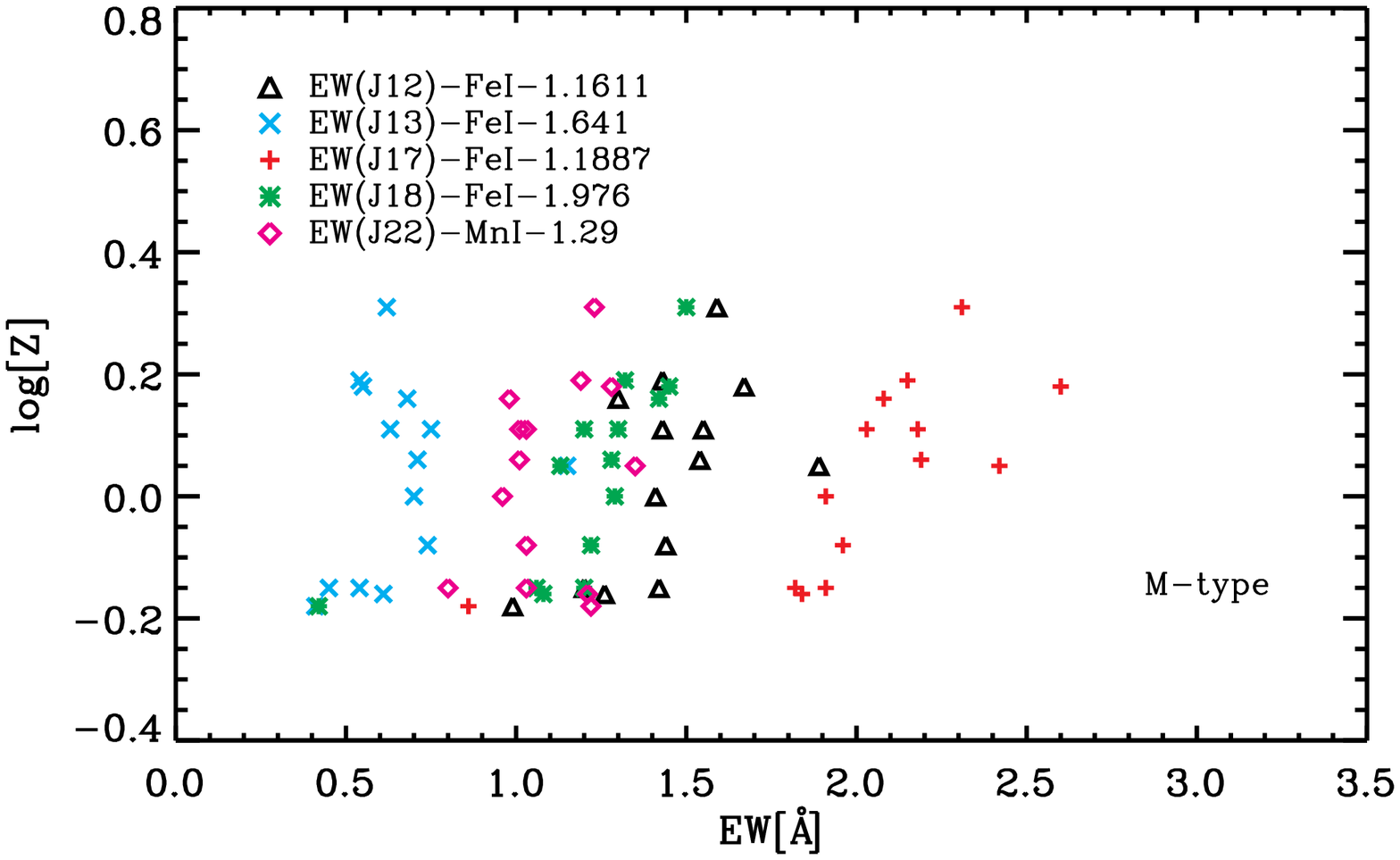}}
\resizebox{0.9\hsize}{!}{\includegraphics[angle=0]{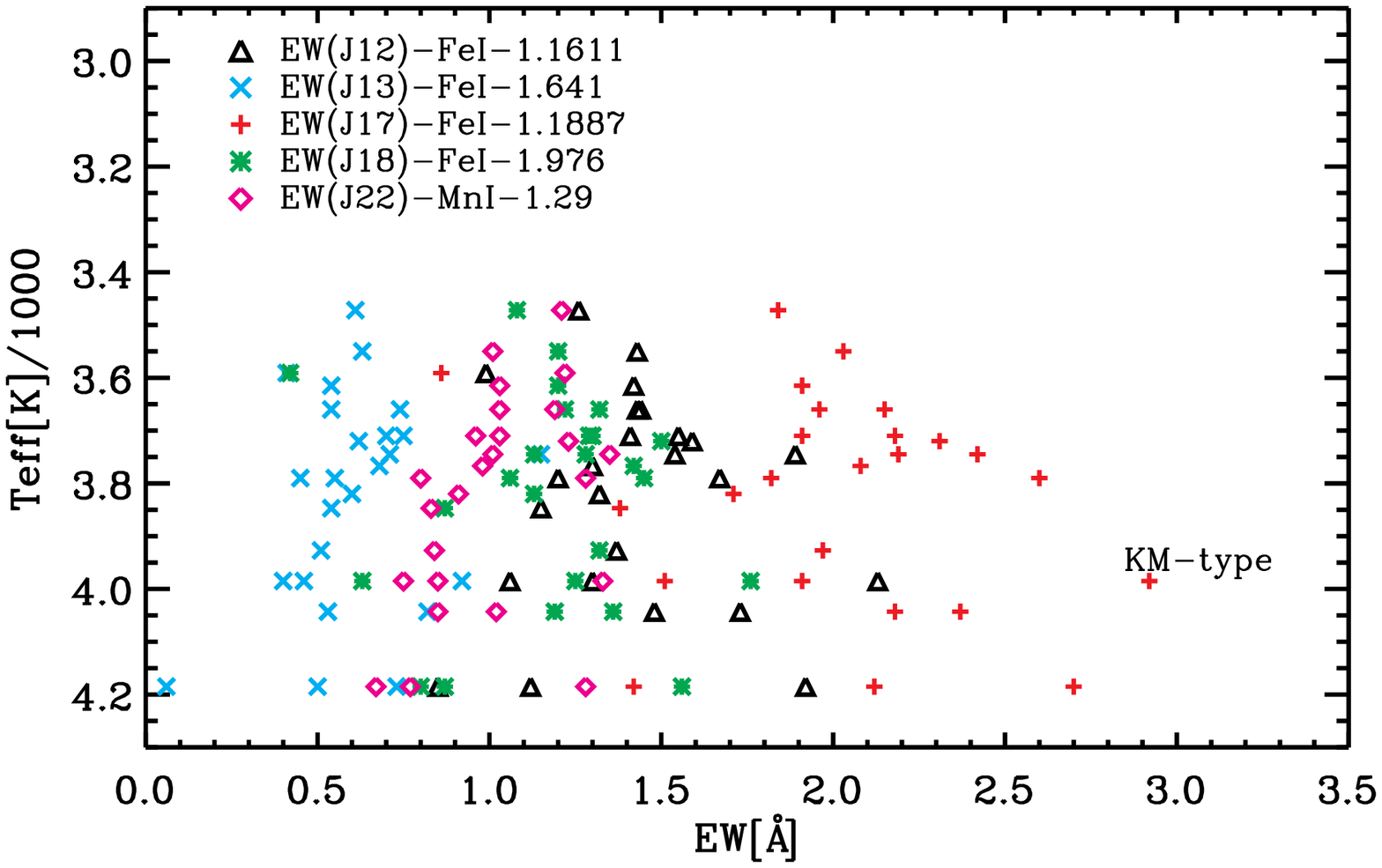}}
\caption{ \label{zplot}  
{\it Top Panel :} Log(Z) values for  Galactic RSGs  
(see text) vs. the measured \ew s of possible 
indicators of metallicity (the J12, J13, J17, J18, and J22 lines), 
as described in the legend; only known K-type RSGs ( $> 3800$ K) are plotted.
{\it Middle Panel:} as in the top panel, only known M-type RSGs ( $< 3800$ K) 
are plotted.
{\it Bottom Panel:} Stellar effective temperatures 
(\Teff\ from Table \ref{aliaskrsg}) of the RSGs plotted in 
the above panels vs. \ew s.} 
\end{center}
\end{figure}

The FeI-1.1611, FeI-1.1641,  FeI-1.1887, FeI-1.976  lines
of \citet{davies10} correspond to the J12, J13, J17, and J18 lines in 
Table  \ref{inddef}, respectively. 
Theoretically, the  strengths of  these lines,  
with the exception of FeI1.1641 (J17), 
are expected to increase with increasing metallicity and to 
be almost independent of temperature; 
observationally, no clear relation is  found  between the EWs 
and the metallicity. This is likely due to 
the spectral resolution of SpeX and the narrow range of
Galactic metallicity of RSGs.
When separating the stars in K-type and M-type stars,
the EWs of the FeI-1.1641 lines (J17) and FeI-1.976 (J18) appear larger 
in stars with supersolar metallicity; 
as shown in Fig. \ref{zplot}, stars with \ewJusette $>$ 2 \AA\ are supersolar.
We found also an interesting behavior of the \ew\ of the J22 line, 
which is an  unblended line from Mn I 
at 1.29 \um; for RSGs below log(Z) $< 0.16$, 
\ewJdd\ values appear approximately constant  ($0.96\pm0.04$ \AA);
above that Z, \ew s are larger  ($1.15\pm0.07$ \AA), 
similarly to the findings of \citet[][]{feltzing07} for thin Disc stars.

\section{ Notes on individual stars}
\label{notes}

Optical spectral types from \citet{keenan89} 
are  reproduced within one spectral type.
EWs from  Na I, Mg I,  Si I, and CO at 1.62 \um\ 
yield spectral types in better agreement 
with the optical ones than those estimated only from 
the CO band heads at 2.29 \um.
For example,  2MASS J18082651$-$1833078 
(HD~165782) is an optical K0 Ia star; 
from the CO at 2.29 \um, we estimated a K2.5 type,
while by using  Na I, Mg I,  Si I, and CO at 1.62 \um, 
we obtain  a K0.5 type.
2MASS J07473853$-$1559263 (HD~63302) is reported in optical works
as a K1-2 supergiant.
By using the \ewcodue\ we infer an M0 type, 
while by using  Na I, Mg I,  Si I, and CO at 1.62 \um, 
we obtain  a K2.5 type.
2MASS J17143885+1423253 (HD~156014) is a known  M5 Ib-II star,  a possible AGB \citep{moravveji13};
the  CO at 2.29 \um\ yields a K4.5 type (RSG) or M5.5 (giant); 
from \tcombi\ we infer an M3. 
2MASS J20285059+3958543/RW Cygn is optically an M3, 
but its infrared \ew s are more typically of an M1.5 star.
2MASSJ 22543171+6049388/MY Cep is reported as an 
M7.5 I star from \citet{fawley74}, but
we estimated  an M3.\\

For stars previously classified via CO band heads 
or TiO  at 0.8 \um, we determined spectral types
later than those reported in the literature, up to 5 spectral types,
and, therefore,  cooler temperatures.
For example, 2MASS J18424927$-$0445095  was classified  
by \citet{negueruela12}  and listed as a  K1 Iab star.
From the CO band heads at 2.29 \um\ we derived an  M1 star, while 
from the average infrared temperature an M3 type. 
We observed this star twice to be sure of the acquisition.
In Fig. \ref{JKovsTall}, the literature temperatures and \tcombi\
are plotted versus the dereddened $J-$\Ks\ colors, $(J-$\Ks$)_o$, 
which are not affected by TiO bands, putting in evidence those stars
with discrepant spectral types in {\bf the} literature.

2MASS J17143885+1423253/HD 156014/$\alpha$ Her is 
an outlier data point in our \ewmg\ versus \ewcodue\ 
diagram  in Fig. \ref{MgI5}. It is
classified 
as an M5 Ib-II by \citet{keenan89}
and as a M5 I by \citet{levesque05}.
The location on the \ewcodue\ versus \ewmg\ 
giant sequence, weak CN band heads, 
are suggestive of a lower luminosity class (III-II).
The star is variable (SRC) and \citet{yamashita67} lists 
different classes for different epochs (IIIa, IIab, Ib).
\citet{moravveji13} modeled  $\alpha$ Her
as  a thermal pulsing AGB of about 2 \Msun.

2MASS J18345133$-$0713162  and  2MASS J18345840$-$0714247,
listed in   Table \ref{aliaskrsg},
were marked as possible RSGs in the cluster 
Alicante 8 near RSGC1 by \citet{negueruela10}.
Because of the strong TiO bands and water absorption,
their spectra resamble those of Mira AGBs.
Furthermore, these stars have low radial velocity ($9.37$ and $-11.79$ \kms), 
while the RSGC clusters are at \Vlsr\ from 90 to 120 \kms\ 
\citep[e.g.][]{davies07,davies08,origlia15}. For 
2MASS J18345133$-$0713162 we estimated Mbol=$-1.27$ mag, typical of AGB.

2MASS J00403044+5632145/HD 3712/
$\alpha$ Cas is   listed in our  Table  
\ref{aliasgiants} of known giants;
 \citet{keenan89}  reported {\bf it}
as a K0 IIIa star.
In previous years, from 1938 to 1968, it had  been classified as 
 class II-III \citep[e.g.,][]{meisel68}.
The \ewCaT\ is $\approx 9$ \AA, 
and in the \ewcodue\ versus \ewmg\  diagram
{\bf it } is located near two  RSGs, with a Mg I line at 1.71 \um\ significantly 
fainter than {\bf in} other giants.
2MASS J00403044+5632145 has  log(L/\Lsun)(Gaia EDR3)=2.83. 
\citet{stock18} estimated a similar luminosity and a  93\% probability that it is a star
in the horizontal branch phase with log(g)=1.7 and $[Fe/H]=-0.1$ dex
\citep[e.g.,][]{anderson12},
and \citet{kervella19}  report it as a possible binary system.

\begin{figure*}
\begin{center}
\resizebox{0.48\hsize}{!}{\includegraphics[angle=0]{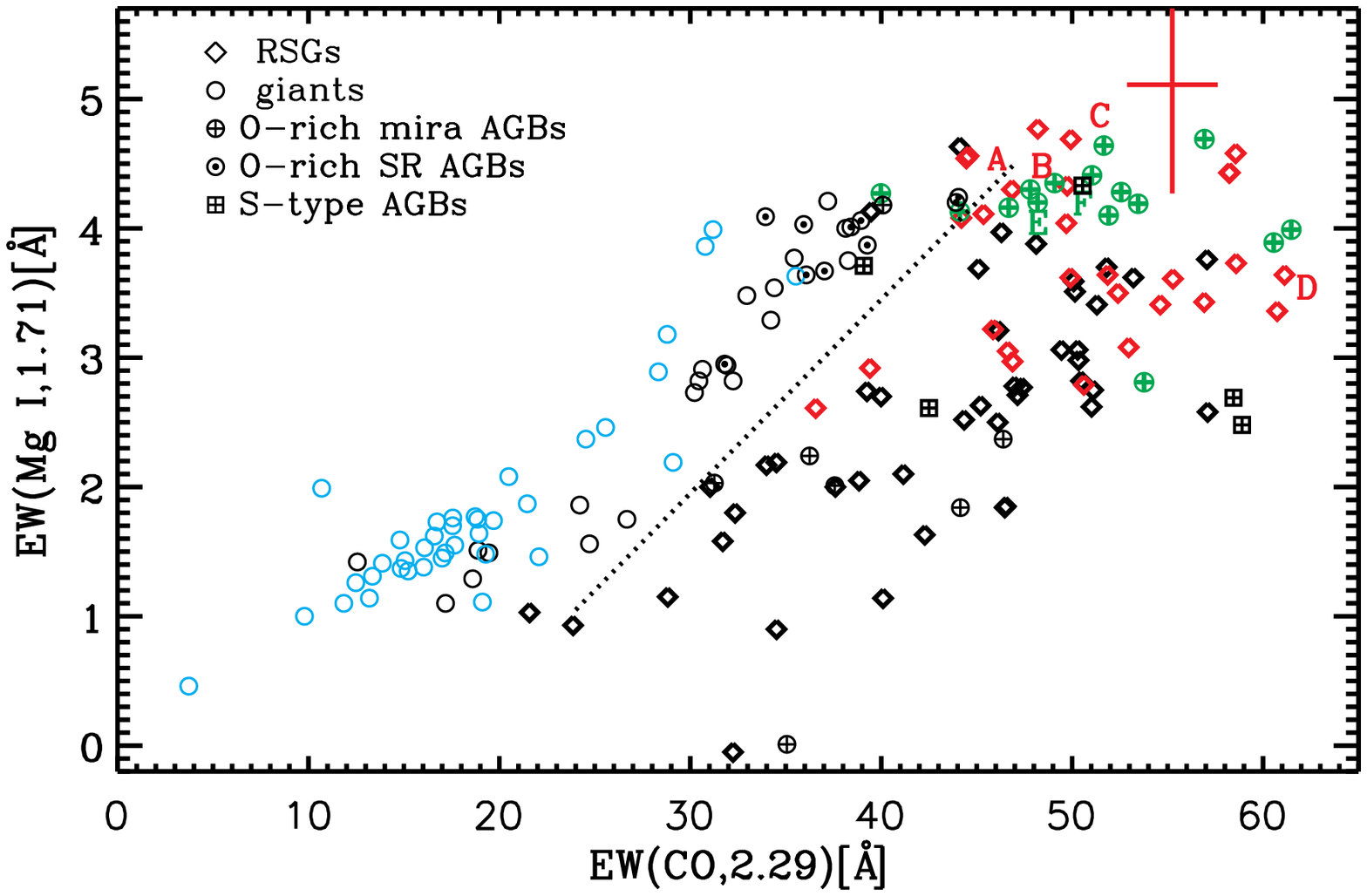}}
\resizebox{0.48\hsize}{!}{\includegraphics[angle=0]{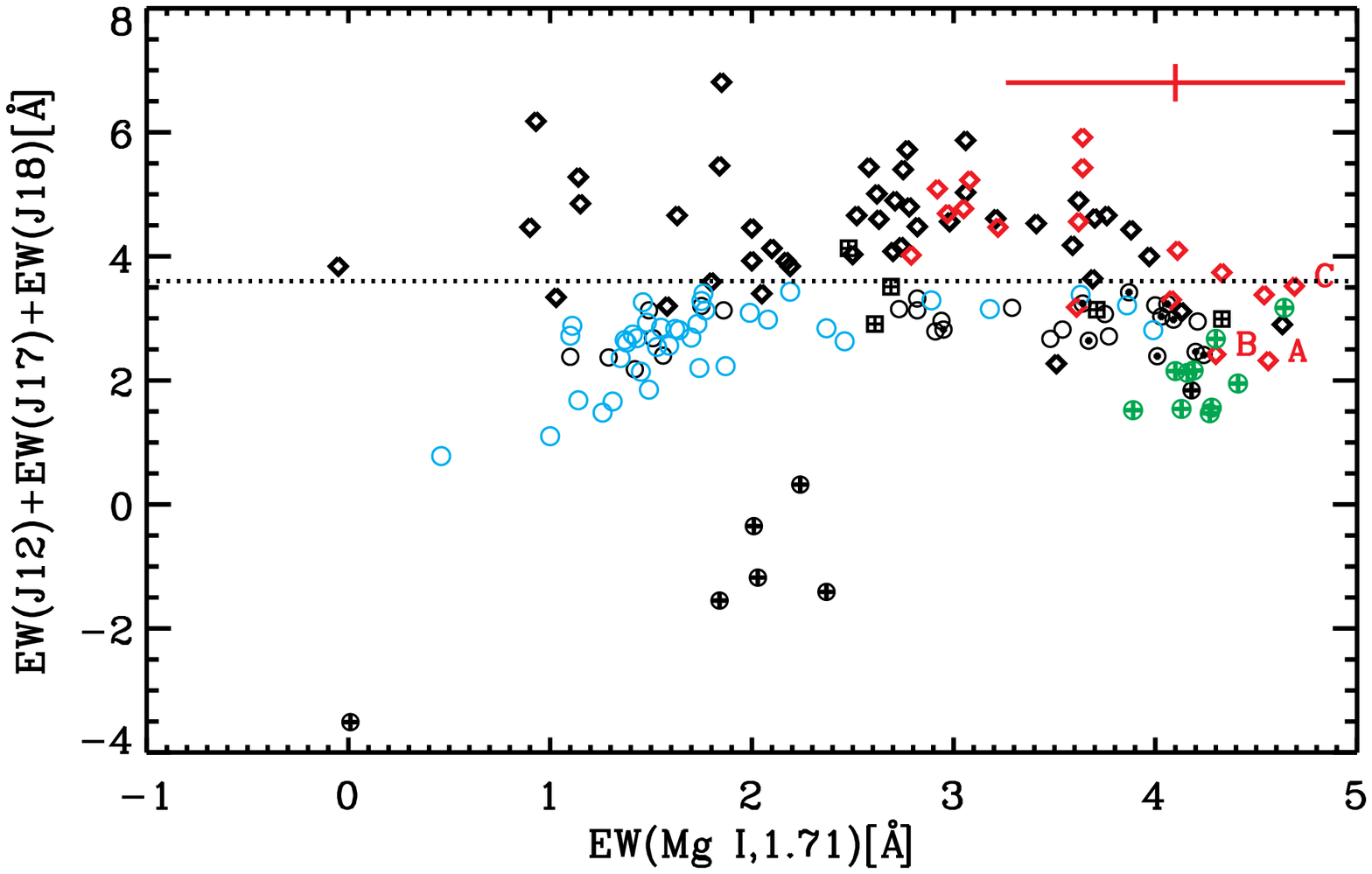}}
\end{center}
\begin{center}
\resizebox{0.48\hsize}{!}{\includegraphics[angle=0]{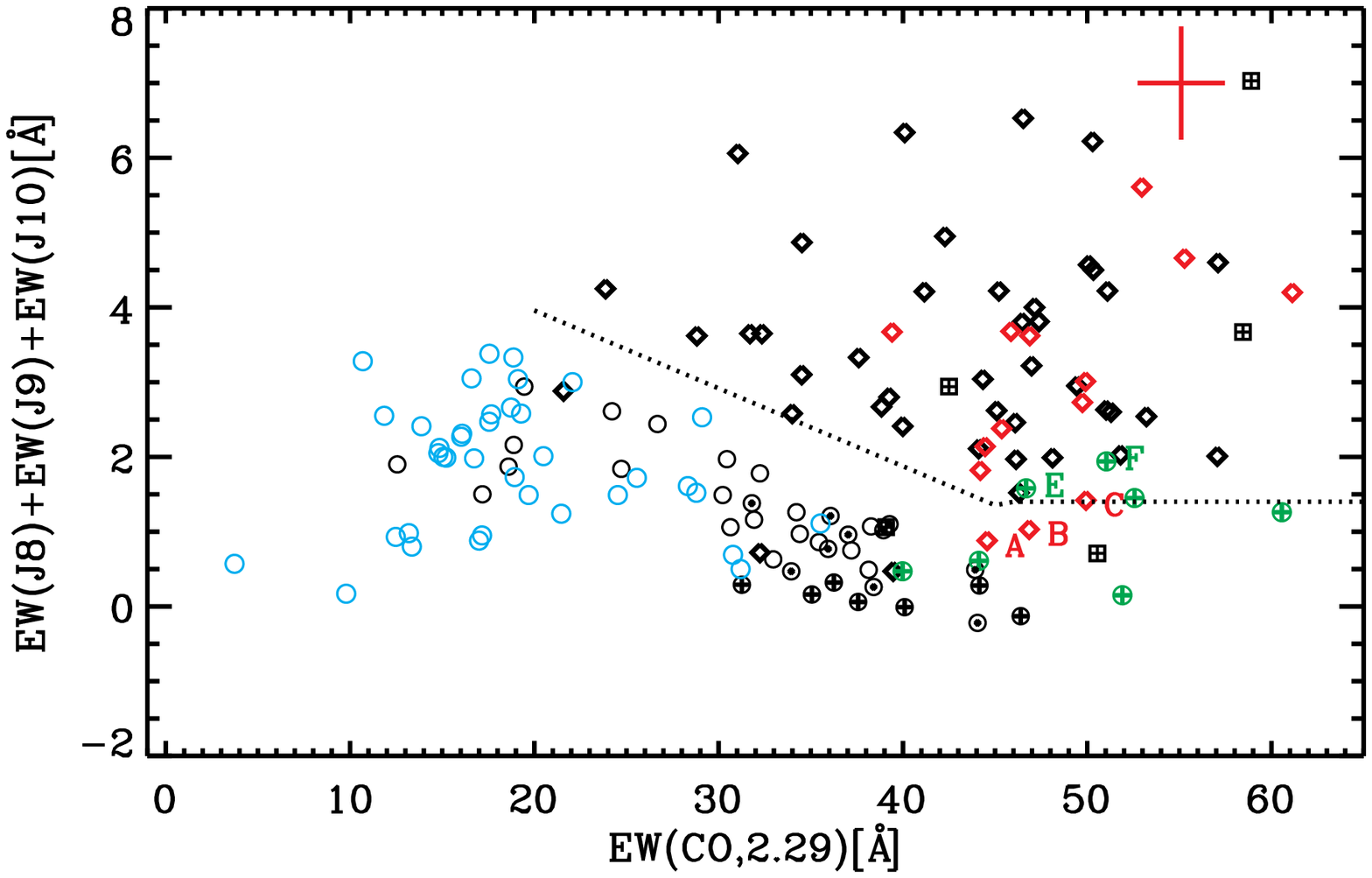}}
\resizebox{0.48\hsize}{!}{\includegraphics[angle=0]{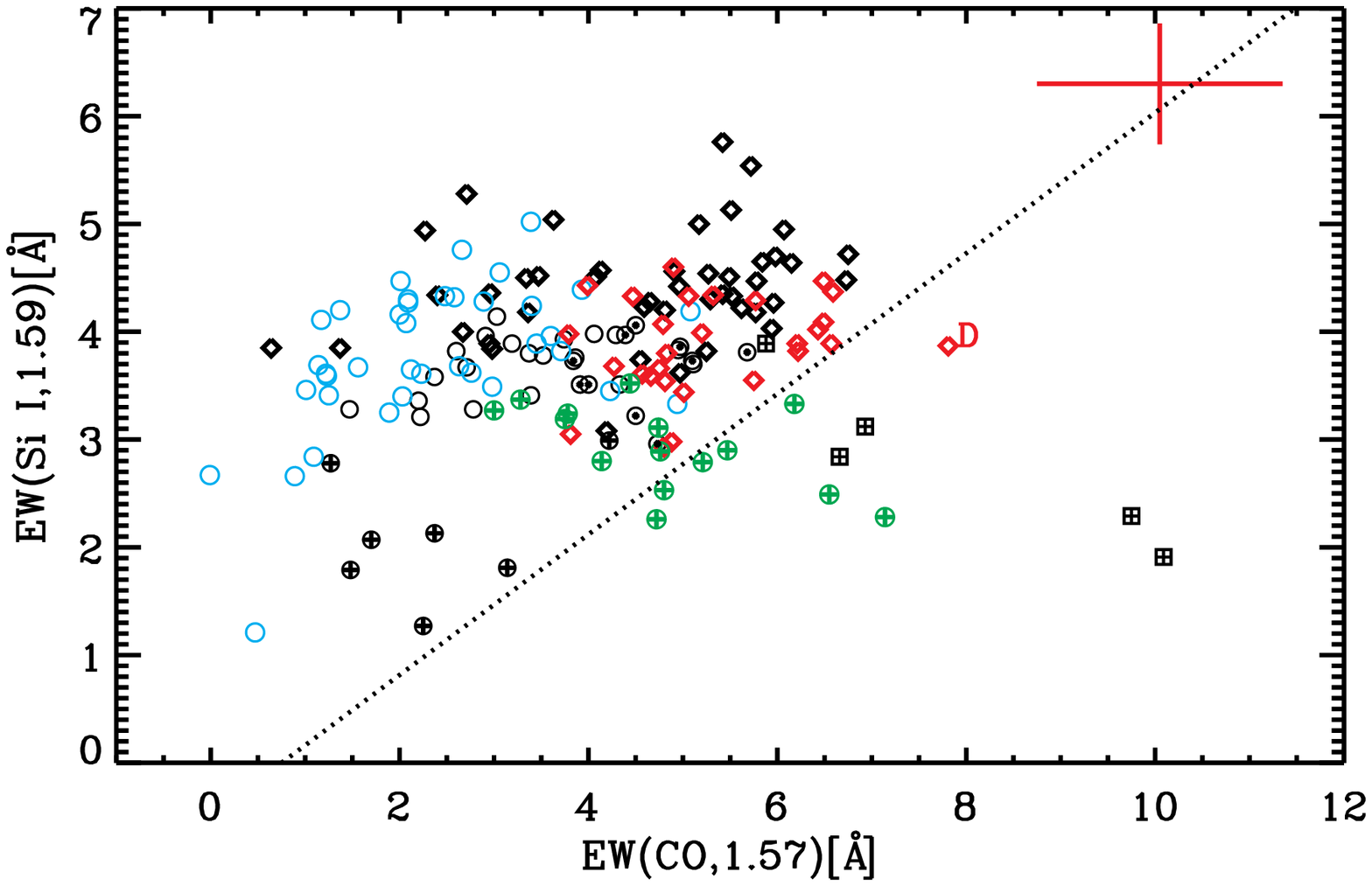}}
\end{center}
\begin{center}
\resizebox{0.48\hsize}{!}{\includegraphics[angle=0]{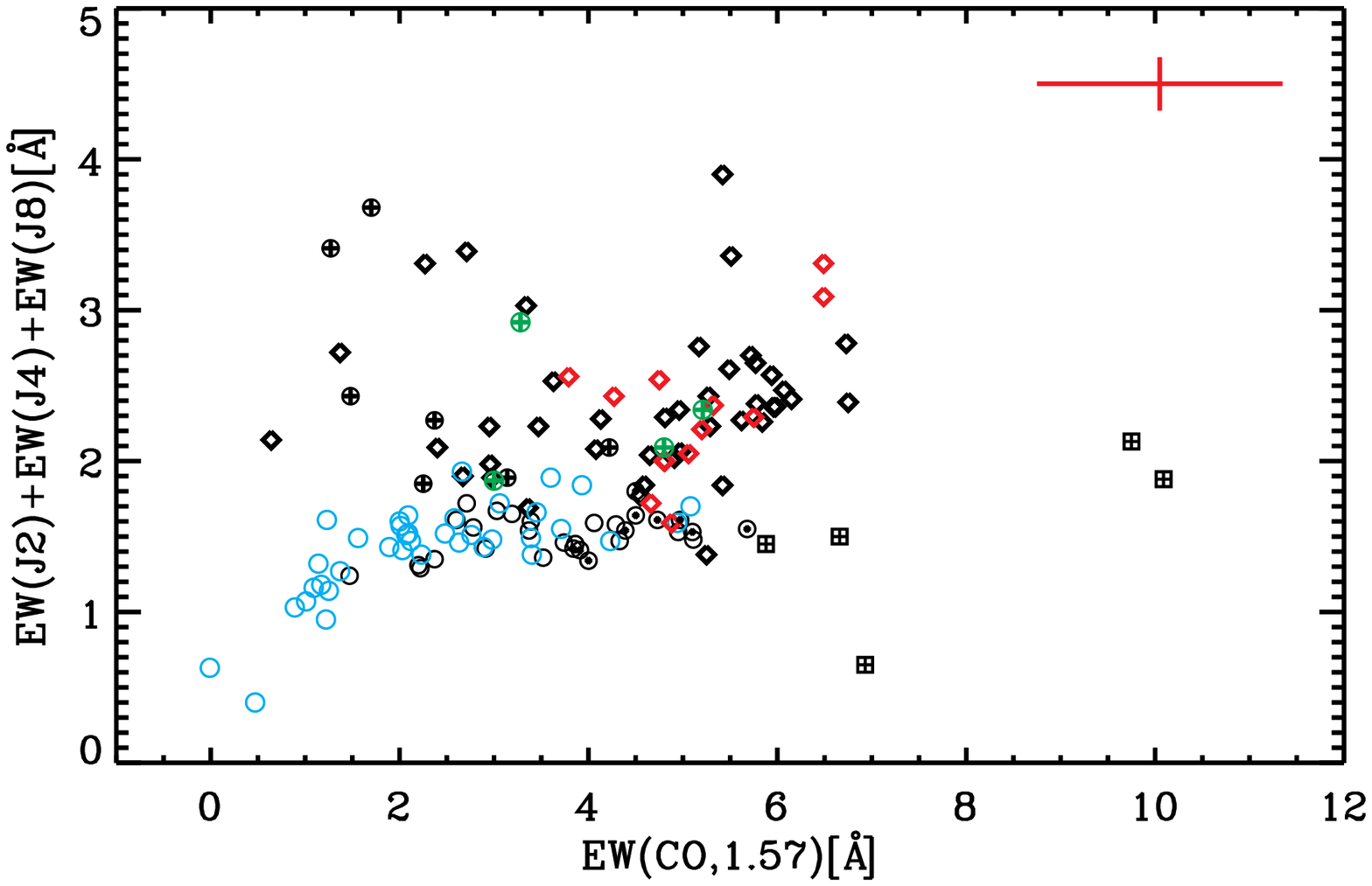}}
\resizebox{0.48\hsize}{!}{\includegraphics[angle=0]{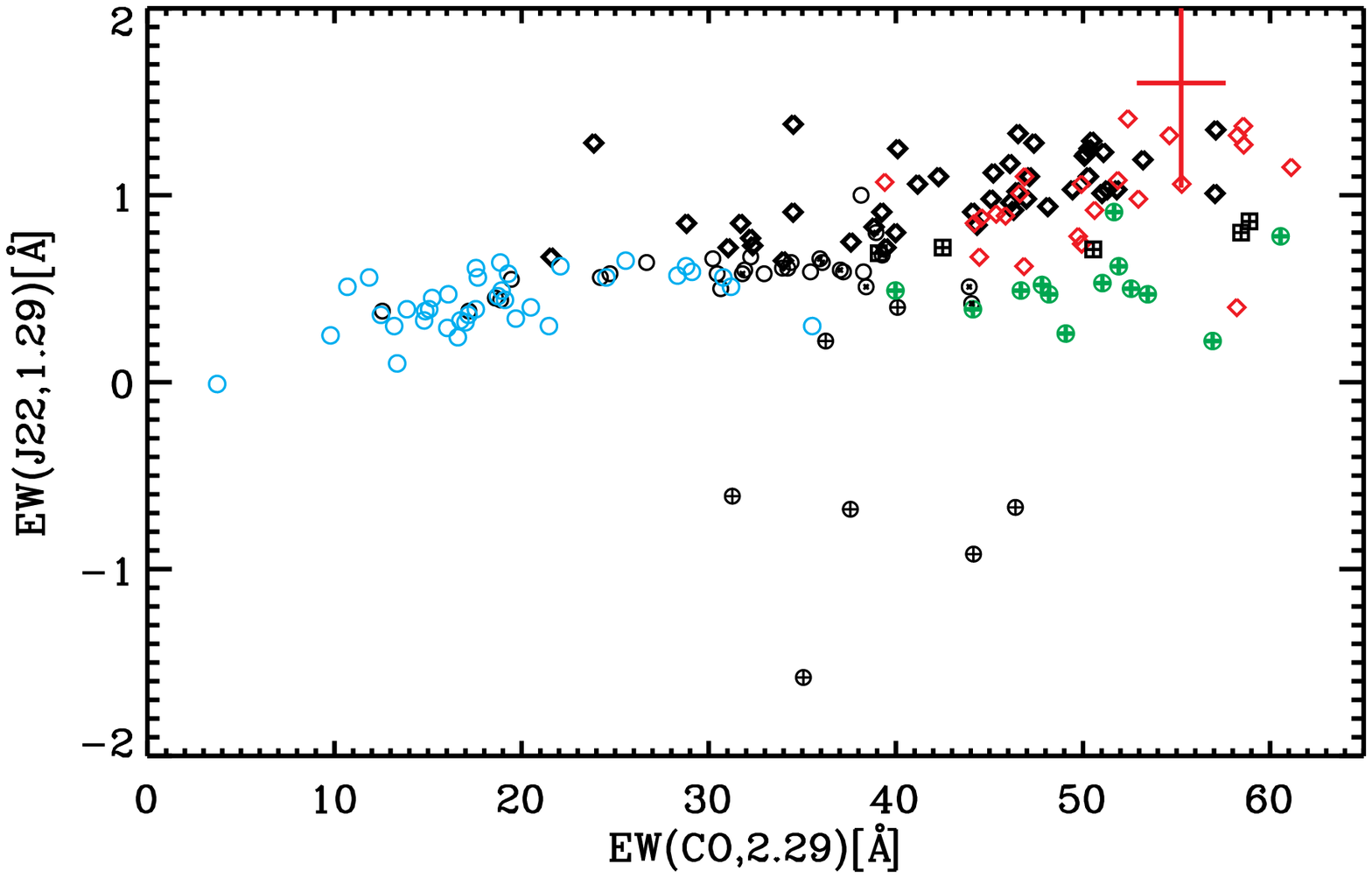}}
\end{center}
\caption{ \label{fig.ewcandidate} \label{fig.ewmira}
Useful diagrams to classify evolved 
late-type stars as  presented
in Figs. \ref{CO4}, \ref{figalJplus}, and \ref{comboCN} (same symbols).  
The \ew\ values of observed candidate RSGs with no water absorption ($< 6.6\%$)
from Table \ref{aliasnewrsg} are over-plotted with red diamonds. 
The locations of the peculiar 2MASS J18444023$-$0315329 (A), 
2MASS J18104421$-$1929072 (B), 2MASS J18334070$-$0750531 (C), and 
  2MASS J18360390$-$0740042 (D), 
which are discussed in the text, are marked.
The \ew\ values of the targeted late-type stars  
in Table \ref{aliasnewrsg}
found to have water absorption ($> 6.6\%$),
and 2MASS J18345133$-$0713162 (E) 
and  2MASS J18345840$-$0714247 (F) from   Table \ref{aliaskrsg}
are over-plotted with green encircled crosses.} 
\end{figure*}

\section{Applications of the new spectroscopic diagnostics}
\label{classtarget}
We observed 42 candidate RSGs, which had been identified
with a 2MASS-GLIMPSE color selection, as described in \citet{messineo17}
and \citet{messineo12}.
In our previous work, low-resolution $H$-band and $K$-band 
observations (R=1000)
enabled us  to conclude that at least 50\% of this sample 
was made of RSGs later than M0 type.

One of the targets (2MASS J18355534$-$0738197) 
turned out  to be of   F3-F6 type.

The remaining 41 candidates are late-type stars (K and M).
By comparing  their spectral features
with those of known RSGs in Table \ref{aliaskrsg}, 
we  spectroscopically 
classify the candidate stars of Table \ref{aliasnewrsg},
by assigning spectral types and luminosity classes.
We used the locations on the  Mg I versus CO diagram, 
the absence of strong water absorption, 
and, for bluer sources detected from 0.9 to 1.1 \um, 
the strengths of the  \ewJudJusJuo, 
along with strong \ewJoJnJd,  
and the absence of TiO band heads at 1.1 \um.
For  sources detected only long ward of 1.3 \um,
we used the \ewJduJdd, and 
strength of the Si I line at 1.59 \um.
The infrared \ew s enable us to spectroscopically classify 
50\% (21 stars) of the late-type targets 
as secure M I and one K5 I.  

The \ew s of the candidate RSGs  
are shown in Fig. \ref{fig.ewcandidate}.

The spectra of the targets  2MASS J18104421$-$1929072, 
2MASS J18334070$-$0750531, and 2MASS J18444023$-$0315329  
are anomalous (Fig. \ref{rsgc4s1fig}).
They have strong TiO band heads at 1.1 \um\
(like in the spectra of known O-rich Mira AGB stars),
but without water absorption. They  also have  VO absorption.
Furthermore, we see strong TiO band heads at 0.92-0.94 \um\ and weak ZrO 
band heads at  0.930-0.936 \um.\\
The spectrum of 2MASS J18360390$-$0740042 displays  a small  \ewmg\ 
and  a small \ewsiucn,  but
 a strong \ewco, resembling the spectrum  of an S-type.

32\% (13 stars) of the late-type targets have  
visible water absorption in their spectra.
Their features are consistent with being Mira AGBs.
For example, they have 
strong CO at 2.29 \um, {\bf a} weak Si I line at 1.59 \um,  
and their \ewJudJusJuo\ values are below the 
threshold for RSGs (see Fig. \ref{fig.ewmira}).
Among them, eight spectra display  TiO band heads 
at 1.1 \um, and three stars also show  strong 
TiO band heads at 0.92-0.94 \um.
 We only cast some doubts
 on the classification of 
2MASS J18105174$-$1947374, 
and 2MASS J18251767$-$1218360 because,
besides water and strong TiO band heads at 1.1 \um, 
they also have  visible CN band heads at  1.09-1.1 \um.\\
For three  stars (2MASS J18341160$-$0940545, 
2MASS J18570375+0152049, and 2MASS J18253811$-$1306250), 
the {\bf analyzed spectra  
do not cover the  TiO bands at 1.1 \um\ and 0.92 \um.}
For the spectra of 2MASS J18391007$-$0655526 
and 2MASS J19293750+1758006, 
the coverage and signal-to-noise,
allow us  to detect only VO absorption, which suggests 
Mira AGBs.

The remaining three unclassified  stars are
2MASS J18363595$-$0718472, 2MASS J18353576$-$0811451,
 2MASS J18153105$-$1744228  and have late M-types.
2MASS J18363595$-$0718472  is an O-rich star with strong VO absorption 
and visible CN band heads, without strong water absorption
(the CaT and TiO at 0.88-0.92 \um\ spectral regions were not extracted); 
it has  stronger Mg I at 1.71 \um\ than most RSGs with 
similar CO band heads at 2.29 \um.
In the spectrum of 2MASS J18153105$-$1744228, there is no visible 
water absorption, but
 the {\bf plot} of Mg I at 1.71\um\ versus CO  shows strong Mg I at 1.71 \um.
In the spectrum of 2MASS J18353576$-$0811451, the first measured index is J21,
there is no water absorption, but,
in contrast to the majority of RSGs,
it has a weak Si I at 1.59 \um\ and a weak \ewJdd.

\begin{figure}
\begin{center}
\resizebox{1.\hsize}{!}{\includegraphics[angle=0]{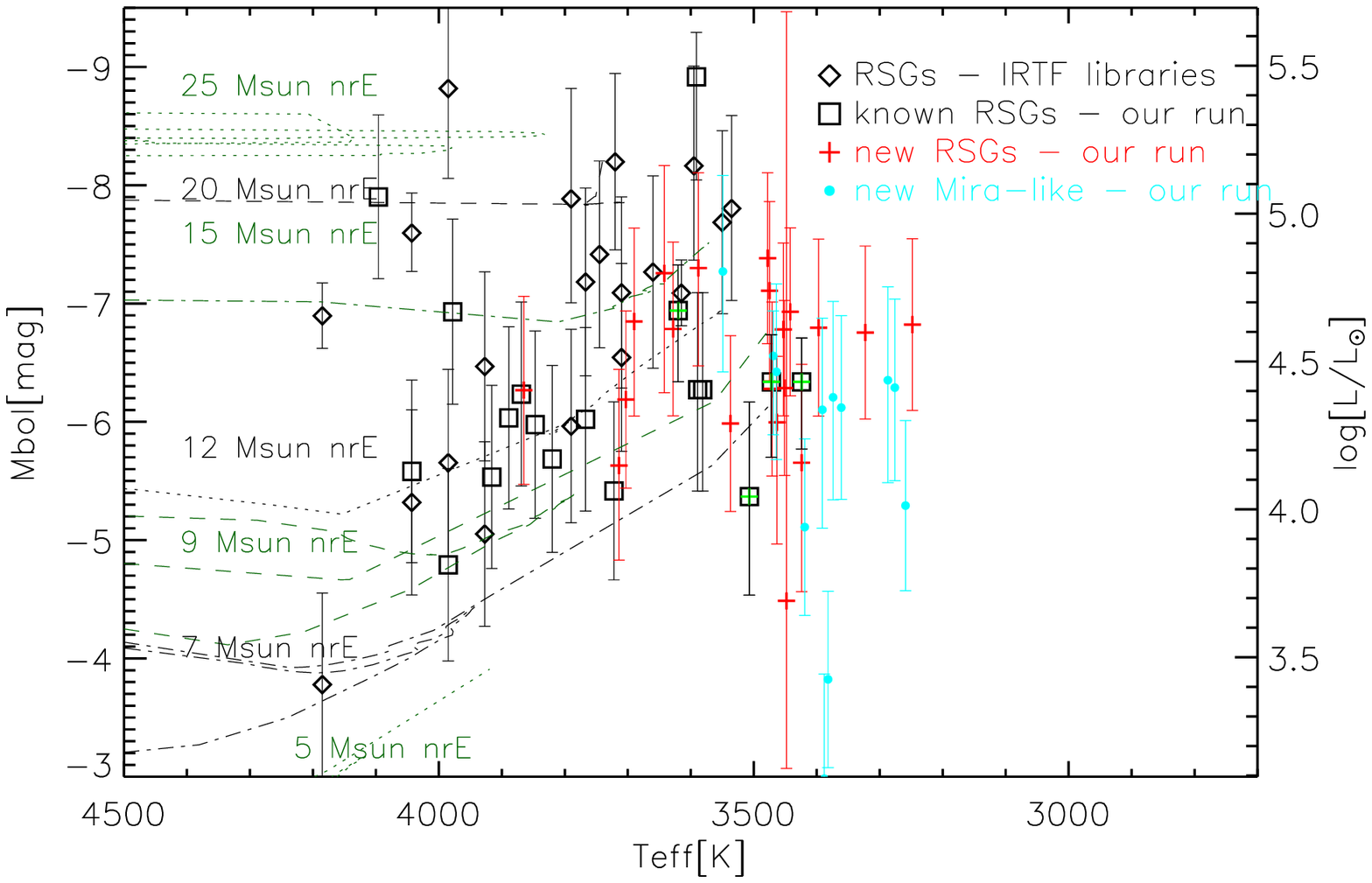}}
\end{center}
\caption{ \label{fig.lum} {\it Top panel:} Bolometric magnitudes vs. \Teff.
\Teff\  are from previous works if optically derived (see Table \ref{aliaskrsg}), 
or our \tcombi. 
For comparison, we add some  non-rotating stellar tracks 
with solar metallicity by \citet{ekstrom12}. 
The green dotted line at the bottom indicates
a track of a 5 \Msun\  star (90-110  Myr).
the black dotted-dashed curve that of   
a 7 \Msun\ star (42-49  Myr);
the green long-dashed curve marks a 9 \Msun\ track  (27-30  Myr); 
the black dotted curve a 12 \Msun\ track  (16-18  Myr); 
the green dotted-dashed curve shows a 15 \Msun\ track (14-15  Myr); 
the black long-dashed  curve  a 20 \Msun\ track  (8.5  Myr), 
and the top green dotted line that of a 25 \Msun\  (6.7  Myr). 
For known RSGs, the absolute \Mbol\ magnitudes are  obtained 
with parallactic distances from Gaia EDR3, when their  fractional errors are $<25 \%$;  
in order to include a sample of faraway known late-M supergiants,
for larger parallactic errors and \Vlsr$>70$ \kms, near-kinematic distances are used.
For the newly detected RSGs, which have only one  parallax within 
the desired constraints, we use the near-kinematic distances.
Plotted data points and symbols are as explained in the legenda.
The four known late-M RSGs 
2MASS J18392036$-$0601426 (RSGC2-14),  
2MASS J18391505$-$0605191 (RSGC2-17)
2MASS J18451939$-$0324483 (RSGC3-S6), 
and 2MASS J18333780$-$0921380 (MMF2014-44) are marked with a green cross.
}
\end{figure}

\subsection{Luminosities and distances of candidate RSGs}
In order to check the spectroscopic classification 
of the 41 candidate RSGs,
we estimated their bolometric magnitudes, which   
are listed in Table \ref{aliasnewrsg} (see also Sect. \ref{seclum}).
In the Gaia EDR3 catalog, parallaxes are available 
for 19 candidates, with only two data points with fractional
errors smaller than 25\%.
Therefore, we estimated their near-kinematic distances by using 
the \Vlsr\ values and the Galactic rotation curve of
\citep{reid14}. For each star, we adopted the   \Vlsr\ values 
from the Gaia DR2 catalog \citet{gaiadr3}, when available, 
or the values from \citet{kharchenko07}
otherwise, those measured from our SpeX measurements.  
We estimated the errors on distances  by
assuming a  SpeX velocity uncertainty of 22 \kms,
which is the $\sigma$ of the differences between the
Gaia DR2 \Vlsr\ values and the \Vlsr\ from SpeX data.  
Most of the  candidates are  at distances larger than 3.5 kpc 
from us (near-kinematic distances).

By considering the 18 known RSGs in Table \ref{aliaskrsg} 
with good Gaia EDR3 distances (parallax fractional error 
smaller than 25\%) and  kinematic distance errors $< 900$ pc,
we obtained a mean difference between the Gaia distances 
and kinematic distances of $-334$ pc with a $\sigma$ {\bf of} 710 pc. 
For  RSGs in Table \ref{aliaskrsg} reported in \citet{messineo19} 
and \citet{levesque05} as members of stellar associations, 
the mean difference between the stellar association distances 
by \citet{levesque05} and the Gaia individual distances is 
$-107$ pc with a $\sigma$= 333 pc; while the mean difference 
between the stellar association distances 
and the kinematic distances is $176$ pc with  $\sigma$= 858 pc.

20 (out of 21) candidates that spectroscopically 
meet the properties of supergiants, 
are shown  in the \Mbol\ versus \Teff\ diagram of Fig. \ref{fig.lum}.
Their log(L/\Lsun) estimated from the EWs differs from 
the photometric estimates by +0.18 with  $\sigma$=0.50.
While eight have spectral types from K5 to M4, which are  
typical of  RSGs  \citep{messineo19},  nine have  M5 types, 
but their luminosities  are still
compatible with that of RSGs within  the margin of  uncertainty. 
Three have extremely late types (M5.5-M6) and are located 
in the  area where  massive AGBs and super AGBs may reside
\citep[5-9 \Msun,e.g.,][]{messineo19,shetye19}.
Stellar evolution  and  threshold masses of AGBs and RSGs
is another chapter of astronomy still {\it ``in fieri''}.
Therefore, detections of these stars are  {\bf highly desirable.}
Precisely because going  from luminosity classes 
{\bf to physical properties like internal types of nuclear 
burning and stellar structures, high-resolution investigation 
of s-process products is required}
\citep[e.g.,][]{shetye20}, 
and larger statistics. In 2022, Gaia DR4 will provide  
improved distances and spectral types for millions of Galactic stars.

\section{Summary of analyzed infrared diagnostics for RSGs}
\label{result}

The spectroscopic issue we have investigated here is the building up of 
a  reference frame of \ew s at a medium resolution that allows us
to distinguish Galactic giants and RSGs from $K$- to $M$-types, using
the IRTF spectral library.

We have confirmed that with the IRTF/SpeX spectra at R=2000 it is possible 
to separate red giants and supergiants in several diagrams
and to identify contaminant AGBs.

\begin{itemize}
\item
By using the \ewcodue\  values and of \ewmg\  values,
it is possible to distinguish giants from RSGs  down to K0 type
 (\ewcodue$\approx 22-30$ \AA,  as shown in Fig. \ref{CO4}). 
About 95\% of known RSGs show Mg I lines
significantly weaker than those of giants. 
So far, at medium resolution, the  infrared identification of RSGs  
has been based uniquely on the CO band heads at 2.29 \um, 
which has allowed us  to detect M-type RSGs 
with  \ewcodue$\ga$ 45 \AA\ (M0).

\item
A distinctive feature of Mira AGBs is a strong water vapour absorption in $H$ and $K$-band.
This occurs rarely in RSGs.
Only 6.5\% of the spectra of  previously known RSGs is featured with 
water absorption above $>6.6\%$, mimicking O-rich Mira AGBs
(see Appendix Fig. C). 

\item
In addition to the Si I at 1.59 \um\ already mentioned by \citet{origlia93},
the J12, J17, J18, J21, J22 are found to be good indicators 
of luminosity classes  (see Figs. \ref{lumsepa}
and \ref{figEWlum-teff}). 
By combining the indices, for example, J12 and J21, 
stars with \ewJduJdd $> \approx 2.4$ \AA\ are found to be uniquely RSGs.
These are undoubtedly optimal indicators of  luminosity class 
capable of identifying supergiants  
from K to late M-types in $J$-band.
O-rich Mira AGB stars behave in a similar way to giants.

\item 
The J9 and J10 lines  in $Y$-band are dominated by CN band heads and are
visibly stronger in RSGs than in giants 
of similar temperatures  (Fig. \ref{comboCN}). 
O-rich Mira AGB stars and normal giants have \ew s smaller than RSGs. 

\item
The spectra of contaminant S-type stars have weaker 
Si I lines at 1.59 \um, 
Si I lines at 1.0588 \um\ (J4), 
and Si I at 1.078 \um\ (J8) than those of RSGs
 (Fig. \ref{figStype}).

\item
Spectral types consistent with those from the optical classification 
are obtained by comparison with  the Na I at 2.21, 
Mg I at 1.71 \um, Si I at 1.64 \um,  
CO at 1.62 \um\ of well-known RSGs  (Fig. \ref{measureTeff}).
Typically they are accurate  within 1-2 spectral types.

\item
When sorting the spectra by the  estimated spectral types/temperatures, we find
a nice sequence   of dominant molecules with spectral types.
For example, ZrO and VO appear in RSGs with spectral types 
later than M3 ( Appendix Figs. B and C).

\end{itemize}

\begin{itemize}

\item
Only five K0-2 RSGs are in the IRTF library and one in the extended
IRTF library. \\
We observed 21 other known RSGs, 20 of which with quoted K-types.
Unfortunately, only 11 were found to be K-types.
2MASS J22362757+5825340 (W Cep) is a  K0 I, while 2MASS J22495897+6017567 and 
2MASS J21440431+5342116 are K1-K2 I.

\item
We used these RSGs' features to spectroscopically classify
41 newly targeted late-type stars.
We found that 28  out of 41  candidates (68\%)
have a low  water content 
($<6.6\%$), and we  assigned  21 candidates (51 \%) 
 to class I because of their location in the
\ewmg\ versus \ewcodue\ diagram, 
of their \ewJudJusJuo\ and \ewJoJnJd\ values, which are 
 typical of RSGs. 

13 out of 41  (32\%) have water absorption ($>6.6\%$),
of which 23\% have also strong TiO features at 0.93,  
and 62\% have TiO features at 1.1 \um.
The 1.1 band head is always seen in Mira AGBs.
Their locations in the diagrams are consistent 
with those of known O-rich Mira stars. 

\item
In conclusion, our 72 observed targets 
consist of 9 giants, 21 already known RSGs, and 42 candidate RSGs.
Of the 21  previously known RSGs, 19 are  reclassified 
as spectroscopic RSGs, while two stars 
(2MASS J18345840$-$0714247, and 2MASS J18345133$-$0713162) 
appear to be  Mira AGBs. Among the candidate RSGs, we are able 
to confirm 21 RSGs.

\end{itemize}

These infrared results are particularly useful for defining new and 
focused studies of the inner Galaxy, where
optical spectroscopy is strongly hampered by  dust obscuration, 
with \Av\ extinction up to 40 magnitudes, and where
most of the Galactic mass resides.

Furthermore, we have located an unblended  line useful to map the 
Galactic abundance of  Mn I at 1.29 \um.
Only six stars (out of the 96 in Table 5) are particularly 
rich in Mn I (three RSGs from the IRTF library, 
three from our observations).


\begin{acknowledgements}
We are thankful to the IRTF staff  for helping us with the observations.
MM is thankful to Dr. Anthony Brown for the catalog of RSGs.
{\bf We thank the anonymous referee for her/his careful reading and constructive comments.}
This work has made use of data from the European Space Agency (ESA) 
mission {\it Gaia} (\url{http://www.cosmos.esa.int/gaia}), 
processed by the {\it Gaia} Data Processing and Analysis
Consortium (DPAC, \url{http://www.cosmos.esa.int/web/gaia/dpac/consortium}). 
Funding for the DPAC has been provided by national institutions, 
in particular the institutions participating in the {\it Gaia} 
Multilateral Agreement. This publication makes use of data products 
from the Two Micron All Sky Survey, which is a joint project of 
the University of Massachusetts and the Infrared Processing
and Analysis Center/California Institute of Technology, 
funded by the National Aeronautics and Space
Administration and the National Science Foundation. 
This research has made use of the VizieR catalog access tool, 
CDS, Strasbourg, France, and SIMBAD database. 
This research has made use of NASA’s Astrophysics Data 
System Bibliographic Services.  
R.P.K. acknowledges
support by the Munich Excellence Cluster Origins Funded by the Deutsche 
Forschungsgemeinschaft (DFG, German Research Foundation) under 
the German Excellence Strategy EXC-2094 390783311.
This work was partially supported by the National Natural 
Science Foundation of China (NSFC-11773025, 11421303), and USTC grant KY2030000054. 
\end{acknowledgements}

\clearpage

\begin{appendix}
\def\thefigure{\thesection}
\def\thetable{\thesection}

\section{Relation between luminosities, temperatures, and gravities of RSG stars.} 

The fundamental theoretical quantities of stellar astrophysics
are the effective temperature \Teff, the gravity log(g),
the luminosity log(L/\Lsun), and the initial chemical composition. \\
The \Teff\ values can be estimated
using dereddened colors and/or  spectral types 
(assuming a temperature scale).
We have used $HK$ atomic lines (Ca I, Na I, Si I, Mg I) as 
well as the CO band heads at 1.62 \um\ to estimate the \Teff\ values (Sect. \ref{sectemperature}).
Luminosities have  been estimated by photometric measurements 
and by assuming a distance (e.g. the Gaia parallactic distances).
For Galactic RSGs, which span a small range of metallicity 
(mostly between $-$0.5 and +0.5), the  stellar gravity can be  estimated 
quite straightforwardly by establishing a relation between \Teff\ or log(L/\Lsun) and log(g)
with stellar tracks.

\begin{figure*}
\begin{center}
\resizebox{\hsize}{!}{\includegraphics[angle=0]{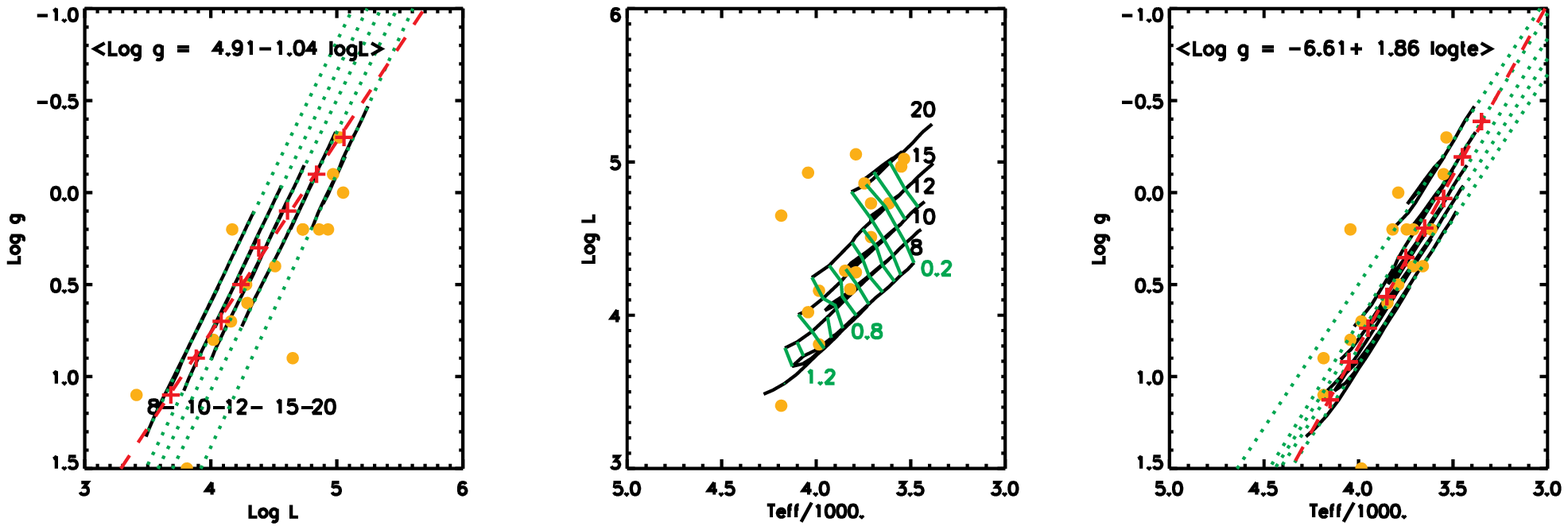}}
\resizebox{\hsize}{!}{\includegraphics[angle=0]{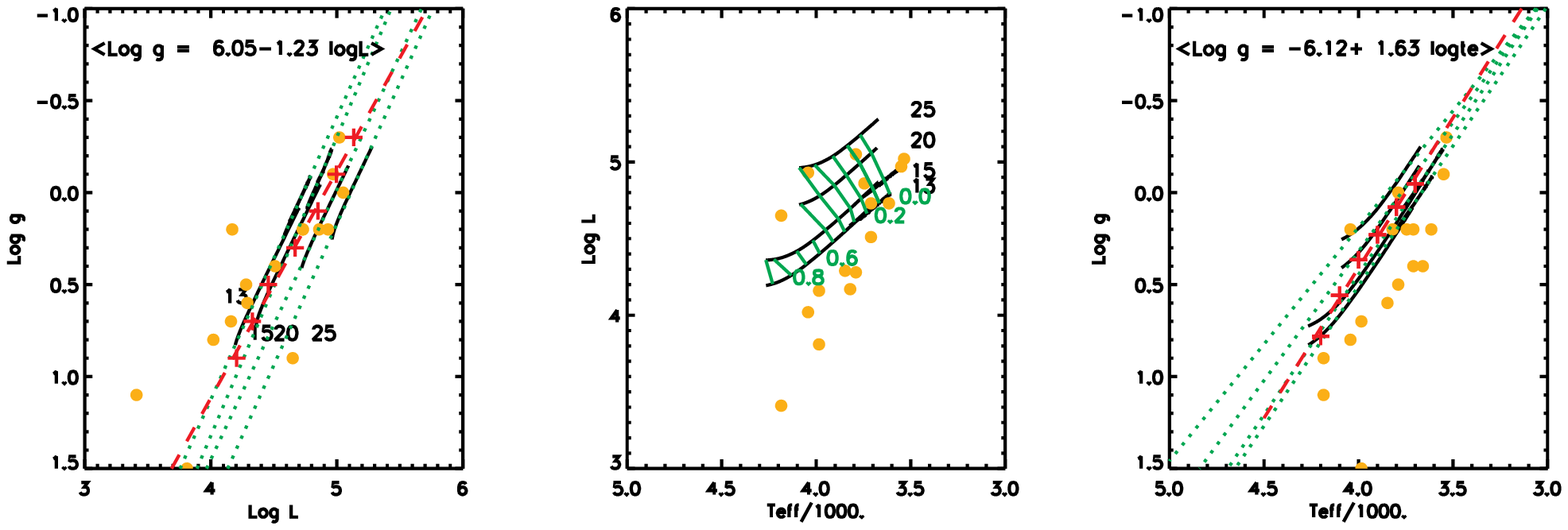}}
\caption{\label{fig.bertelliG} 1.  
{\it Top Panels:} For stars of 8, 10, 12, 15, and 20 \Msun,
the RSG portions of the stellar tracks  with solar metallicity by \citet{bertelli09}
were extracted. {\it In the top left panel}, their log g values are plotted vs. 
the log L values (black segments). They can be approximated with 
linear curves (green dotted lines). The red crosses represent the average locus
per bin of luminosities and the red dashed line is their fit. 
Orange filled circles mark parameters of RSGs collected from literature
 and listed in Table \ref{litpar}.
{\it In the top middle panel}, The Log L values are plotted vs. the \Teff\ values.
In black the stellar track segments, and in green the connecting segments
 of points at constant gravity. Orange filled circles are data points from
 Table \ref{litpar}.
 {\it In the top right panel}, the log g values are plotted vs. 
the \Teff\ values (black segments). They are approximated with 
linear curves (green dotted lines). The red crosses represent the average locus
per bin of temperatures, which are fitted  with the red dashed line. 
Orange filled circles are data points from
 Table \ref{litpar}.
{\it Bottom panels:} For stars of 13, 15, 20, and 25 \Msun,
the RSG portions of the stellar tracks  with solar metallicity by \citet{chieffi13}
were extracted. Plotted diagrams are as  above.
}
\end{center}
\end{figure*}

We used  the RSG portions of the stellar tracks from 8 to 25 \Msun\
with solar metallicity by \citet{bertelli09}
and by \citet{chieffi13}. In Fig. \ref{fig.bertelliG}1, 
the log(g) values are plotted versus the log(L/\Lsun)
values, as well as versus the \Teff\ values.
During the RSG phase,
while the star ascends the red part of the track,
there is one linear relation between log(L/\Lsun) and log(g) for any given initial mass.
By increasing the initial mass
we shift this relation (segment) to a higher luminosity and 
lower gravity. By simply considering the average locations of these segments
for stars from 8 to 20 \Msun, we derived an average relation 
between the  log(L/\Lsun) (estimated with observable quantities) and 
log(g) (the theoretical quantity to be measured).
Similar equations are derived from 
the log(g) values and  the \Teff\ values. 

The obtained log(g) values from \citet{bertelli09} are  accurate to within 0.5 dex (the span expected due to different initial masses),
and independent of the knowledge of the initial mass of the star
(which is not an observable quantity). These are the derived equations:
 $${\rm log~g} = 4.91(\pm 0.13) -1.04(\pm0.03) \times {\rm log~L;}$$
$${\rm log~g} = -6.61(\pm 0.08) +1.86(\pm0.02) \times {\rm log~T_{\rm eff.}}$$

We found gravity estimates from stellar spectra for 17  RSGs included in this work (Table \ref{litpar}).
The equations from the \citet{bertelli09} well describe the  observed data points;
when using the average relations inferred from the tracks of \citet{bertelli09}, 
we found a general offset of $-0.28$
and a $\sigma$ of 0.28 dex.

The theoretical relations (log(g), Teff, log(L/\Lsun)
depend on the assumed code, for example, on the opacities, the mixing length scale $\alpha$, 
the inclusion of overshooting. Tracks become cooler for larger value of $\alpha$ \citep[e.g.,][]{chun18}. 
\citet{bertelli09} adopted $\alpha=1.68$, while \citet{chieffi13} use
$\alpha$=2.3 \citep{straniero97}.

\begin{table*}
\caption{\label{litpar} Parameters of known RSGs collected from literature.}
\begin{tabular}{llrrrrrrr}
\hline
\hline
2MASS   & Alias & $[$Fe/H$]$ & Ref. & log(g) & Ref. & log(g)$^*$ \\
\hline
         00511639+6148196  &                   HD 4817  &     0.0  & 7  &    0.6  & 8&      0.3$\pm$    0.2   &   \\
         01195361+5818308  &                 HD 236697  &     0.0  & 9  &    0.4  &10&      0.1$\pm$    0.1   &   \\
         01333316+6133296  &                 BD+60 265  &     0.2  & 1  &   $..$  & 0&   $-$0.0$\pm$    0.0   &   \\
         01431110+4831002  &                  HD 10465  &  $-$0.1  & 1  &   $..$  & 0&   $-$0.0$\pm$    0.1   &   \\
         02214241+5751460  &                  HD 14404  &     0.1  & 2  &    0.2  & 3&   $-$0.1$\pm$    0.1   &   \\
         02220690+5636148  &                  HD 14469  &     0.1  & 2  &   -0.1  &10&   $-$0.1$\pm$    0.1   &   \\
         02222428+5706340  &                  HD 14488  &    $..$  & 0  &   -0.3  &10&   $-$0.2$\pm$    0.1   &   \\
         05271022+2955158  &                  HD 35601  &     0.1  & 1  &    0.2  &10&      0.0$\pm$    0.0   &   \\
         05551028+0724255  &                  HD 39801  &     0.2  & 1  &    0.4  & 4&      0.0$\pm$    0.0   &   \\
         06121911+2254305  &                 HD 042543  &     0.2  & 1  &    0.0  &10&   $-$0.1$\pm$    0.2   &   \\
         06224788+2759118  &                  HD 44391  &     0.0  & 4  &    1.1  & 4&      0.7$\pm$    0.6   &   \\
         06300229+0755159  &                 HD 045829  &     0.1  & 1  &    0.8  & 5&      0.4$\pm$    0.3   &   \\
         07410262-3140591  &               CD-31~4916   &  $-$0.2  & 1  &    0.2  & 4&   $-$0.0$\pm$    0.0   &   \\
         07473853-1559263  &                  HD 63302  &     0.0  & 1  &    0.2  & 4&   $-$0.1$\pm$    0.2   &   \\
         18082651-1833078  &                 HD 165782  &     0.2  & 6  &    0.9  & 6&      0.1$\pm$    0.0   &   \\
         18375890-0652321  &                RSGC1-F13   &  $-$0.2  & 3  &   $..$  & 0&      0.0$\pm$    0.0   &   \\
         18392036-0601426  &                RSGC2-14    &  $-$0.2  & 3  &   $..$  & 0&      0.0$\pm$    0.0   &   \\
         19392533+1634163  &                 HD 185622  &     0.1  & 1  &   $..$  & 0&      0.2$\pm$    0.1   &   \\
         19461557+1036475  &                   gam Aql  &  $-$0.2  & 7  &    1.5  & 8&      0.5$\pm$    0.4   &   \\
         19481183+2245463  &                 HD 187238  &     0.0  & 1  &    0.7  & 8&      0.3$\pm$    0.2   &   \\
         19501193+2455240  &                 HD 339034  &     0.3  & 1  &   $..$  & 0&   $-$0.3$\pm$    0.4   &   \\
         20185752+3900151  &               BD+38 4003   &  $-$0.2  & 8  &    0.2  & 8&      0.3$\pm$    0.3   &   \\
         20285059+3958543  &                   RW Cyg   &     0.3  & 1  &   $..$  & 0&   $-$0.2$\pm$    0.3   &   \\
         21433045+5846480  &                 HD 206936  &     0.1  & 7  &   $..$  & 0&      0.0$\pm$    0.0   &   \\
         22230701+5557477  &                 HD 212466  &     0.2  & 1  &   $..$  & 0&   $-$0.7$\pm$    0.8   &   \\
         22562598+4944006  &                 HD 216946  &  $-$0.2  & 1  &    0.5  & 8&      0.3$\pm$    0.2   &   \\
\hline
\end{tabular}
\begin{list}{}
\item{\bf References:} 1=\citet{davies10}, 2=\citet{davies09}, 
3=\citet{gazak15}, 4= \citet{arentsen19}, 5=\citet{prugniel11},
6=\citet{stevens17}, 7=\citet{anderson12}, 8=\citet{pastel16}, 
9=\citet{ivanov04}, 10=\citet{dicenzo19}.\\
{\bf $^*$ log(g) are estimated} with the tracks of \citet{bertelli09}.
\end{list}
\end{table*}

\section{Molecular diagnostics for late-types and 
 contamination of  O-rich Mira AGB}
\label{molecules}

Infrared spectra of RSGs may display  broad features in absorption
due to molecular absorption, for example from TiO, VO, ZrO, ${\rm H_2O}$, 
and CN molecules.
A list of identified molecules in the SpeX spectra is provided
in the catalog of \citet{rayner09}. For the VO bands,
we used the list of \citet{joyce98}.
An illustrative work is also that of \citet{alvarez00}, where the authors
describe the occurrence of  molecular features with stellar types,
chemistry type, and envelope types.
Here we only annotate molecular diagnostics that 
are useful for spectral classification 
and luminosity classification of RSGs.

\begin{itemize}
\item
{\it  In our sample, there is no contamination of C-rich AGB stars.}\\
C-rich AGBs are easily discernible at infrared wavelengths because
of their strong CN, $C_2$, and CO lines. 
For example, typically the \ewJdieci\ (CN band head) 
values of RSGs range from 1-4 \AA,
while ranges from 5 to 10 \AA\ in C-rich stars \citep{rayner09}.
There are no C-rich stars among our targets.

\item
{\it Contamination: O-rich Mira AGBs and S-type AGBs}\\
The \ewmg\ versus \ewcodue\ diagram  effectively distinguishes  
RSGs from normal giant stars. 
However, it maintains  large contamination of  O-rich Mira AGBs
and hybrid S-type AGB stars, 
which are stars  that go from O- to C-rich composition 
\citep{wright09,joyce98,rayner09}.
Mira AGBs have very broad  \ewcodue, 
but they do not obey the \ewco-temperature relation \citep{blum03}. 
O-rich Mira AGBs are usually identified by
the strength of the water absorption.
\end{itemize}

The SpeX spectra with their large baseline 
provide us with further additional molecular 
considerations useful for distinguishing RSGs 
and O-rich Mira stars and S-type AGBs.

\begin{itemize}
\item
Independently  of their temperature,
the atmospheres of RSGs are rich in CN molecules; 
for example, strong  CN band heads
appear  around 0.91 \um\ and 1.09 \um. \\

The four CN band heads at 1.0875 \um, 1.0929 \um, 1.0966 \um, 
and 1.0999 \um\
appear distinguishable  in the spectra of any RSG --
also in the spectra of the rare M7.5I 2MASS J22543171+6049388 (MY-Cep) --
with band heads significantly
stronger than those in O-rich Mira stars (\ewJoJnJd),
as shown in Fig. \ref{comboCN} 
and already described in Sec. \ref{secJlum}.

\item
Absorption bands  of water vapor are a strong feature 
of O-rich Mira stars, but are weak and rare in RSGs.
A list of the spectral regions with water absorption
is given by \citet{rayner09}.
We measured the strength of the water absorption
in the region 1.75-2.05 as done by \citet{blum03} 
and \citet{messineo17}, and the results are 
listed in Table \ref{obstargets}.
In the spectrum of RSGC1-F13 (M2 I) we measured 0.\%, 
in 2MASS J22543171+6049388 (MY-Cep, M7.5 I)  5.7\%, 
in 2MASS J05551028+0724255/HD~39801 (M2 I) 5.9\%,  
and in  2MASS J22495897+6017567/PER406 (K2 I) 7.3\%.
However, in the spectra of  2MASS J18345133$-$0713162 
(\spcombi = M5.5) 
and 2MASS J18345840$-$0714247 (\spcombi = M5), 
we measure 
10.4\%, 18.2\%, which are 
particularly high values, typically found in Mira AGBs.

Spectra of late-type stars 
with continuum absorption  higher than 6.6\% should
be considered as possible O-rich Mira AGBs and are marked with 
a flag in Tables \ref{aliaskrsg},
\ref{aliasnewrsg}, and \ref{aliasgiants}.
A small percentage (6.5\%) of spectra of
known RSGs shows water absorption.

A steep absorption, likely due to water \citep{jonsson92,joyce98},
also appears at $\approx 1.34$ \um\ in the spectra of MY-Cep (M7.5I), 
2MASS J18375890$-$0652321/RSGC1-F13 (M2I), 
and  2MASS J22495897+6017567/PER406 (K1I).\\

\item 
TiO bands (0.82-0.86, 0.88-0.90, 0.92-0.94 \um)
are seen in the spectra of M-type RSGs, 
as well as in those of O-rich Mira AGB stars.\\

The TiO band heads at 0.88-0.90 \um\ start
to be visible in the spectra of M0-type RSGs, and are strong
in stars later than M2.5. 
TiO at 0.8859 \um\ is often used to infer spectral 
types of M-type RSGs \citep[e.g.][]{negueruela16}.

In late-M RSGs TiO bands at 0.92-0.94 \um\  are usually weak,  the dominant
feature in this range being the ZrO absorption.
On the contrary, TiO bands at 0.92-0.94 \um\ are strong 
in Mira AGB stars, as shown  in Figs. \ref{zrofig}1 and \ref{rsgc4s1fig}2.\\

Sharp TiO band heads at 1.103 and 1.116 \um\ 
appear in the spectra of all 
but one  O-rich  Mira AGB star \citep{rayner09,joyce98}.
In the late Mira AGB (IRAS 14303-1042, M8-9III), the 
TiO feature is  no  longer visible 
because there is  strong water absorption.
On the contrary, TiO band heads around 1.1 \um\
do not appear in the spectra of known RSGs.\\

\item
Continuum absorption due to ZrO at 0.930-0.936 \um\ 
appears to be visible in the spectra of all  RSG with 
spectral type M3 and is strong in M5-M6, as shown  in Fig. \ref{zrofig}. \\

The spectrum of  2MASS J17143885+1423253/HD 156014 (M5 Ib-II AGB) shows strong ZrO 
band heads.\footnote{ The star, however, falls in the giant sequence
in the \ewmg\ versus \ewcodue\ diagram and in the 
\ewJoJnJd\ versus \ewcodue. Its TiO bands are displayed in  
Fig.\ \ref{zrofig}.}\\

Unfortunately, absorption by ZrO also characterizes  
the spectra of all S-type AGBs 
\citep{rayner09,wright09,joyce98}. 
However, in contrast  to S-type stars, 
RSGs  with late spectral types (M3-M7.5) have ZrO along with strong 
TiO at 0.82-0.86, 0.88-0.90 \um, as shown  in Fig. \ref{zrofig}. \\

\item
If the ZrO and TiO regions are not covered, 
the spectra of S-type AGB stars and RSGs  can  be
 distinguished by using the ratios of their CO band heads at 
1.57 \AA\ and Si I absorption lines at 1.59 \AA. Indeed, the 
Si I line at 1.59 \um\ is visibly weaker in S-type AGBs 
than in RSGs (see Fig. \ref{figStype}).
High-resolution studies of Si I lines can open 
new frontiers in Galactic spectroscopy 
as shown by \citet{thorsbro20}.
The J2, J4, and J8 lines  have similar behavior to the 
Si I line at 1.59 \um.

\begin{figure*}
\begin{center}
\resizebox{0.9\hsize}{!}{\includegraphics[angle=0]{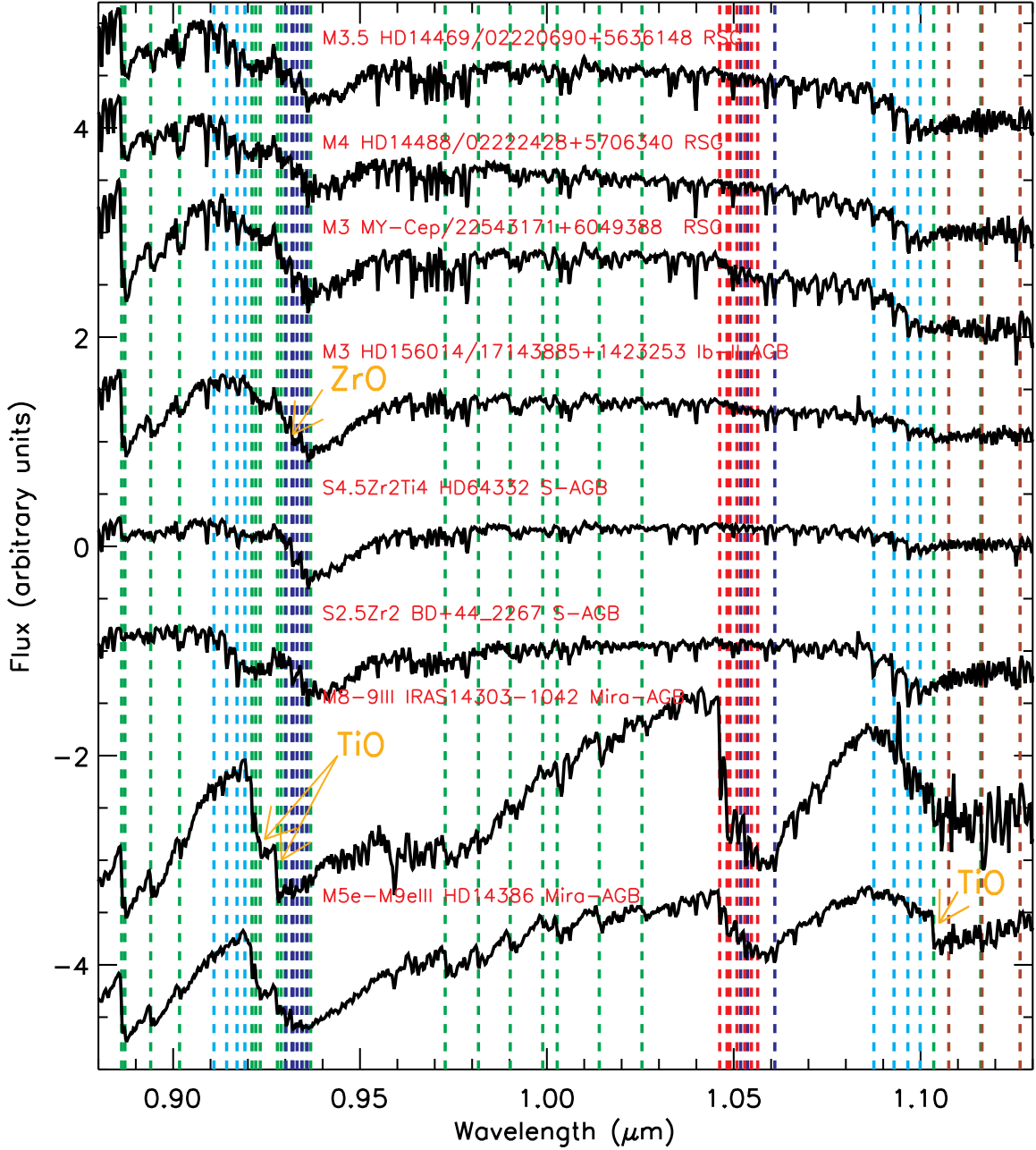}}
\caption{1. \label{zrofig}  Examples of  spectra taken from the IRTF library from 0.88 to 1.2 \um\ showing
the TiO band heads (green vertical lines), CN band heads (cyan vertical lines),
and ZrO band heads (blue vertical lines), VO band heads (in red).
The source names, spectral types, and classes are annotated 
above each spectrum. For RSGs,  our spectral types from $H$-band and $K$-band and
those from optical literature are marked, as in Table \ref{aliaskrsg}. 
ZrO band heads are evident in  late-type RSGs and S-type AGBs (see orange arrow), 
but not in O-rich Miras. TiO band heads are strong in Miras (see orange arrow). }
\end{center}
\end{figure*}

\begin{figure*}
\begin{center}
\resizebox{0.9\hsize}{!}{\includegraphics[angle=0]{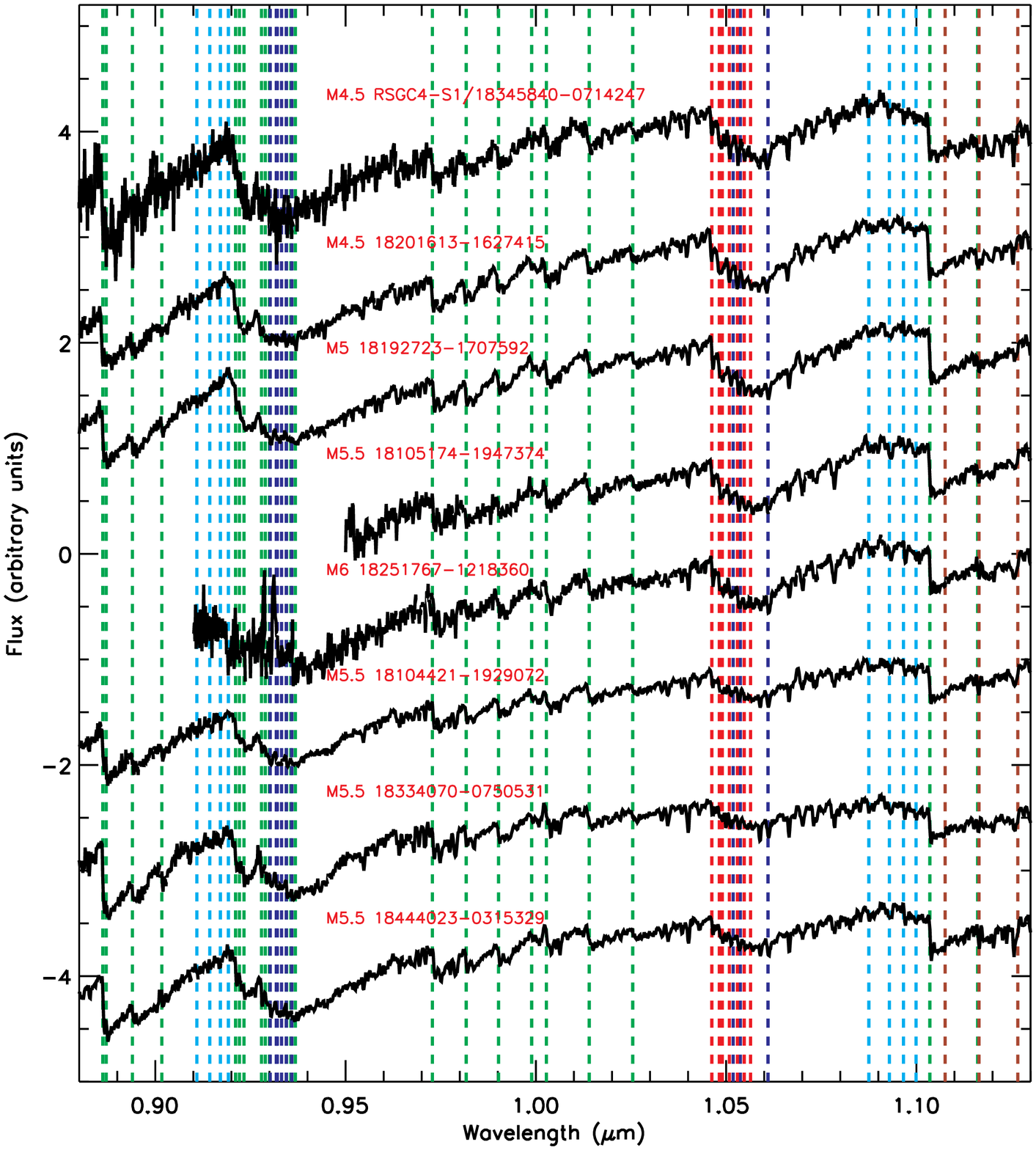}}
\caption{2. \label{rsgc4s1fig} Spectrum from 0.88 to 1.2 \um\ 
of  2MASS J18345840$-$0714247  from Table \ref{aliaskrsg} and  similar
spectra of  targets  from Table \ref{aliasnewrsg}. All but 
2MASS J18104421$-$1929072, 
2MASS J18334070$-$0750531, and 2MASS J18444023$-$0315329 
have water absorption.
These spectra resemble those of O-rich Mira AGBs. 
Vertical lines mark molecular band heads.
TiO band heads are in green, CN band heads in cyan,
ZrO band heads in blue, and the  VO band heads in red.
The source names and spectral types are annotated 
above each spectrum. The spectral types are those adopted in Table \ref{aliaskrsg}.
}
\end{center}
\end{figure*}

\item
VO absorption at 1.046-1.056 \um\ with a strong round 
signature of the continuum   
occurs in the  spectra of O-rich Mira AGBs,
as well as in the spectrum of   
 2MASS J18375890$-$0652321/RSGC1-F13 (M2I), 
2MASS J18451939$-$0324483/RSGC3-S6 (M4.5 I).
More typically, in the spectra of known 
RSGs with spectral type M3 or later, 
a weak VO mark  appears 
(as a change of  the stellar continuum slope).
\end{itemize}

\section{Spectral atlas and indices} \label{appendixspectra}

The figures with the obtained spectra are available only electronically
(Figs. \ref{fig.krsg}1, \ref{fig.new}2, \ref{fig.sp.mira}3, \ref{fig.sp.giant}4).
Only the wavelength range where the  signal-to-noise of the stellar trace 
is above 50 is plotted.

In Table \ref{obstargets} we list the measured EWs of the 72  stars in our
observational program and 24 from the IRTF libraries of \citet{rayner09} and \citet{villaume17}.
Only indices relative to spectral segments with a signal to noise 
larger than 50 are retained.

\begin{figure*}
\begin{center}
\resizebox{\hsize}{!}{\includegraphics[angle=90]{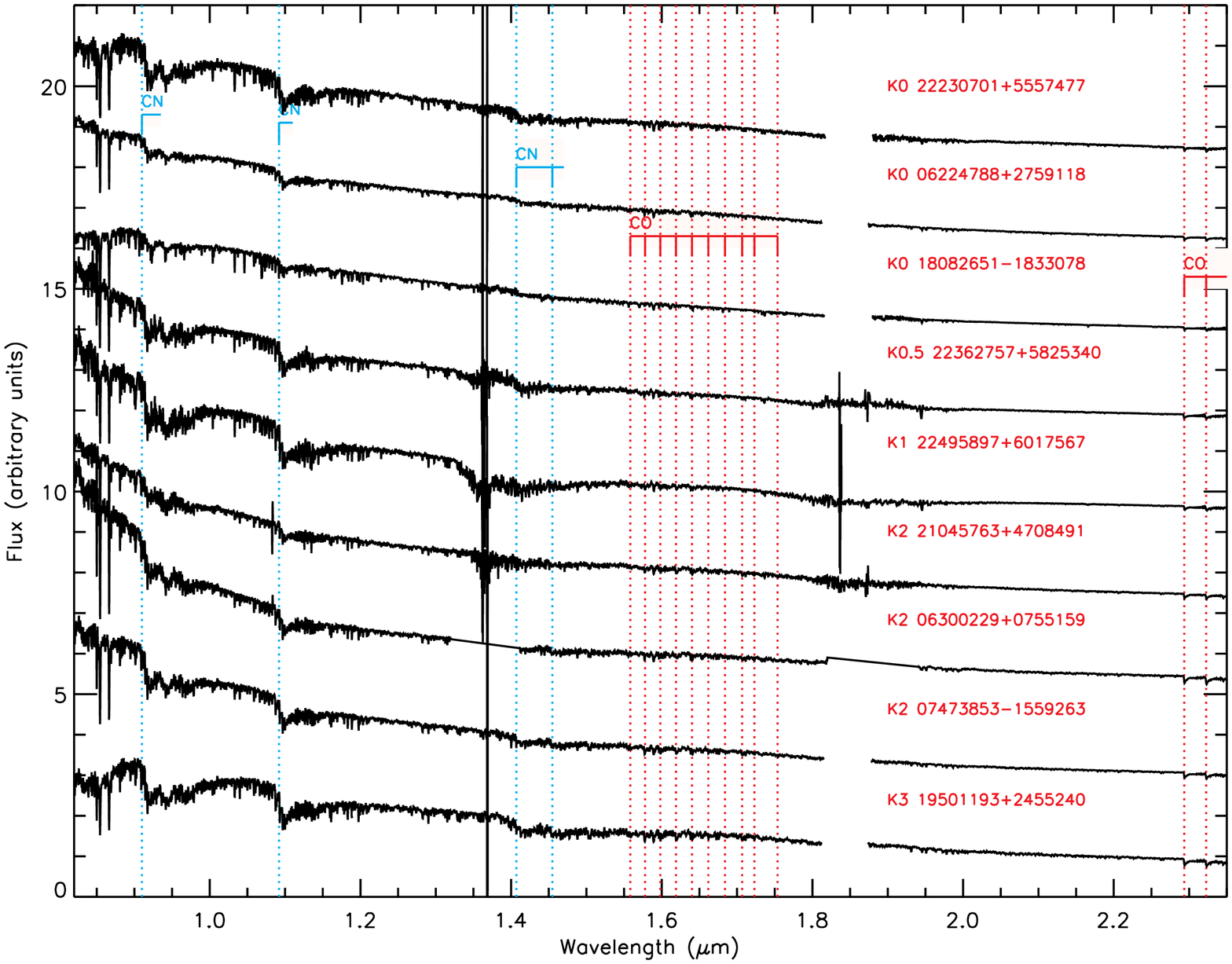}}
\caption{\label{fig.krsg} 1. IRTF spectra of  known RSGs
from Table  \ref{aliaskrsg}. 
 2MASS J17143885+1423253/HD 156014/$\alpha$ Her (M5 Ib-II, AGB) 
 is also plotted.
The spectra are sorted by adopted spectral type.}
\end{center}
\end{figure*}

\addtocounter{figure}{-1}
\begin{figure*}
\begin{center}
\resizebox{\hsize}{!}{\includegraphics[angle=90]{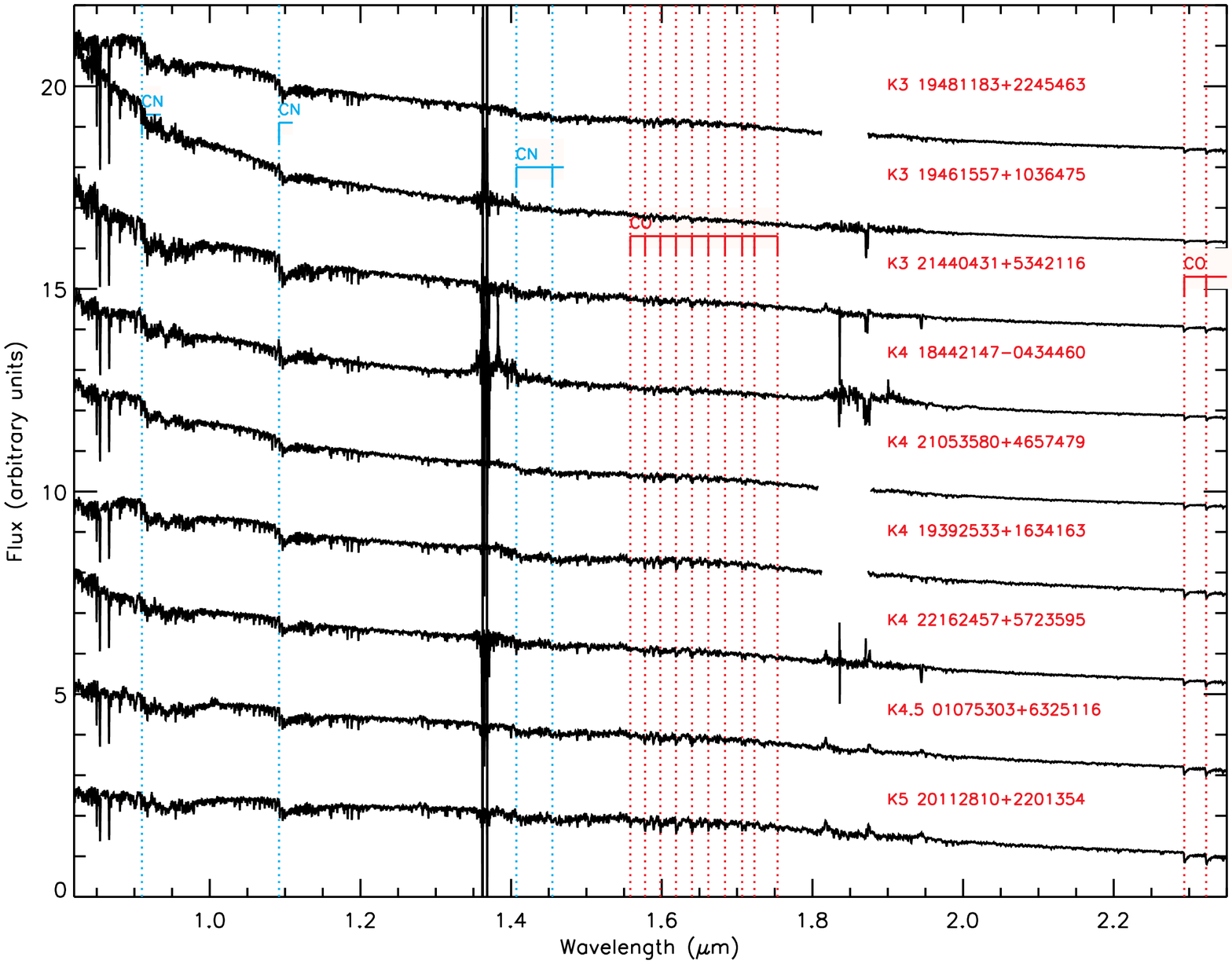}}
\caption{Continuation of Fig. \ref{fig.krsg} 1. IRTF spectra of known RSGs.}
\end{center}
\end{figure*}

\addtocounter{figure}{-1}
\begin{figure*}
\begin{center}
\resizebox{\hsize}{!}{\includegraphics[angle=90]{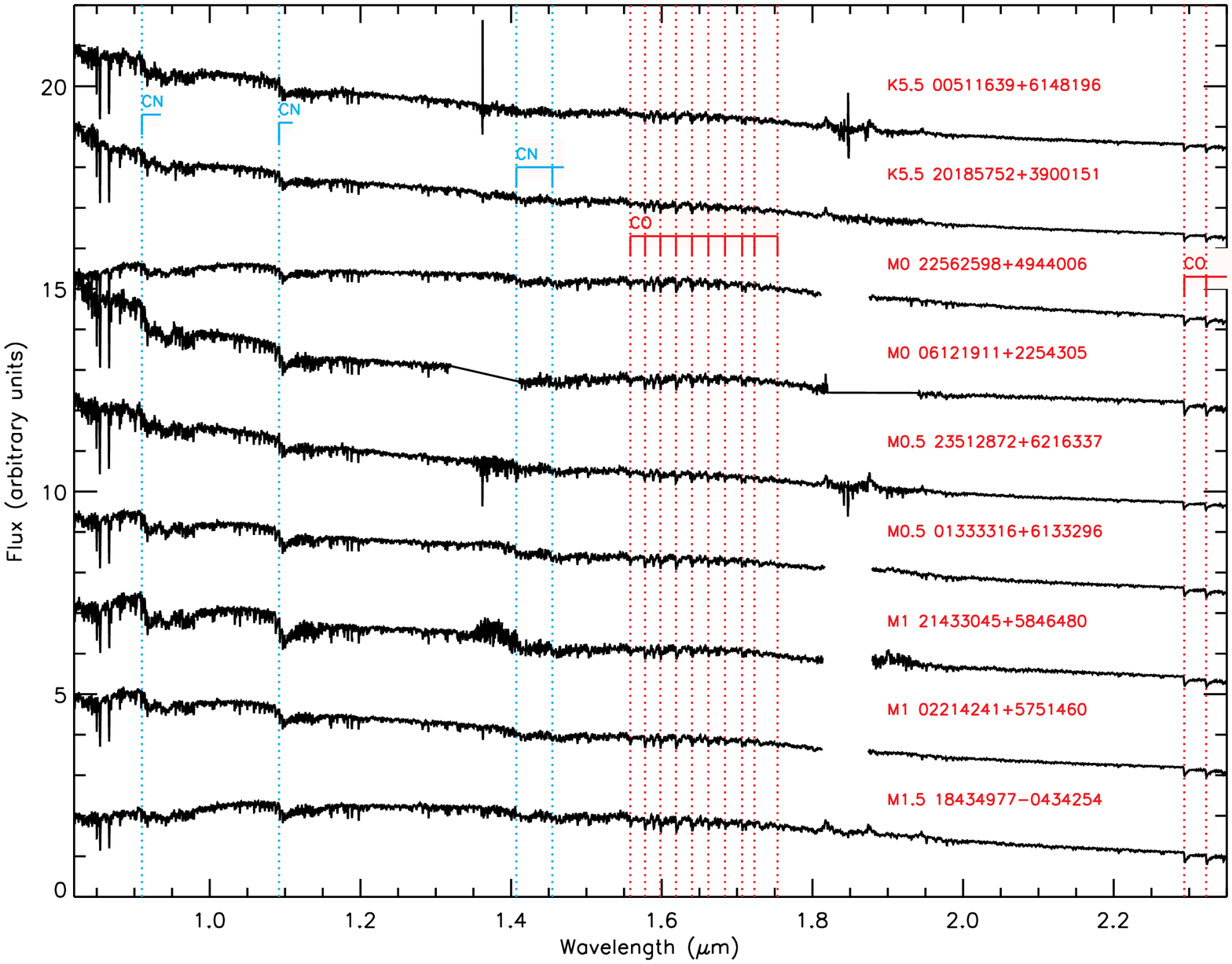}}
\caption{Continuation of Fig. \ref{fig.krsg} 1. IRTF spectra of known RSGs.}
\end{center}
\end{figure*}

\addtocounter{figure}{-1}
\begin{figure*}
\begin{center}
\resizebox{\hsize}{!}{\includegraphics[angle=90]{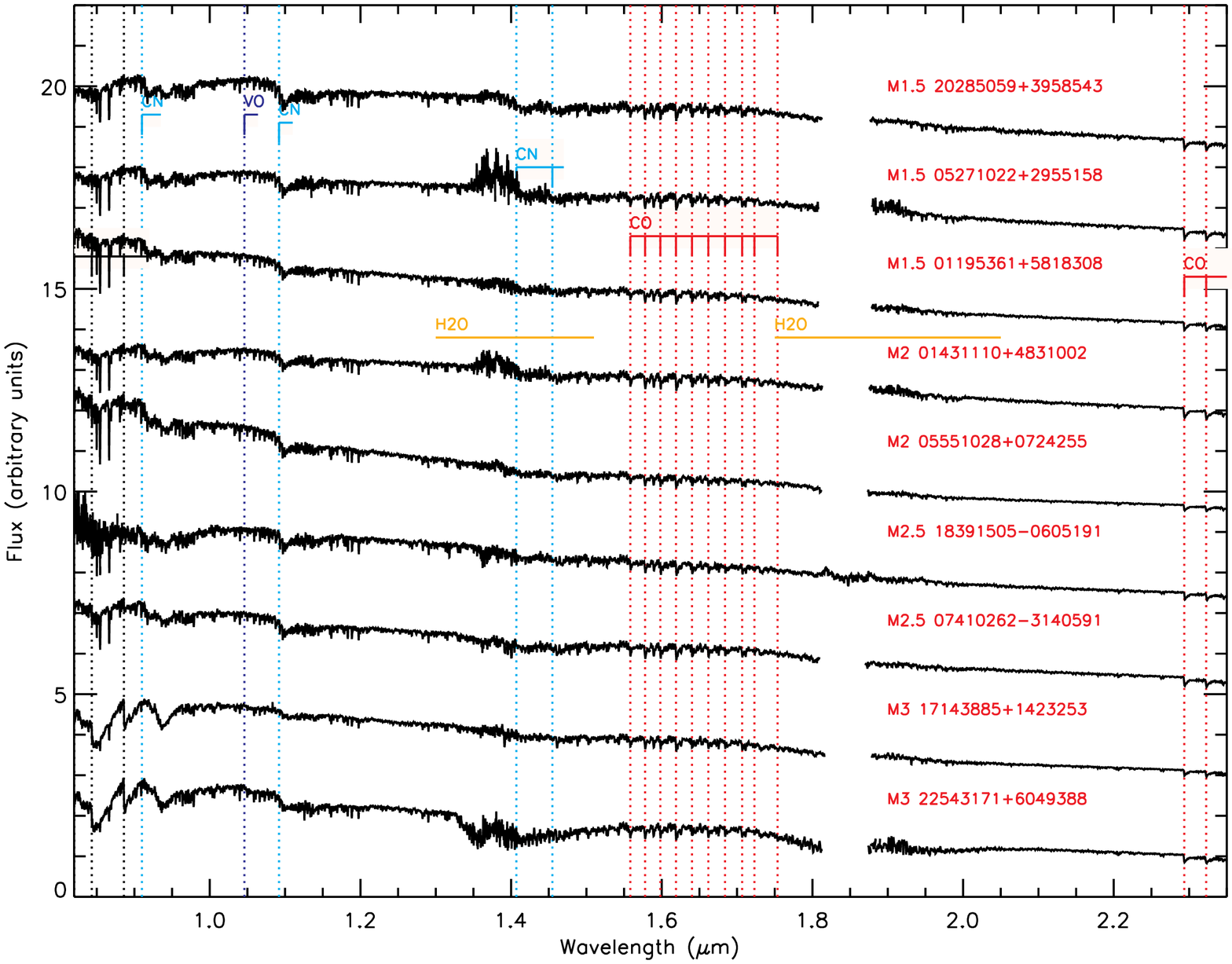}}
\caption{Continuation of Fig. \ref{fig.krsg} 1. IRTF spectra of known RSGs.}
\end{center}
\end{figure*}


\addtocounter{figure}{-1}
\begin{figure*}
\begin{center}
\resizebox{\hsize}{!}{\includegraphics[angle=90]{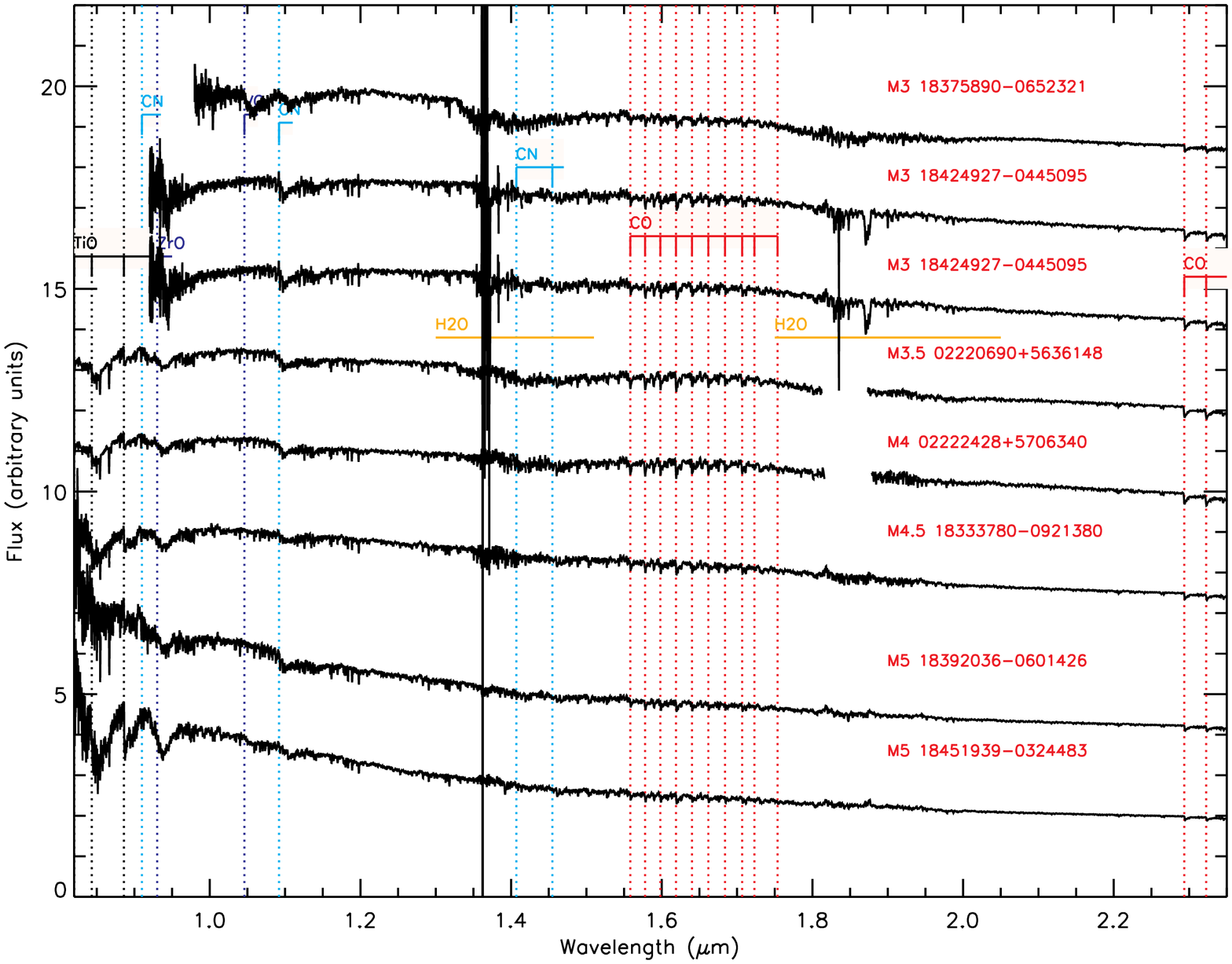}}
\caption{Continuation of Fig. \ref{fig.krsg} 1. IRTF spectra of known RSGs.}
\end{center}
\end{figure*}

\begin{figure*}
\begin{center}
\resizebox{\hsize}{!}{\includegraphics[angle=90]{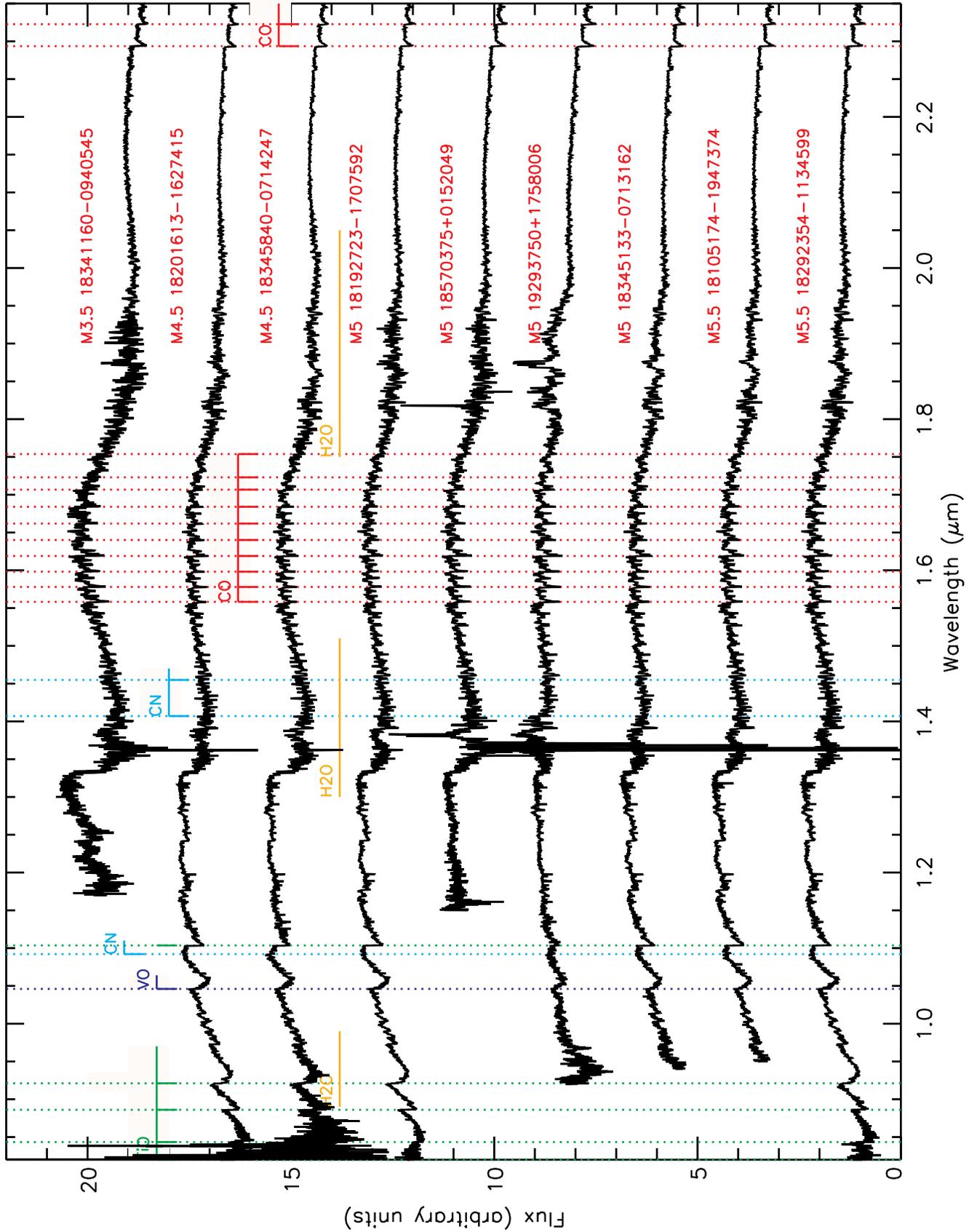}}
\caption{\label{fig.sp.mira} 2. IRTF spectra where the 
measured water absorption was found above 6\%.
13 from Table \ref{aliasnewrsg} and  two  from Table \ref{aliaskrsg} 
(2MASS J18345133$-$0713162  and  2MASS J18345840$-$0714247).
The spectra are sorted by adopted spectral type.}
\end{center}
\end{figure*}

\addtocounter{figure}{-1}
\begin{figure*}
\begin{center}
\resizebox{\hsize}{!}{\includegraphics[angle=90]{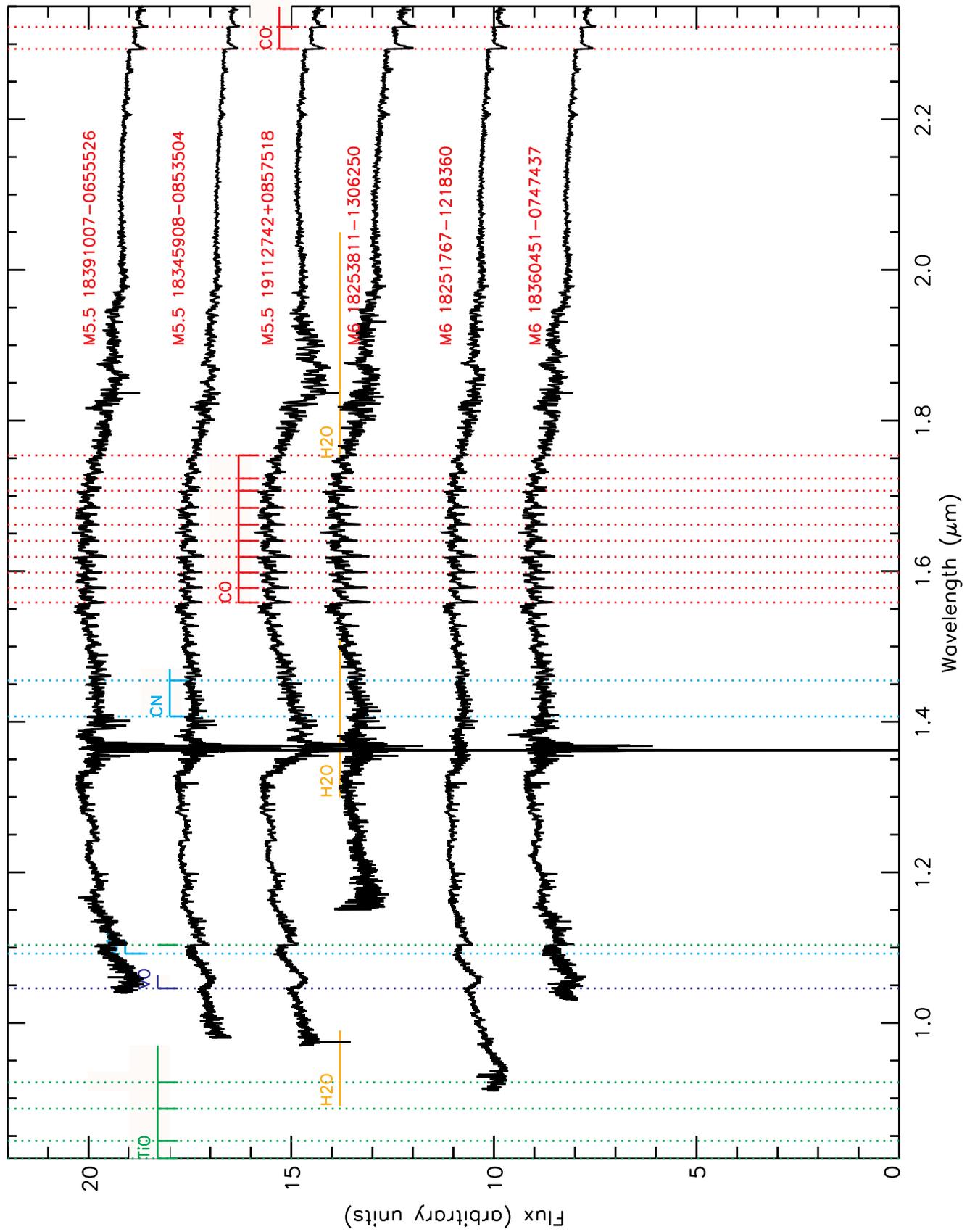}}
\caption{Continuation of Fig. \ref{fig.sp.mira} 2. 
Spectra with water absorption.}
\end{center}
\end{figure*}

\begin{figure*}
\begin{center}
\resizebox{\hsize}{!}{\includegraphics[angle=90]{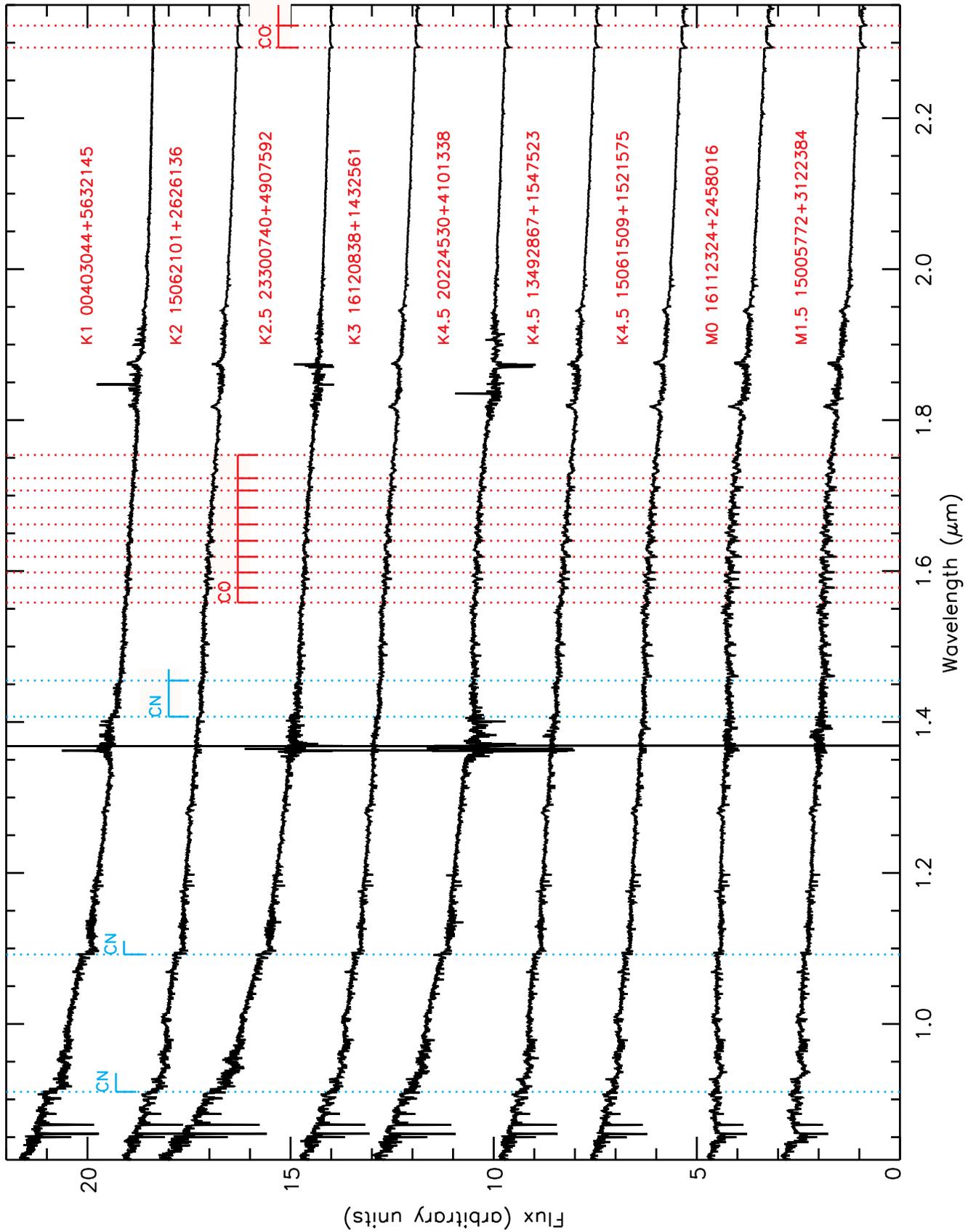}}
\caption{\label{fig.sp.giant} 3. IRTF spectra of giant stars
from Table \ref{aliasgiants}.
The spectra are sorted by adopted spectral type.}
\end{center}
\end{figure*}


\begin{figure*}
\begin{center}
\resizebox{\hsize}{!}{\includegraphics[angle=90]{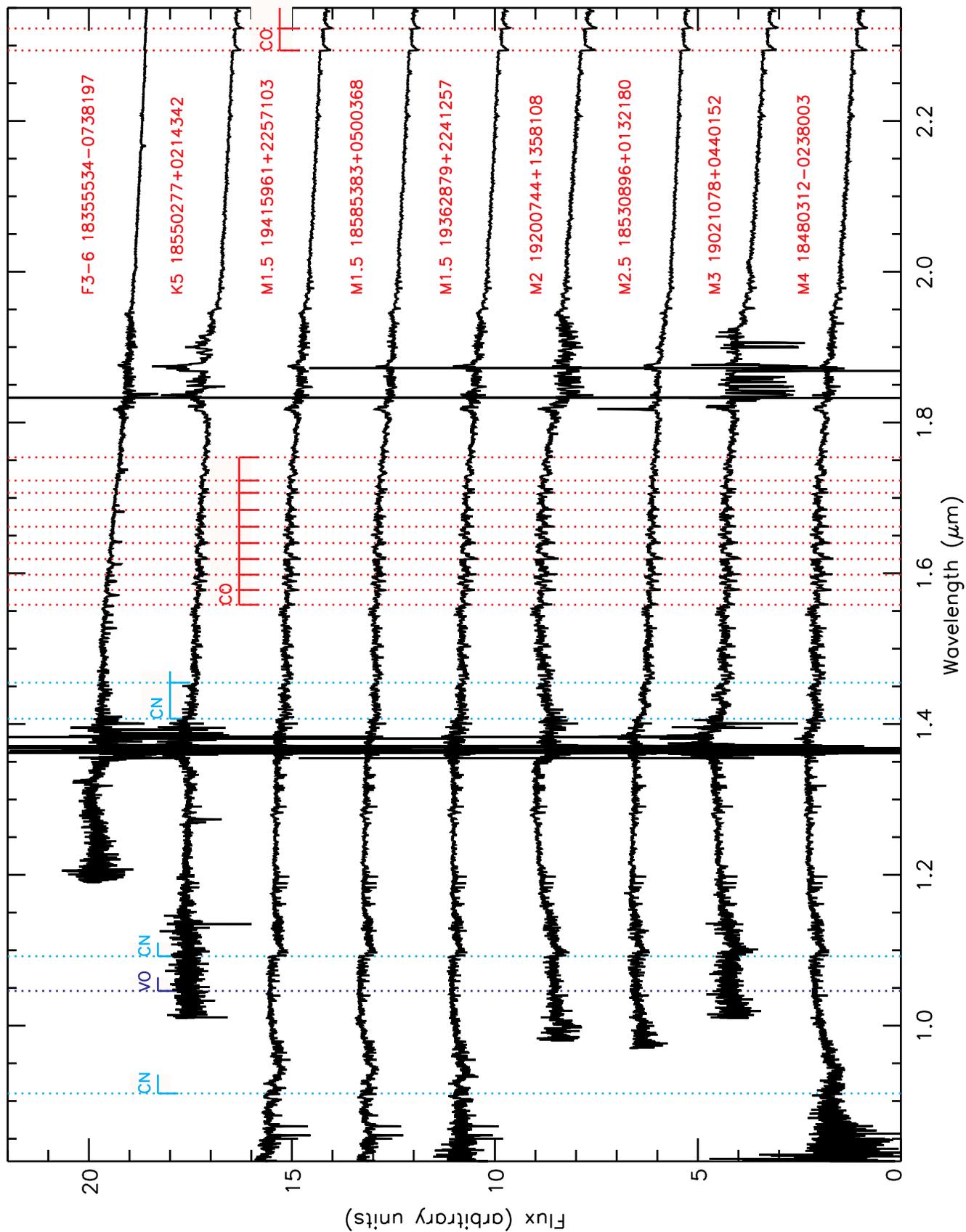}}
\caption{ \label{fig.new} 4. IRTF spectra of observed unknown targets
with \indwater\ smaller than 6\%  from Table \ref{aliasnewrsg}.
The spectra are sorted by adopted spectral type}.
\end{center}
\end{figure*}


\begin{figure*}
\begin{center}
\resizebox{\hsize}{!}{\includegraphics[angle=90]{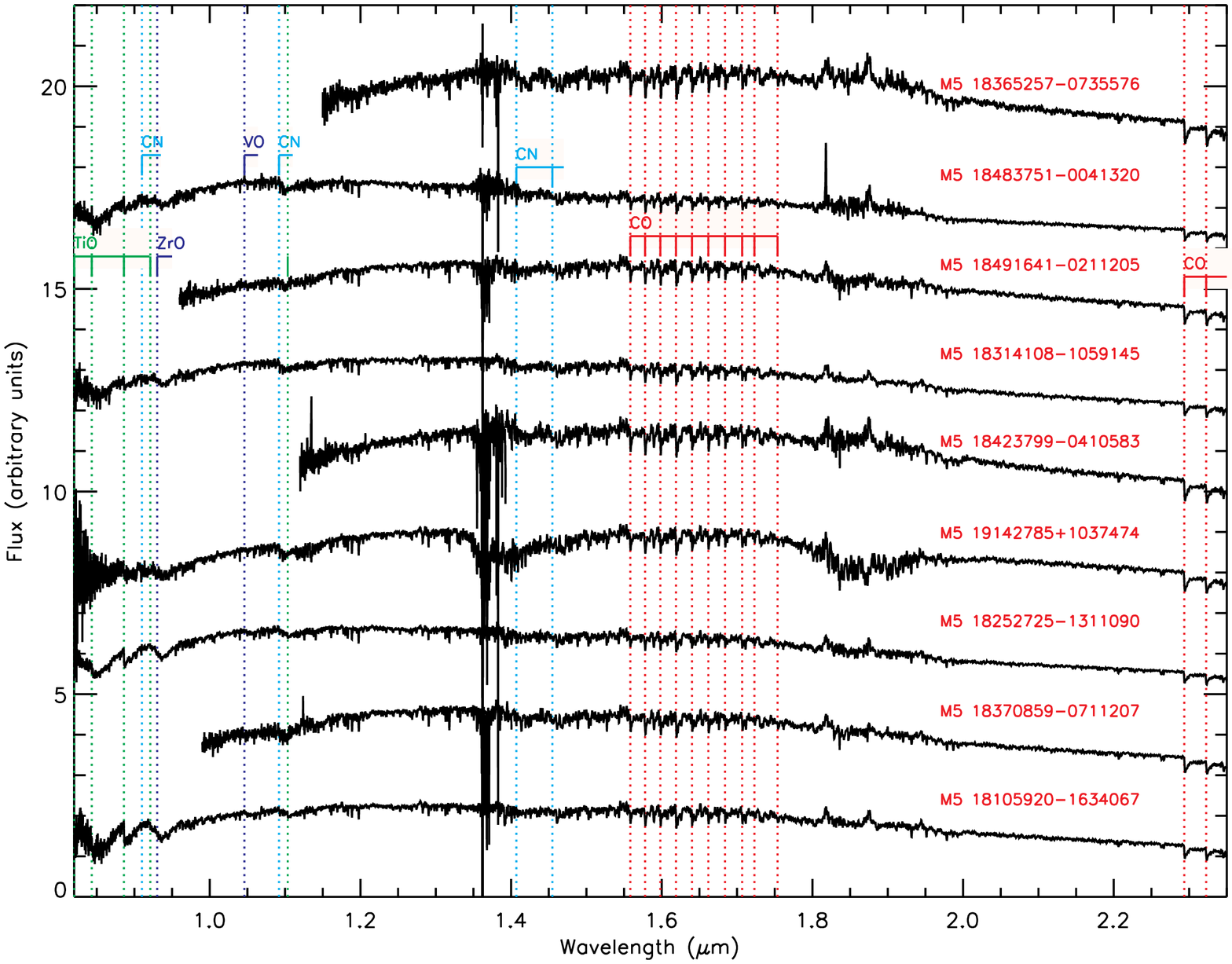}}
\caption{Continuation of Fig. \ref{fig.new} 4 (new targets).}
\end{center}
\end{figure*}


\addtocounter{figure}{-1}
\begin{figure*}
\begin{center}
\resizebox{\hsize}{!}{\includegraphics[angle=90]{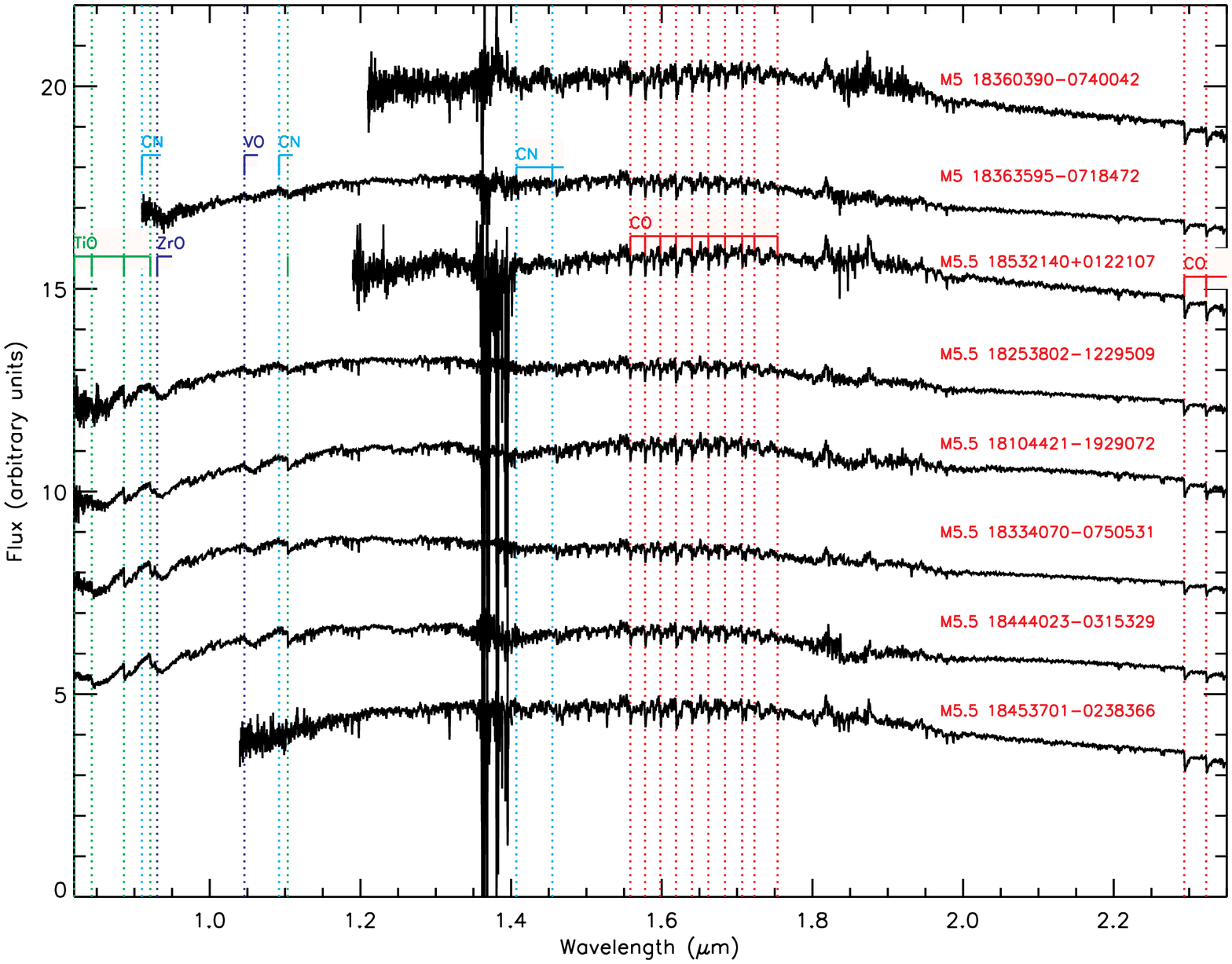}}
\caption{Continuation of Fig. \ref{fig.new} 4 (new targets).}
\end{center}
\end{figure*}


\addtocounter{figure}{-1}
\begin{figure*}
\begin{center}
\resizebox{\hsize}{!}{\includegraphics[angle=90]{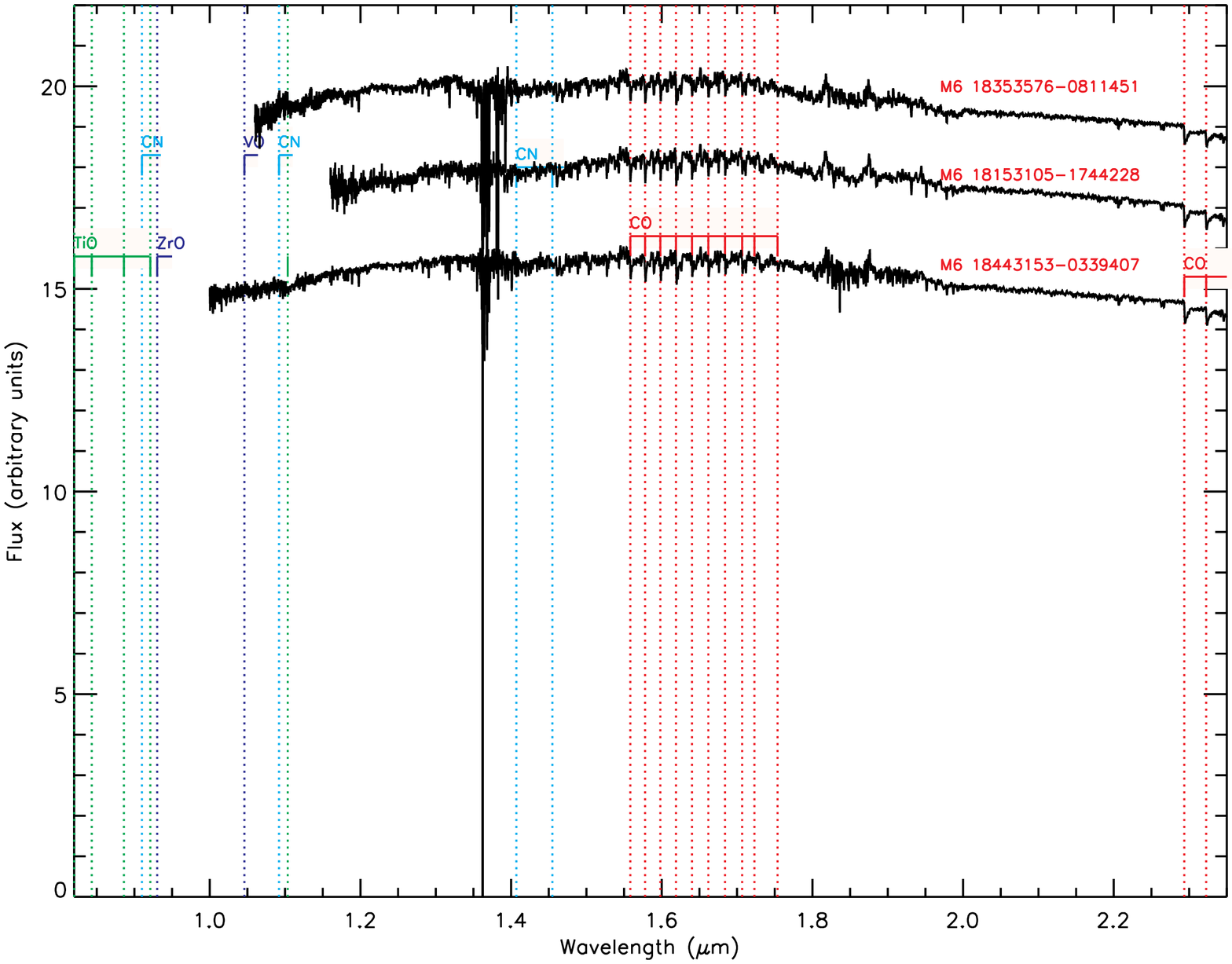}}
\caption{Continuation of Fig. \ref{fig.new} 4 (new targets).}
\end{center}
\end{figure*}

\begin{table*} 
\vspace*{-.5cm} 
\caption{\label{obstargets} Spectral indices of  K-M I stars from the IRTF libraries and newly observed K-M stars.}
{\tiny
\renewcommand{\arraystretch}{0.8}
\begin{tabular}{@{\extracolsep{-.06in}}llllrrrrrrrrrrrrrrrrrrrr}
\hline
\hline
ID    & Date-obs &\ewcau         & \ewcad            & \ewcat             &  \ewmg  &  \ewJuno & \ewJdue & \ewJtre &\ewJquattro&\ewJcinque&\ewJsei&\ewJsette&\ewJotto&\ewJnove&         \\
      &          & [\AA]  &  [\AA] & [\AA]  &  [\AA]  & [\AA]  &[\AA]  &  [\AA]  &  [\AA]  &    [\AA]  &      [\AA]            & [\AA]              & [\AA]               & [\AA]              \\  
      &          & [0.8498 \um]     & [0.8542 \um]       & [0.8662 \um]        & [0.8807  \um] & 1.0331 & 1.0346 & 1.0400  & 1.0588     & 1.0665 &1.0730 & 1.0755 & 1.0789 & 1.0874 &             \\ 
\hline
    00403044+5632145 &   2020$-$06$-$20 &    1.79 $\pm$    0.30 &   4.73 $\pm$    0.30 &   4.26 $\pm$    0.30 &   0.87 $\pm$    0.10 &   0.29 $\pm$    0.10 &   0.31 $\pm$    0.09 &   0.37 $\pm$    0.06 &   0.43 $\pm$    0.10 &   0.37 $\pm$    0.08 &   0.25 $\pm$    0.45 &   0.29 $\pm$    0.37 &   0.50 $\pm$    0.15 &   0.82 $\pm$    0.12 &  \\
    00511639+6148196 &   2020$-$06$-$20 &    1.68 $\pm$    0.40 &   5.44 $\pm$    0.40 &   5.09 $\pm$    0.39 &   1.24 $\pm$    0.18 &   0.49 $\pm$    0.06 &   0.44 $\pm$    0.05 &   0.70 $\pm$    0.07 &   0.53 $\pm$    0.08 &   0.75 $\pm$    0.10 &   0.73 $\pm$    0.55 &   0.34 $\pm$    0.40 &   0.72 $\pm$    0.18 &   0.73 $\pm$    0.22 &  \\
    01075303+6325116 &   2020$-$06$-$20 &    2.31 $\pm$    0.26 &   6.52 $\pm$    0.27 &   6.23 $\pm$    0.27 &   1.65 $\pm$    0.17 &   0.73 $\pm$    0.06 &   0.59 $\pm$    0.05 &   0.72 $\pm$    0.09 &   0.76 $\pm$    0.06 &   0.65 $\pm$    0.05 &   0.78 $\pm$    0.22 &   0.55 $\pm$    0.16 &   0.88 $\pm$    0.07 &   1.28 $\pm$    0.14 &  \\
\hline
\end{tabular}
}
\begin{list}{}
\item {\bf Notes:}
Errors in the equivalent widths are calculated following the 
prescription for weak lines 
by \citet{vollmann06}.~
\end{list}
\end{table*}

\addtocounter{table}{-1}
\begin{table*} 
\vspace*{-.5cm} 
\caption{Continuation of Table \ref{obstargets}.}
{\tiny
\renewcommand{\arraystretch}{0.8}
\begin{tabular}{@{\extracolsep{-.06in}}lrrrrrrrrrrrrrrrrrrrr}
\hline
\hline
ID    & Date-obs &\ewJdieci     &\ewJundici        &\ewJdodici      & \ewJutre            &  \ewJuquattro    &\ewJucinque          & \ewJusei       &\ewJusette&\ewJuotto     &\ewJunove &\ewJdzero &\ewJdu & \ewJdd &\\
      &          & [\AA]             & [\AA]                & [\AA]                & [\AA]                 & [\AA]          & [\AA]               & [\AA]                & [\AA]   &  [\AA]      &  [\AA]     &  [\AA]     &  [\AA]     &  [\AA]     &\\  
      &          & [1.0931 \um]&  [1.1596 \um]                 &[1.1613 \um]        & [1.1642 \um]          &[1.1694 \um]    &[1.1786 \um]         & [1.1832 \um]  & [1.1888 \um]  & [1.1977 \um] &[1.2037 \um]&[1.2086 \um]&[1.2826 \um]&[1.2904 \um] &\\
\hline
    00403044+5632145 &   2020$-$06$-$20 &    1.21 $\pm$    0.12 &   0.68 $\pm$    0.08 &   0.79 $\pm$    0.10 &   0.35 $\pm$    0.07 &   0.15 $\pm$    0.20 &   0.40 $\pm$    0.06 &   0.64 $\pm$    0.11 &   0.69 $\pm$    0.20 &   0.36 $\pm$    0.08 &   0.26 $\pm$    0.09 &   0.24 $\pm$    0.07 &   0.97 $\pm$    0.11 &   0.64 $\pm$    0.39 &  \\
    00511639+6148196 &   2020$-$06$-$20 &    1.22 $\pm$    0.22 &   0.74 $\pm$    0.06 &   1.15 $\pm$    0.07 &   0.54 $\pm$    0.04 &   0.62 $\pm$    0.14 &   0.81 $\pm$    0.10 &   1.08 $\pm$    0.11 &   1.38 $\pm$    0.20 &   0.87 $\pm$    0.06 &   0.39 $\pm$    0.10 &   0.46 $\pm$    0.05 &   1.54 $\pm$    0.14 &   0.83 $\pm$    0.38 &  \\
    01075303+6325116 &   2020$-$06$-$20 &    2.05 $\pm$    0.14 &   0.80 $\pm$    0.08 &   1.30 $\pm$    0.11 &   0.63 $\pm$    0.07 &   0.70 $\pm$    0.20 &   0.83 $\pm$    0.06 &   1.20 $\pm$    0.13 &   1.70 $\pm$    0.25 &   1.13 $\pm$    0.10 &   0.71 $\pm$    0.09 &   0.58 $\pm$    0.03 &   1.85 $\pm$    0.13 &   1.06 $\pm$    0.48 &  \\
\hline
\end{tabular}
}
\end{table*}

\addtocounter{table}{-1}
\begin{table*}
\vspace*{-.5cm} 
\caption{Continuation of Table \ref{obstargets}.}
{\tiny
\renewcommand{\arraystretch}{0.8}
\begin{tabular}{@{\extracolsep{-.06in}}lrrrrrrrrrrrrrrrrrr}
\hline
\hline
ID    & Date-obs & \ewJdtre& \ewJdquattro & \ewJdcinque& \ewcoucs   &   \ewsiucn      & \ewcousd            & \ewsiusq              & \ewal                 & \ewmg               & \ewna                &  \ewca               & \ewcodue      & \indwater\\
      &          & [\AA] &      [\AA]      &  [\AA]     &  [\AA] &          [\AA]     & [\AA]                & [\AA]                & [\AA]                 & [\AA]               & [\AA]                & [\AA]                & [\AA]             & [\%]\\  
      &          &[1.2979 \um]& [1.3127 \um]&  [1.3154 \um] &  [1.5783 um] &[1.5890 \um]  & [1.61975 \um]        & [1.6445 \um]         & [1.6762  \um]         & [1.7115 \um]        & [2.2075 \um]         & [2.2635 \um]         & [2.2935 \um]      &\\  
\hline
    00403044+5632145 &   2020$-$06$-$20 &    0.56 $\pm$    0.23 &   0.51 $\pm$    0.08 &   0.76 $\pm$    0.08 &   0.59 $\pm$    1.12 &   3.38 $\pm$    0.32 &   1.35 $\pm$    0.38 &   1.56 $\pm$    0.35 &  $-$0.01 $\pm$    0.79 &   0.45 $\pm$    0.58 &   2.31 $\pm$    0.30 &   1.08 $\pm$    0.71 &  20.32 $\pm$    1.90 &     5.3 \\
    00511639+6148196 &   2020$-$06$-$20 &    0.58 $\pm$    0.27 &   0.08 $\pm$    0.18 &   0.38 $\pm$    0.18 &   3.36 $\pm$    1.67 &   4.18 $\pm$    0.46 &   5.14 $\pm$    0.45 &   3.15 $\pm$    0.41 &   0.40 $\pm$    0.68 &   2.05 $\pm$    0.47 &   4.04 $\pm$    0.30 &   2.69 $\pm$    0.60 &  38.84 $\pm$    1.73 &     2.1 \\
    01075303+6325116 &   2020$-$06$-$20 &    0.35 $\pm$    0.16 &   0.57 $\pm$    0.15 &   0.79 $\pm$    0.15 &   3.47 $\pm$    1.75 &   4.52 $\pm$    0.44 &   4.95 $\pm$    0.45 &   2.86 $\pm$    0.42 &   0.57 $\pm$    0.70 &   2.10 $\pm$    0.58 &   3.56 $\pm$    0.31 &   2.70 $\pm$    0.66 &  41.16 $\pm$    1.69 &     2.5 \\
\hline
\end{tabular}
}
\end{table*} 

\end{appendix}

\end{document}